\documentclass[12pt, a4paper]{report}

\usepackage{fancyhdr}
\fancypagestyle{plain}{%
\fancyhead{} 
}

\usepackage[latin1]{inputenc}
\usepackage{amsmath}
\usepackage{amsfonts}
\usepackage{amssymb}
\usepackage{makeidx}
\usepackage[dvips,dvipsnames,usenames]{color}

\usepackage{graphicx}

\usepackage{varioref}

\usepackage{multirow}
\usepackage{epsfig}
\usepackage{rotating}
\usepackage{colortbl}

\usepackage{mathrsfs}
\usepackage{eucal}

\usepackage{amsmath,amssymb}

\usepackage{packages/feynmp}

\usepackage{packages/tautitle}

\usepackage{caption}
\usepackage{keyval}
\usepackage{subfig}

\newcommand{\Subref}[1]{\protect\subref{#1}}

\usepackage[dvips, breaklinks={true}, bookmarks, colorlinks={true}, pdfpagemode={NON}, pdffitwindow=true, pdfstartview={FitH}, linkcolor={BlueViolet}, menucolor={BlueViolet}, citecolor={Mahogany}, filecolor={OliveGreen}, pagecolor={Mahogany}, urlcolor={MidnightBlue}, pdftitle={Iftach Sadeh - M.Sc.}, pdfauthor={Iftach Sadeh}, pdfsubject={M.Sc.}, pdfkeywords={HEP, ILC, Luminosity, Clustering}]{hyperref}

\usepackage{packages/breakurl}

\usepackage[all]{hypcap}

\labelformat{equation}{\textup{(#1)}}
\labelformat{enumi}{\textup{(#1)}}

\let\orgautoref\autoref
\providecommand{\Autoref}
        {\def\equationautorefname{Equation}%
         \def\figureautorefname{Figure}%
         \def\subfigureautorefname{Figure}%
         \def\sectionautorefname{Section}%
         \def\subsectionautorefname{Section}%
         \def\subsubsectionautorefname{Section}%
         \def\Itemautorefname{Item}%
         \def\tableautorefname{Table}%
         \orgautoref}

\renewcommand{\autoref}
        {\def\equationautorefname{Eq.}%
         \def\figureautorefname{Fig.}%
         \def\subfigureautorefname{Fig.}%
         \def\sectionautorefname{Sect.}%
         \def\subsectionautorefname{Sect.}%
         \def\subsubsectionautorefname{Sect.}%
         \def\Itemautorefname{item}%
         \def\tableautorefname{Table}%
         \orgautoref}

\providecommand{\autorefs}
        {\def\equationautorefname{Eqs.}%
         \def\figureautorefname{Figs.}%
         \def\subfigureautorefname{Figs.}%
         \def\sectionautorefname{Sects.}%
         \def\subsectionautorefname{Sects.}%
         \def\subsubsectionautorefname{Sects.}%
         \def\Itemautorefname{items}%
         \def\tableautorefname{Tables}%
         \orgautoref}

\newcommand{\abs}[1] {\lvert #1 \lvert}

\newcommand{\mc}[1]{\mathcal{#1}}

\renewcommand{\vec}[1]{\boldsymbol{#1}}

\begin{document}


\title{\bf{Luminosity Measurement at the International Linear Collider}}

\author{
\large\bf{Iftach Sadeh}\thanks{
The research work for this thesis was carried out in the Experimental High Energy Physics Group of the Tel Aviv University under the supervision of Prof.\ {\large\bf{Halina Abramowicz}} and Prof.\ {\large \bf{Aharon Levy}}.
}
}
\date{April 2008}

\maketitle 
\setcounter{page}{1}	\pagenumbering{roman}

\begin{abstract}

The International Linear Collider (ILC) is a proposed electron-positron collider with a center-of-mass energy of 500~GeV, and a peak luminosity of $2 \cdot 10^{34}~\mathrm{cm}^{-2}\mathrm{s}^{-1}$. The ILC will complement the Large Hadron Collider, a proton-proton accelerator, and provide precision measurements, which may help in solving some of the fundamental questions at the frontier of scientific research, such as the origin of mass and the possible existence of new principles of nature.

The linear collider community has set a goal to achieve a precision of $10^{-4}$ on the luminosity measurement at the ILC. This may be accomplished by constructing a finely granulated calorimeter, which will measure Bhabha scattering at small angles. The Bhabha cross-section is theoretically known to great precision, yet the rate of Bhabha scattering events, which would be measured by the luminosity detector, will be influenced by beam-beam effects, and by the inherent energy spread of the collider. The electroweak radiative effects can be calculated to high precision and partially checked with events with final state photon radiation  by distinguishing between the observable energy deposits of electrons and of photons in the luminosity calorimeter, using a clustering algorithm.

In order to achieve the design goal, the geometrical parameters of the calorimeter need to be reevaluated. This must be done in a generalized manner, so as to facilitate future modifications, the need for which is foreseen, due to expected changes in the detector concept.

This work demonstrates that the clustering approach is viable, and that a luminosity calorimeter may be designed to match the precision requirements on the luminosity measurement.

\end{abstract}

\tableofcontents  \newpage
\pagenumbering{arabic}	\setcounter{page}{1}

\chapter{Introduction \label{introductionCH}}

\section{The International Linear Collider}

\subsection{Physics Case of the ILC}

The triumph of $20^{\mathrm{th}}$ century particle physics was the development of the Standard Model. Experiments determined the particle constituents of ordinary matter, and identified four forces binding matter and transforming it from one form to another. This success leads particle physicists to address even more fundamental questions, such as the existence of undiscovered principles of nature, the unification of the four forces, the nature of dark matter and dark energy, and the possible existence of extra dimensions. The International Linear Collider (ILC)~\cite{introductionBIB6} is expected to play a central role in an era of revolutionary advances with breakthrough impact on many of these fundamental questions~\cite{introductionBIB1}.

\begin{figure}[htp]
\begin{center}
\includegraphics[width=.99\textwidth]{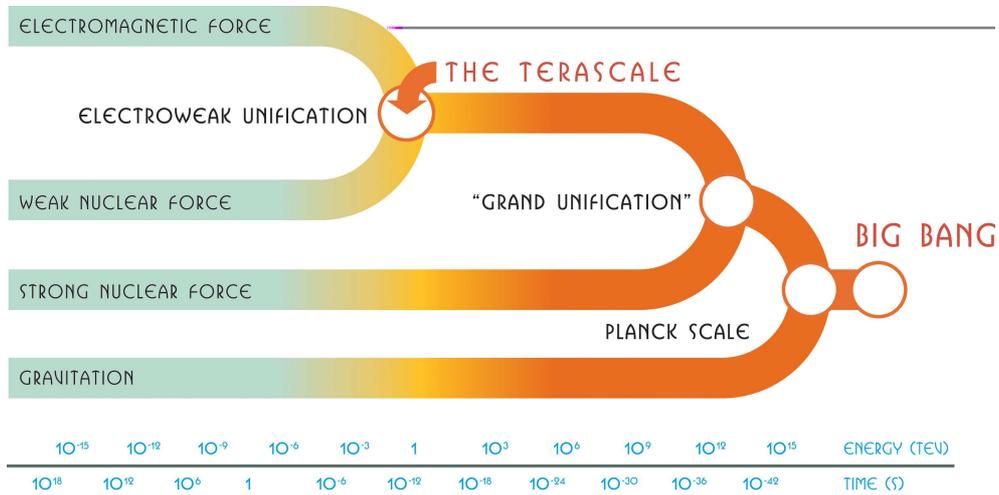}
\caption{\label{energyScaleFlowFIG}The electromagnetic and weak nuclear forces unify at the Terascale.}
\end{center}
\end{figure}

The Standard Model includes a third component beyond particles and forces that has not yet been verified, the Higgs mechanism that gives mass to the particles. Many scientific opportunities for the ILC involve the Higgs particle and related new phenomena at Terascale energies. The Standard Model Higgs field permeates the universe, giving mass to elementary particles, and breaking a fundamental electroweak force into two, the electromagnetic and the weak forces, as is illustrated in \autoref{energyScaleFlowFIG}. Due to quantum effects, the Higgs cannot be stable at Terascale energies. As a result, new phenomena, such as a new principle of nature called supersymmetry, or extra space dimensions, must be introduced. If the Higgs exists, its properties need to be investigated, and if not, then the mechanism responsible for the breaking of the electroweak force must be explained.

During the next few years, experiments at the CERN Large Hadron Collider (LHC), a 14~TeV proton-proton collider, will have the first direct look at Terascale physics. If there is a Higgs boson, it is almost certain to be found at the LHC and its mass measured by the ATLAS and CMS experiments~\cite{introductionBIB2,introductionBIB3}. If there is a multiplet of Higgs bosons, there is a good chance the LHC experiments will see more than one. However, it will be difficult for the LHC to measure the spin and parity of the Higgs particle, and thus to establish its essential nature. On the other hand, the ILC, an $e^{+}e^{-}$ collider, can make these measurements accurately, due to the point-like structure of the beam particles. If there is more than one decay channel of the Higgs, the LHC experiments will determine the ratio of branching fractions (with an accuracy roughly of $7-30\%$); the ILC will measure these couplings to quarks and vector bosons at the few percent level, and thus reveal whether the Higgs is the simple Standard Model object, or something more complex.

If supersymmetry is responsible for stabilizing the electroweak unification at the Terascale and for providing a light Higgs boson, signals of superpartner particles should be seen at the LHC. The task would then be to deduce the properties of this force, its origins, its relation to the other forces in a unified framework, and its role in the earliest moments of the universe. This will require precise measurements of the superpartner particles and of the Higgs particles, and will require the best possible results from the LHC and the ILC in a combined analysis.

Alternative possible structures of the new physics include phenomena containing extra dimensions, introducing connections between Terascale physics and gravity. One possibility is that the weakness of gravity could be understood by the escape of gravitons into the new large extra dimensions. Events with unbalanced momentum caused by the escaping gravitons could be seen at both the LHC and the ILC. Additionally, the ILC could confirm this scenario by observing anomalous electron-positron pair production caused by graviton exchange. The advantage of the LHC is a large energy coverage. However, in all cases, the ILC is a better tool for understanding the background to new physics.

\subsection{Design of the ILC}

The ILC is designed to achieve the specifications listed in the ILCSC Parameter Subcommittee Report~\cite{introductionBIB7}. The three most important requirements are

\begin{list}{-}{}
\item
an initial center-of-mass energy, $\sqrt{s}$, of up to 500 GeV, with the ability to upgrade to 1 TeV ,
\item
a peak luminosity of $2 \cdot 10^{34}~\mathrm{cm}^{-2}\mathrm{s}^{-1}$ with $75\%$ availability, resulting in an integrated luminosity in the first four years of 500~$\mathrm{fm}^{-1}$ at $\sqrt{s}=500$~GeV, or equivalent at lower energies, and
\item
the ability to scan the energy range $200 < \sqrt{s} < 500$~GeV.
\end{list}

\noindent Additional physics requirements are electron beam polarization $> 80\%$, an energy stability and precision $\le 0.1\%$, an option for $\sim 60\%$ positron beam polarization, and alternative $e^{-}e^{-}$ and $\gamma \gamma$ collisions.

\begin{figure}[htp]
\begin{center}
\includegraphics[height=0.95\textheight]{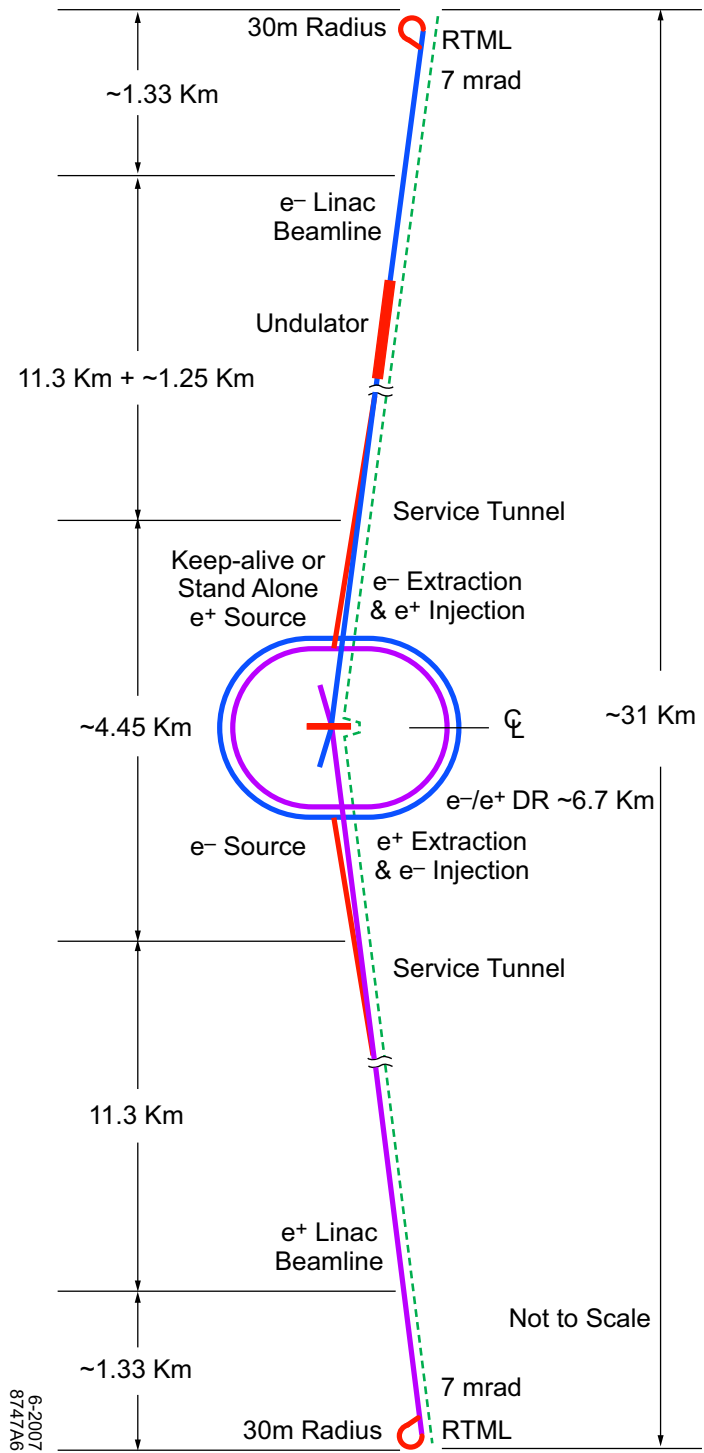}
\caption{\label{ILCdesignFIG}Schematic design of the ILC for a center of mass energy $\sqrt{s} = 500$~GeV.}
\end{center}
\end{figure}

The ILC Reference Design Report~\cite{introductionBIB8} describes a collider that is intended to meet these requirements. \Autoref{ILCdesignFIG} shows its schematic design. In this concept the ILC is based on 1.3~GHz superconducting radio-frequency cavities operating at a gradient of 31.5~MV/m~\cite{revisedDetectorModelBIB4}. The collider operates at a repetition rate of 5~Hz with a beam pulse length of roughly 1~msec. The site length is 31 km for $\sqrt{s} = 500$~ GeV, and would have to be extended to reach 1~TeV. The beams are prepared in low energy damping rings that operate at 5~GeV and are 6.7~km in circumference. They are then accelerated in the main linacs which are $\sim$~11~km per side. Finally, the beams are focused down to very small spot sizes at the collision point with a beam delivery system that is $\sim$~2.2~km per side.

\section{The Detector Concept}

Presently, two detectors are considered, with the pull-and-push scheme. With\-in the next half a year, letters of intent are expected. The European high energy committee, which have been working on the fifth version (v5) of the so-called ``Large Detector Concept'' (LDC)~\cite{introductionBIB25}, has recently joined forces with the Japanese and American communities to promote the International Large Detector (ILD)~\cite{introductionBIB26} concept. Central to the detector design are a micro-vertex detector, a time projection chamber (TPC) tracking device, and calorimetry with very fine granularity, in order to reconstruct the particle flow for best jet-energy resolution. A schematic representation of the LDC(v5) detector concept is presented in \autoref{forwardRegionLayout1FIG}.

\subsubsection{The Forward Region of the ILC}

The instrumentation of the forward region aims at measuring the luminosity, providing electron veto at small angles, as well as a fast beam monitoring system, and complementing the hermeticity of the full detector.

\begin{figure}[htp]
\begin{center}
\includegraphics[width=.95\textwidth , height=0.4\textheight]{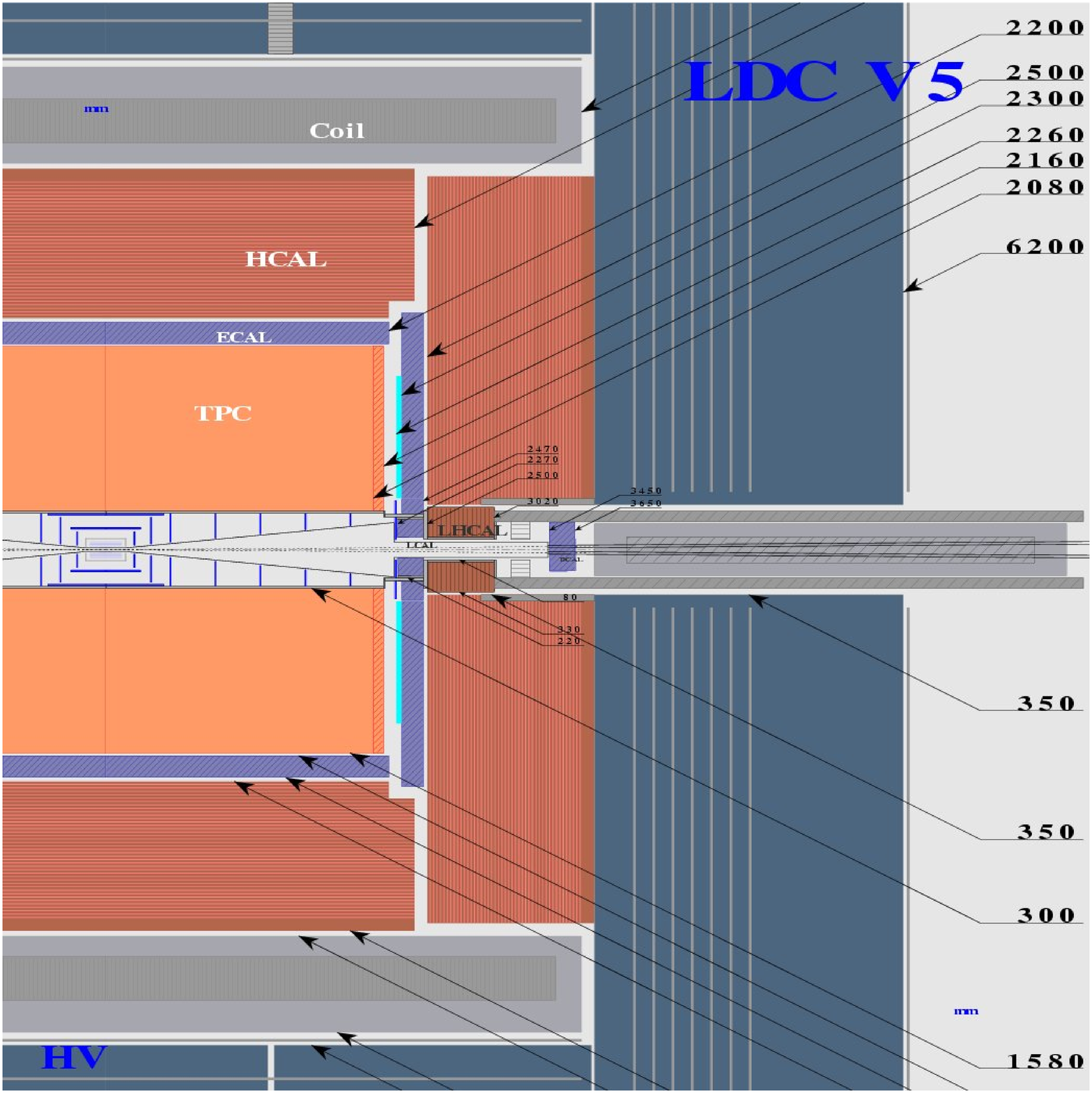}
\caption{\label{forwardRegionLayout1FIG}Schematic design of the LDC(v5) detector concept.}
\vspace{20pt}
\includegraphics[width=.95\textwidth , height=0.4\textheight]{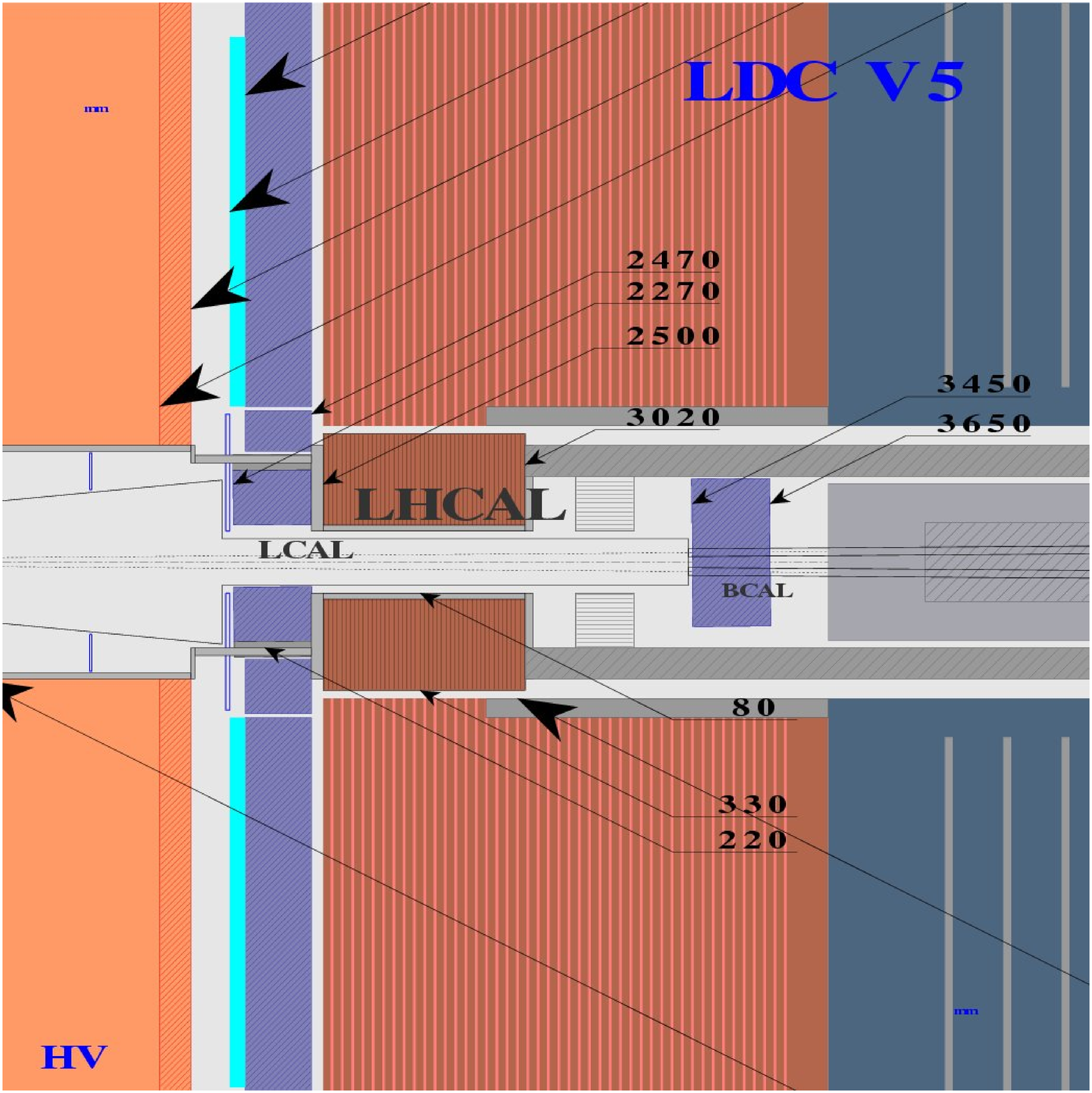}
\caption{\label{forwardRegionLayout2FIG}Schematic design of the forward region of the ILC in the LDC(v5) detector concept.}
\end{center}
\end{figure}

The following sub-systems comprise the forward region of the ILC detector: the luminosity detector (LumiCal) for precise measurement of the Bhabha event rate; the beam calorimeter (BeamCal) and the beamstrahlung photons monitor (GamCal), for providing a fast feed-back in tuning the luminosity. BeamCal is also intended to support the determination of beam parameters. Both LumiCal and BeamCal extend the angular coverage of the electromagnetic calorimeter to small polar angles~\cite{introductionBIB5}. The layout of the forward region is depicted in \autoref{forwardRegionLayout2FIG}.

The requirement for LumiCal is to enable a measurement of the integrated luminosity with a relative precision of about $10^{-4}$. The use of Bhabha scattering as the gauge process is motivated by the fact that the cross-section is large and dominated by electromagnetic processes, and thus can be calculated with very high precision. The purpose of the BeamCal is to efficiently detect high  energy electrons and photons produced e.g. in low transverse momentum QED processes, such as Bhabha scattering and photon-photon events. BeamCal is important in order to suppress this dominant background in many searches for new particles predicted in scenarios for physics beyond the Standard Model. In the polar angle range covered by the BeamCal, typically 5 to 45~mrad, high energy electrons must be detected on top of wider spread depositions of low energy $e^{+}e^{-}$ pairs, originating from beamstrahlung photon conversions. The measurement of the total energy deposited by these pairs, bunch by bunch, can be used to monitor the variation in luminosity and provide a fast feedback to the beam delivery system. Moreover, the analysis of the shape of the energy flow can be used to extract the parameters of the colliding beams. This information can be further used to optimize the machine operation. GamCal is used to analyze beamstrahlung photons. It will be positioned at a distance of about 180~m from the interaction point. It will be sensitive to the energy of the beamstrahlung photon and to the size of the beamstrahlung photon cone, which in turn is sensitive to the beam parameters.

\section{Work Scope \label{workScope}}

The focus of this thesis is the design and performance of the luminosity calorimeter. The objective is to demonstrate that it is possible to design a LumiCal, such that the relative error on the luminosity measurement meets the performance requirements. The way to accomplish this is twofold.

On the one hand, it must be proven that the Bhabha process, which is the benchmark process for measuring luminosity at the ILC, may be measured directly. This is important for several reasons. For one, the required precision with which the Bhabha process needs to be know, is in par with the current theoretical uncertainty. Further more, one has to take into account that the energy of the colliding beams is not monochromatic. This is mainly due to beamstrahlung radiation, energy loss by the incoming positron (electron) due to its interaction with the electron (positron) bunch moving in the opposite direction. In addition, the physical Bhabha cross-section is affected by electroweak radiative effects, which have been calculated to a precision of $5.3 \cdot 10^{-4}$ on the $Z$ resonance~\cite{introductionBIB27}. A measurement of the differential Bhabha cross-section itself would serve as a control mechanism for checking the calculations in the region of high momentum photon emission, which is subject to the largest corrections. This is done by way of performing clustering in LumiCal, and measuring the distribution of the (final state) radiative Bhabha photons, on top of the electron distribution.

The second requirement that needs to be addressed is the measurement of the integrated luminosity in LumiCal. The design of LumiCal must balance between oftentimes contradicting constraints. On one hand, one would like to improve the precision of the luminosity measurement. On the other hand, other considerations need to be taken into account, such as minimizing the material budget, ensuring the viability of the readout etc. Due to the fact that R$\&$D efforts are continuing, the detector concept is still fluid. Consequently, the restrictions on the size and positioning of LumiCal may change every few months. It is, therefore, necessary to define a clear procedure, that will allow for adjustment of the design parameters of LumiCal, while keeping its performance within the requirements. A study is presented here, in which LumiCal is optimized to this effect. The purpose is both to arrive at the best design of the calorimeter, under the present constraints imposed by the detector concept, and to show how such an optimization should take place.

This work includes a theoretical introduction (\autorefs{lumiCalCH} and \ref{clorimetryCH}), an overview of the clustering algorithm which has been developed for LumiCal, and of its performance (\autoref{clusteringCH})\footnotemark, an optimization study of the design of LumiCal (\autoref{revisedDetectorModelCH}), and finally, a summary of the results (\autoref{summaryCH}).

\footnotetext{A full description of the clustering algorithm is presented in (\autoref{clusteringAPP}).}

\chapter{Luminosity\label{lumiCalCH}}

To measure the cross-section, $\sigma$, of a certain process we count the number of events, $N$, registered in the detector, and obtain $\sigma$ using the corresponding integrated luminosity, $\mc{L}$, according to the relation

\begin{equation}
\sigma = \frac{N}{\mc{L}} .
\label{luminosityByXsCountingEQ} \end{equation}

\noindent Neglecting other systematic uncertainties, the required precision on the luminosity measurement is given by the statistics of the highest cross-section processes which is measured.

\section{Luminosity Measurement at the ILC}

\subsection{Precision Requirements on the Luminosity Measurement \label{precisionLumiRequirementsSEC}}

At $\sqrt{s} = 340$~GeV the cross-section for $e^{+}e^{-} \rightarrow W^{+}W^{-}$ is about 10~pb, and the one for fermion pairs, $e^{+}e^{-} \rightarrow q^{+}q^{-}$, is about 5~pb, both scaling with $1/s$. In both processes one, therefore, expects event samples of $\mc{O}(10^{6})$ events in a few years of running, which would require a luminosity precision, $\Delta\mc{L}/\mc{L}$, of better than $10^{-3}$ (see \autoref{luminosityRelativeErrStatEQ} below).

\begin{figure}[htp]
\begin{center}
\subfloat[]{\label{GigaZFIG1}\includegraphics[width=0.49\textwidth]{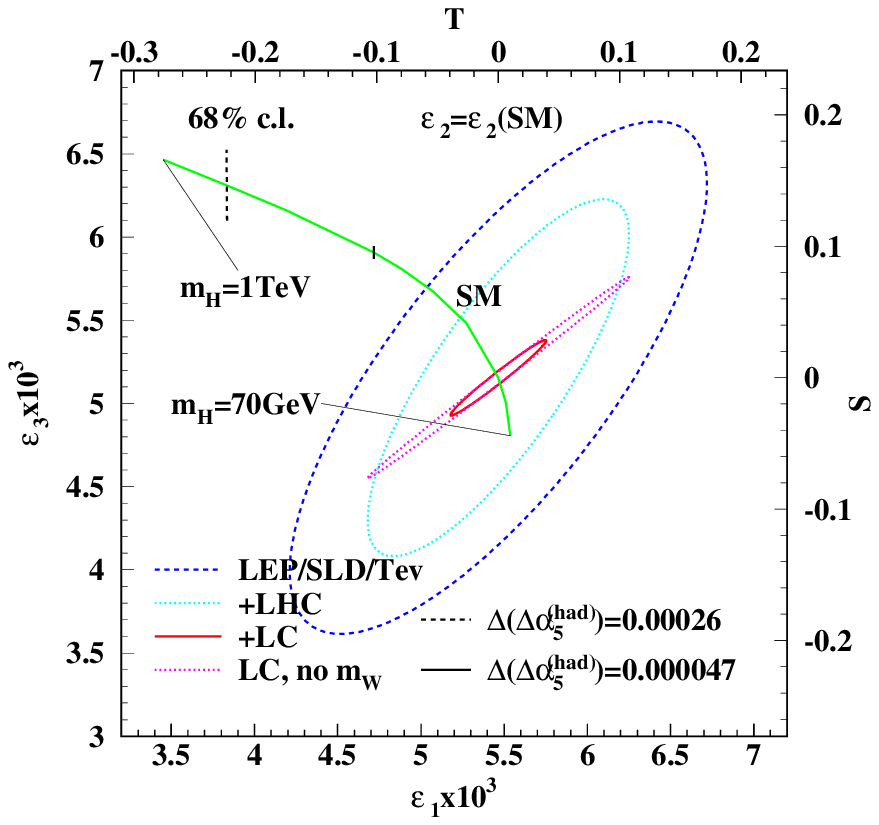}}
\subfloat[]{\label{GigaZFIG2}\includegraphics[width=0.39\textwidth]{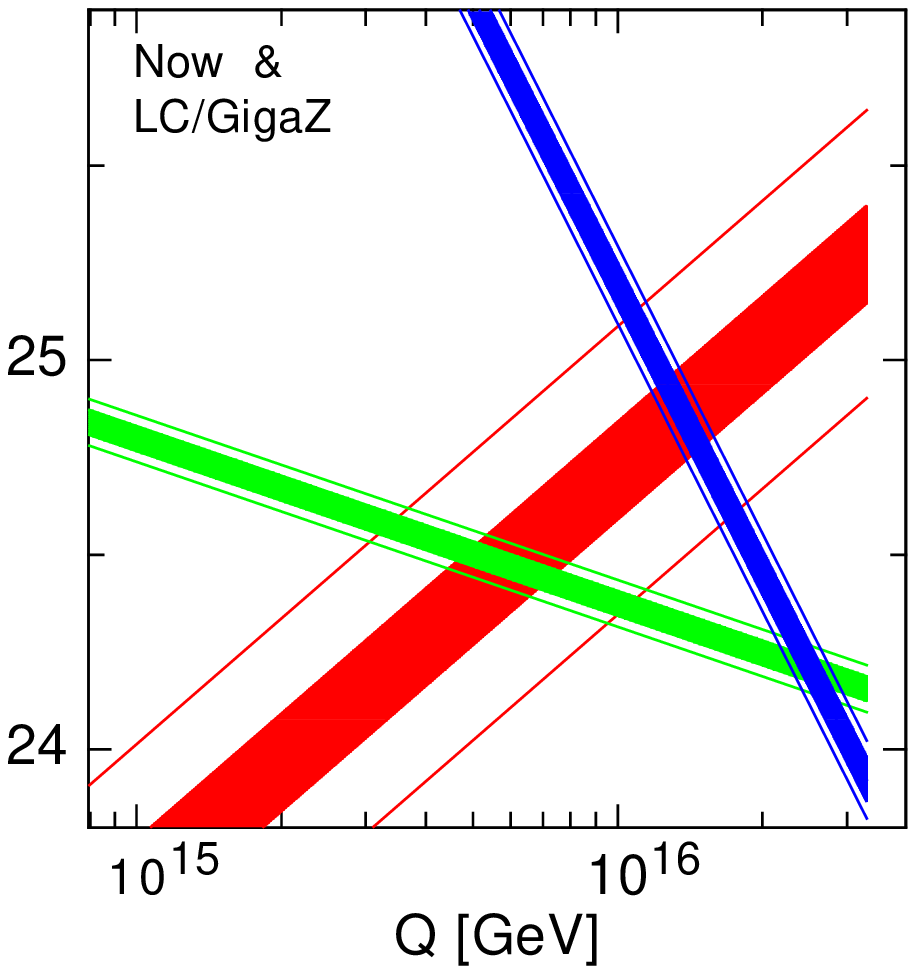}}
\caption{\label{GigaZFIG}\Subref{GigaZFIG1} Presently known precision for $\varepsilon_{1,3}$, expected after the LHC and after the ILC measurements. All curves, apart from the one denoted $~\mathrm{``LC,~no~m_{W}"}$, assume that $\varepsilon_{2}$ is equal to its SM value. \Subref{GigaZFIG2} Unification of couplings now and after GigaZ.}
\end{center}
\end{figure}

The GigaZ program requires running the collider at an energy corresponding to the $Z$ pole. The ILC is designed to reach very high luminosity in this mode of operation, and thus can become a very powerful laboratory for advancing the tests of the SM, which have been performed at LEP/SLC, to a new level of accuracy. The goal of the GigaZ run is a test of the radiative corrections to the $Z$-fermion couplings with extremely high precision. In general these radiative corrections can be parametrized in terms of three parameters,  $\varepsilon_{1,2,3}$~\cite{introductionBIB9}. \autoref{GigaZFIG1} shows the expected precision on $\varepsilon_{1,3}$ under different assumptions ~\cite{introductionBIB10}. These two parameters can be obtained from the $Z$-observables alone while $\varepsilon_{2}$ needs in addition a measurement of the $W$-mass. Another task at GigaZ is the measurement of the strong coupling constant, $\alpha_{s}$, which can be obtained from the ratio of hadronic to leptonic $Z$ decays to a precision of $5-7 \cdot 10^{-4}$. Tests of grand unification are limited by the knowledge of the strong coupling constant, as depicted in \autoref{GigaZFIG2}. Some models, e.g. within string theory, predict small deviations from unification, thus making this measurement very important. GigaZ is also especially interesting if no direct evidence for physics beyond the Standard Model is found. In this case the structure of radiative corrections should be tested without artificial constraints.

In the GigaZ mode, more than $10^{9}$ hadronic $Z$ decays are expected, which would in principle require a luminosity precision of roughly $10^{-5}$. However there are other systematic uncertainties that play a role. These include the selection efficiency for hadronic events and the modification of the cross-section on top of the Breit-Wigner resonance, due to the beam energy spread. Hence, a luminosity precision of $\Delta\mc{L}/\mc{L} \sim 10^{-4}$ seems adequate~\cite{introductionBIB28}.

\subsection{Bhabha Scattering as the Gauge Process \label{bhabhaScatteringIntroSEC}}

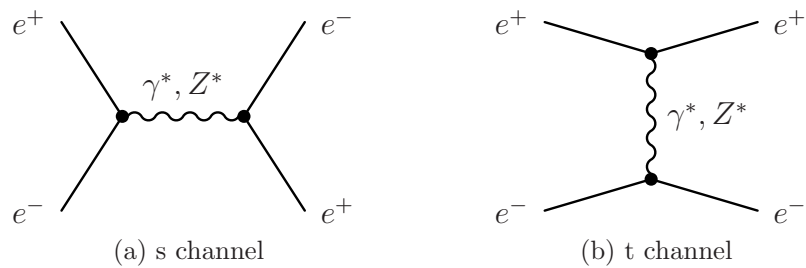
\begin{figure}[htb]
  \begin{fmffile}{pics/introduction/feynBhabha/bhabha}\hfill%
    \subfloat[s channel]{\label{bhabhaScatteringDiagramFIG1}%
      \fmfframe(5,0)(5,2){
        \begin{fmfgraph*}(40,25)
          \fmfleft{i1,i2} \fmfright{o1,o2}
          \fmfv{label=$e^-$,label.angle=180}{i1}   \fmfv{label=$e^+$,label.angle=180}{i2}
          \fmf{plain}{i1,v1,i2}
          \fmf{wiggly,label=$\gamma^\ast,,Z^\ast$}{v2,v1}    \fmfdot{v1,v2}
          \fmf{plain}{o1,v2,o2}
          \fmfv{label=$e^+$,label.angle=0}{o1}   \fmfv{label=$e^-$,label.angle=0}{o2}
        \end{fmfgraph*}
      }}\hfill%
    \subfloat[t channel]{\label{bhabhaScatteringDiagramFIG2}%
      \fmfframe(5,0)(5,2){%
        \begin{fmfgraph*}(35,25)
          \fmfleft{i1,i2} \fmfright{o1,o2}
          \fmfv{label=$e^-$,label.angle=180}{i1}   \fmfv{label=$e^+$,label.angle=180}{i2}
          \fmf{plain,tension=2}{i1,v1,o1}
          \fmf{plain,tension=2}{i2,v2,o2}
          \fmf{wiggly,label=$\gamma^\ast,,Z^\ast$}{v2,v1}    \fmfdot{v1,v2}
          \fmfv{label=$e^-$,label.angle=0}{o1}   \fmfv{label=$e^+$,label.angle=0}{o2}
        \end{fmfgraph*}
      }}\hspace*{\fill}
  \end{fmffile}
  \caption{Feynman diagrams of the $s$- and $t$-channel Born-level  elastic Bhabha scattering, as denoted in the figure.}
  \label{bhabhaScatteringDiagramFIG}
\end{figure}

\autoref{bhabhaScatteringDiagramFIG} shows the elastic Bhabha scattering process, $e^{+}e^{-} \rightarrow e^{+}e^{-}$.


\vspace{20pt}
\unitlength=1mm

Strictly speaking, Born-level elastic Bhabha scattering never occurs. In practice, the process is always accompanied by the emission of electromagnetic radiation, for example

\begin{equation}{
e^{+}e^{-} \rightarrow e^{+}e^{-}\gamma .
}\label{bhabhaScatteringEQ} \end{equation}

In a simplified picture, a Bhabha event may be depicted as occurring in three steps: emission of radiation from the initial particles, Bhabha scattering, and emission of radiation from the final particles. In the angular scattering range considered for the luminosity measurement, one can discard the effects of interference between the initial and final state radiation. It should also be noted that the initial state radiation is mostly emitted in the direction of the beams and travels through the beampipe, thus remaining undetected. The ability to distinguish between a final state radiative photon and its accompanying lepton is determined by the resolving capabilities of the detector, and is a function of the angular separation between the two particles. When the two can be resolved, then the experimental measurements can be compared with the theoretical prediction, and thus the theory can be partly tested.

The Bhabha scattering includes at the Born level $\gamma$ and $Z^{0}$ exchange, both in the $s$- and the $t$-channels. The process may be written in terms of ten contributions, where four terms correspond to pure $\gamma$ and $Z^{0}$ exchanges in the $s$- and the $t$-channels, and the other six correspond to $\gamma - Z^{0}$ and $s - t$ interferences~\cite{introductionBIB15}. Taking into account $\gamma$ exchange only, the Bhabha cross-section may be presented as the sum of three terms,

\begin{equation}{
\frac{d\sigma_{\mathrm{B}}}{d\Omega} =
\frac{\alpha^{2}}{2s} 
\left[ 
\frac{1 + \cos ^{4}(\theta / 2)}{\sin ^{4} (\theta / 2)} 
- 2 \, \frac{\cos ^{4}(\theta / 2)}{\sin ^{2} (\theta / 2)}
+  \frac{1 + \cos ^{2} \theta}{2}
\right] ,
}\label{bhabhaXs1EQ} \end{equation}

\noindent where the scattering angle, $\theta$, is the angle of the scattered lepton with respect to the beam, $\alpha$ is the fine structure constant, and $s$ is the center-of-mass energy squared. The first and last terms correspond to $\gamma$ exchange in the $t$- and $s$-channels, respectively, and the middle term reflects the $s - t$ interference. 

For small angles ($\theta \le 10^{\circ}$), Bhabha scattering is dominated by the $t$-channel exchange of a photon. Discounting the $s$-channel contributions, one can rewrite \autoref{bhabhaXs1EQ} in terms of the scattering angle as:

\begin{equation}{
\frac{d\sigma_{\mathrm{B}}}{d\theta} =
\frac{2\pi \alpha^{2}}{s} \frac{\sin \theta}{\sin ^{4}(\theta / 2)} \approx 
\frac{32\pi \alpha^{2}}{s} \frac{1}{\theta^{3}}.
}\label{bhabhaXs2EQ} \end{equation}

The advantage of Bhabha scattering as a luminosity gauge process is that the event rate exceeds by far the rates of other physical processes, and in addition, the theoretical calculations of the cross-section are under control.

For the determination of the luminosity, the precise calculation of the Bhabha cross-section at small polar angles is needed. Theorists are working currently in several laboratories to improve the accuracy of higher-order electroweak corrections to the Bhabha cross-section~\cite{introductionBIB11,introductionBIB12,introductionBIB13,introductionBIB14}. The current theoretical uncertainty was estimated to be $5.3 \cdot 10^{-4}$ on the $Z$ resonance~\cite{introductionBIB27}, with the prospect of reducing this uncertainty to $2 \cdot 10^{-4}$, matching the need of GigaZ.

\subsection{Beam-Beam Effects at the ILC}

The theoretical uncertainties quoted above are based on the assumption that the energy of the colliding beams is monochromatic. For the ILC, though, this is not the case. The colliding electron and positron bunches at the ILC disrupt one another~\cite{introductionBIB16}. Prior to the Bhabha scattering, the interacting particles are likely to have been deflected by the space charge of the opposite bunch, and their energies reduced due to the emission of beamstrahlung. To take into account the cross-section dependence on $s$, the probability used to produce Bhabha scattering events during the beam-beam collision should be rescaled by $s/s'$, where $s'$ is the effective center-of-mass energy after the emission of beamstrahlung. The variance in $s$ will, in addition, be aggravated by the inherent energy spread of the collider. In general, the collision parameters, such as the size of the collision region and the bunch current, that lead to the highest luminosity, also lead to the largest smearing of the luminosity spectrum, $d \mc{L} / d \sqrt{s}$. Additionally, the energy measurements can be tempered by the presence of beam related backgrounds, such as synchrotron radiation and thermal photons of the residual gas, backscattered off the electron beam.

\begin{figure}[htp]
\begin{center}
\includegraphics[width=0.49\textwidth , height=0.275\textheight]{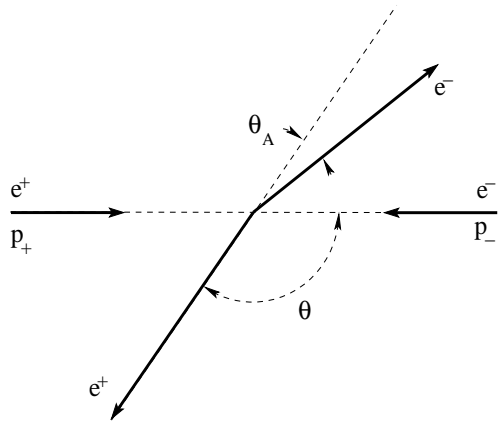}
\caption{\label{acollinearityAngleFIG}Definition of the acollinearity angle, $\theta_{\mathrm{A}}$.}
\end{center}
\end{figure}

The acollinearity angle for the $e^{+}e^{-}$ final state, defined as 

\begin{equation}{
\theta_{\rm{A}} = \theta_{e^{-}} + \theta_{e^{+}} - \pi,
}\label{acollinearityAngleEQ} \end{equation}

\noindent is depicted in \autoref{acollinearityAngleFIG}. Beamstrahlung emissions often occur asymmetrically, with either the electron or the positron loosing most of the energy. Hence the acollinearity of the final state can be significantly enhanced. The final state particles scattered in the acceptance range of LumiCal, following a Bhabha interaction, can typically cross a significant part of the opposite bunch. They can thus be focused by the electromagnetic field from the corresponding space charge, which causes the scattering angle to change.

Both beamstrahlung emissions and electromagnetic deflections vary with the bunch length, the horizontal bunch size, and the energy of the collision, and hence so do the resulting biases on the integrated luminosity. Reconstructing $d \mc{L} / d \sqrt{s}$ from the scattered Bhabha angles is possible~\cite{introductionBIB17,introductionBIB18}. This is done by measuring the acollinearity angle, which is related to the difference in the energies of the electron and positron beams, in the case of small energy and small scattering angle differences. The luminosity spectrum needs to be unfolded from the rates for the observed signal-channels in order to produce cross-sections as a function of energy. This is especially important for such analyses as top-quark and $W$-boson mass measurements~\cite{introductionBIB19}. Knowing $d \mc{L} / d \sqrt{s}$ also provides a good way to measure the amount of beamstrahlung, and thus to predict the corresponding contribution to the bias.

Contrary to the case with beamstrahlung, there is no direct way to control experimentally the bias from the electromagnetic deflections, and so these have to be simulated in order to compensate for their effect.

Since both the beam-beam effects and the collider energy spread depend on the parameters of the collisions, it would be very productive to measure the Bhabha cross-section itself, and thus better control the systematic errors.

\subsection{Relative Error of the Luminosity Measurement}

Several sources (defined below) contribute to the final error of the luminosity measurement,

\begin{equation}
\frac{\Delta\mc{L}}{\mc{L}} = \left( \frac{\Delta\mc{L}}{\mc{L}} \right)_{rec} \oplus \left(\frac{\Delta\mc{L}}{\mc{L}} \right)_{stat}  \oplus \quad ...
\label{luminosityRelativeErrEQ} \end{equation}

The typical signature of Bhabha scattering events is the exclusive presence of an electron and positron, back to back in the detector. A set of topological cuts is applied by comparing the scattering angles of the electron and of the positron, and by constraining the difference between, and the magnitude of, the energy which is collected in each detector arm~\cite{introductionBIB21,introductionBIB22}. The different contributions to the relative error of the luminosity measurement come down to

\begin{equation}
\frac{\Delta\mc{L}}{\mc{L}} = \frac{\Delta N}{N} = \left .  \frac{N_{rec} - N_{gen}}{N_{gen}} \right \rvert  ^{\theta_{max}}_{\theta_{min}},
\label{luminosityRelativeErrCountEQ} \end{equation}

\noindent where $N_{rec}$ and $N_{gen}$ are respectively the number of reconstructed and generated Bhabha events, and $\theta_{min}$ and $\theta_{max}$ are the respective low and high bounds on the fiducial volume (acceptance region) of the detector.

An error on the measurement is, therefore, incurred when events are miscounted. This may happen for several reasons. One of the causes for miscounting events has to do with knowledge of the effectiveness of the cuts for distinguishing between Bhabha events and background; the efficiency and purity of the cuts must be known to good precision in order to avoid counting errors. Another important factor is \textit{migration} of events out of the acceptance region and into it. This may occur if a shower's position in the detector is not reconstructed well, resulting in inclusion of events which were in actuality out of the fiducial volume, or visa-versa. Errors can also result from poor knowledge of the geometrical properties of the detector. For instance, displacement of the two arms of the detector with respect to each other, or with regard to the interaction point, may lead to systematic biases in the position reconstruction.

In the following, the errors which were discussed above are quantified.

\subsubsection{Error in Reconstruction of the Polar Scattering Angle}

The polar angle dependence of the Bhabha cross-section is $d\sigma / d\theta \sim \theta^{-3}$ (\autoref{bhabhaXs2EQ}). This means that the total Bhabha cross-section within the angular range $[\theta_{min} , \theta_{max}]$ is

\begin{equation}
\sigma_{\mathrm{B}} \sim \frac{1}{2} \left( \theta_{min}^{-2} - \theta_{max}^{-2} \right) \sim \frac{1}{2} \theta_{min}^{-2} ,
\label{luminosityRelativeErrRec1EQ} \end{equation}

\noindent where the $\theta_{max}$ dependence can be neglected. The relative error on luminosity is proportional to the relative error on the Bhabha cross-section,

\begin{equation}
\left( \frac{\Delta\mc{L}}{\mc{L}} \right)_{rec} = 2 \frac{\Delta \theta}{\theta_{min}}.
\label{luminosityRelativeErrRec2EQ} \end{equation}

The analytic approximation of \autoref{luminosityRelativeErrRec2EQ} has been shown to hold well in practice~\cite{introductionBIB22}. Its implication is that the polar bias, $\Delta \theta$, and the minimal polar bound of the fiducial volume, $\theta_{min}$, are the two most important parameters that affect the precision of the luminosity measurement. The steep fall of the Bhabha cross-section with the polar angle translates into significant differences in the counting rates of Bhabha events, for small changes in the angular acceptance range.

\subsubsection{Statistical Error of the Number of Expected Bhabha Events}

The probability of observing Bhabha scattering in a given event is determined by the Poisson distribution. The variance of the distribution is then equal to the average number of observed Bhabha scatterings, $N$. The relative statistical error stemming from \autoref{luminosityByXsCountingEQ}, is, therefore

\begin{equation}
\left( \frac{\Delta\mc{L}}{\mc{L}} \right)_{stat} = \frac{\Delta N}{N} = \frac{\sqrt{N}}{N} = \frac{1}{\sqrt{N}}.
\label{luminosityRelativeErrStatEQ} \end{equation}

\Autoref{luminosityRelativeErrStatEQ} is the driving force behind the precision requirements, which were stated in \autoref{precisionLumiRequirementsSEC}.

\subsubsection{Additional Sources of Error}

In congruence to the two major sources of error which were discussed above, several other factors need to be considered in order to keep the design goal of $\Delta\mc{L}/\mc{L} \sim 10^{-4}$. These include controlling the position of the inner radius of the detector on a $\mu \mathrm{m}$ precision level, and the distance between its two arms to within $\mc{O}(100 \, \mu \mathrm{m})$, to name just two. A full account may be found in ~\cite{introductionBIB21}.

\section{The Luminosity Calorimeter \label{lumiCalDescriptionSEC}}

\begin{figure}[htp]
\begin{center}
\includegraphics[width=.99\textwidth]{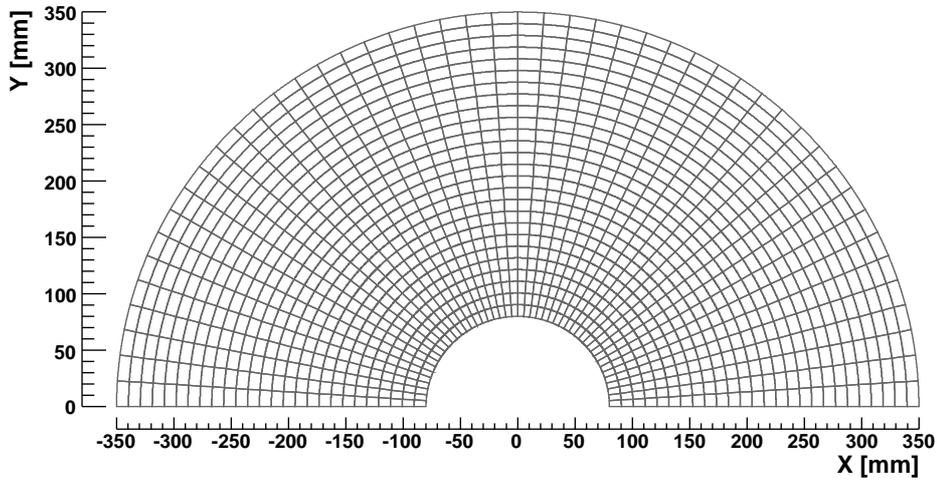}
\caption{\label{sensorPlaneFIG}Half plane of LumiCal silicon sensors (every fourth raidal segment is drawn).}
\end{center}
\end{figure}

LumiCal is a tungsten-silicon sandwich calorimeter. In the present ILD layout the detector is placed $2.27$~m from the interaction point. The LumiCal inner radius is $80$~mm, and its outer radius is $350$~mm, so that its polar coverage is 35 to 153~mrad. The longitudinal part of the detector consists of layers, each composed of $3.5$~mm of tungsten, which is equivalent to 1~radiation length thickness. Behind each tungsten layer there is a $0.6$~mm ceramic support, a $0.3$~mm silicon sensors plane, and a $0.1$~mm gap for electronics. LumiCal is comprised of 30 longitudinal layers. The transverse plane is subdivided in the radial and azimuthal directions.  The number of radial divisions is 104, and the number of azimuthal divisions is 96. \Autoref{sensorPlaneFIG} presents the segmentation scheme of LumiCal.

\begin{figure}[htp]
\begin{center}
\includegraphics[width=.99\textwidth]{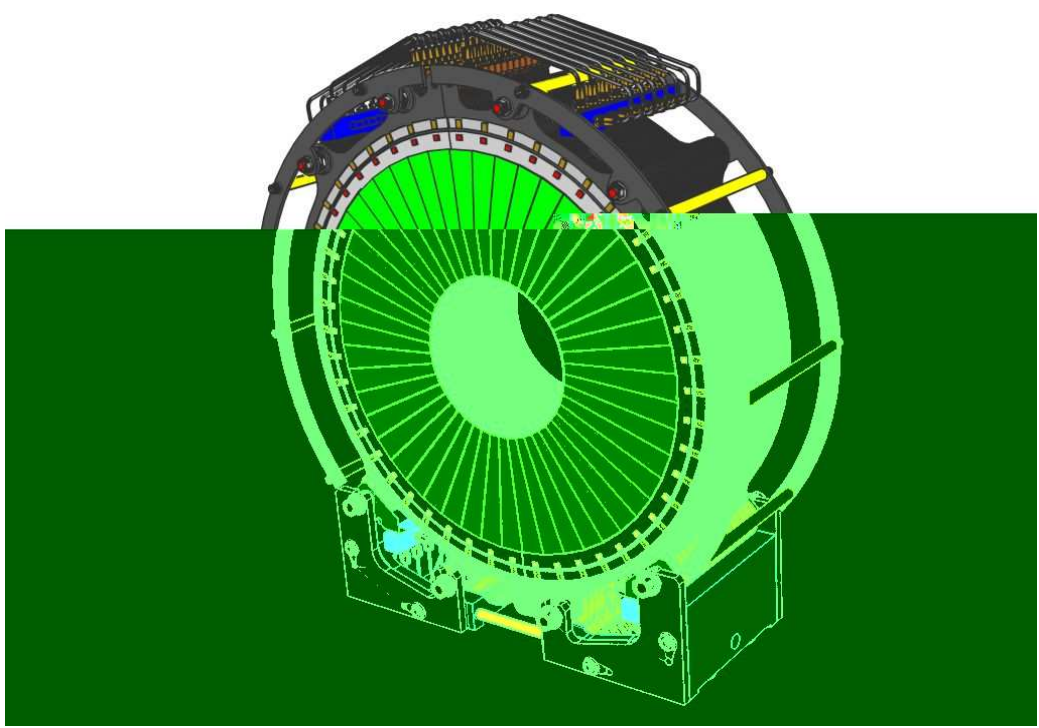}
\caption{\label{mechanicalDesignFIG}Foreseen mechanical design of LumiCal.}
\end{center}
\end{figure}

The two half barrels can be clamped on the closed beam pipe. The position of the two parts of the detector with respect to each other will be fixed by the help of precise pins placed at the top and bottom of each C shaped steel frame~\cite{introductionBIB23}. The latter stabilizes the structure and carries the heavy tungsten disks by the bolts. The gravitation sag of the tungsten absorber can be kept in required tolerance~\cite{introductionBIB24}. The silicon sensors are glued to the tungsten surface with capton foil insulation. Space for readout electronics, connectors and cooling is foreseen at the outer radius of the calorimeter. The sensor plane will be built from a few tiles because the current technology is based on 6-inch wafers, and at the moment it is unclear if and when larger wafers will be available. The tiles of the silicon sensors will be glued to a thick film support ceramic plate or directly to a tungsten surface with some insulation. Reference marks are foreseen on the detector surface for precision positioning. The layout of the sensors and the mechanical design of the calorimeter does not allow for sensors to overlap. To reduce the impact of the gaps, odd and even planes are rotated by half the azimuthal cell pitch $(1.875^{\circ})$. The silicon diodes will be usual planar high resistivity silicon sensors. \Autoref{mechanicalDesignFIG} presents the foreseen mechanical design of LumiCal.

\chapter{Development of EM Showers in LumiCal \label{clorimetryCH}}

\section{Basic Concepts in Calorimetry \label{calorimetryConceptsSEC}}

\subsection{Energy Loss by Electrons and Photons}

High-energy electrons predominantly loose energy in matter by bremsstrahlung, and high-energy photons by $e^{+}e^{-}$ pair production. The characteristic amount of matter traversed for these related interactions is called the \textit{radiation length}, $X_{0}$. It is both the mean distance over which a high-energy electron looses all but $1/e$ of its energy by bremsstrahlung, and $\frac{7}{9}$ of the mean free path for pair production by a high-energy photon~\cite{calorimetryBIB1}. The radiation length is also the appropriate scale length for describing high-energy electromagnetic showers.

\begin{figure}[htp]
\begin{center}
\subfloat[]{\label{electronEnergyLossFIG1}\includegraphics[height=0.4\textheight]{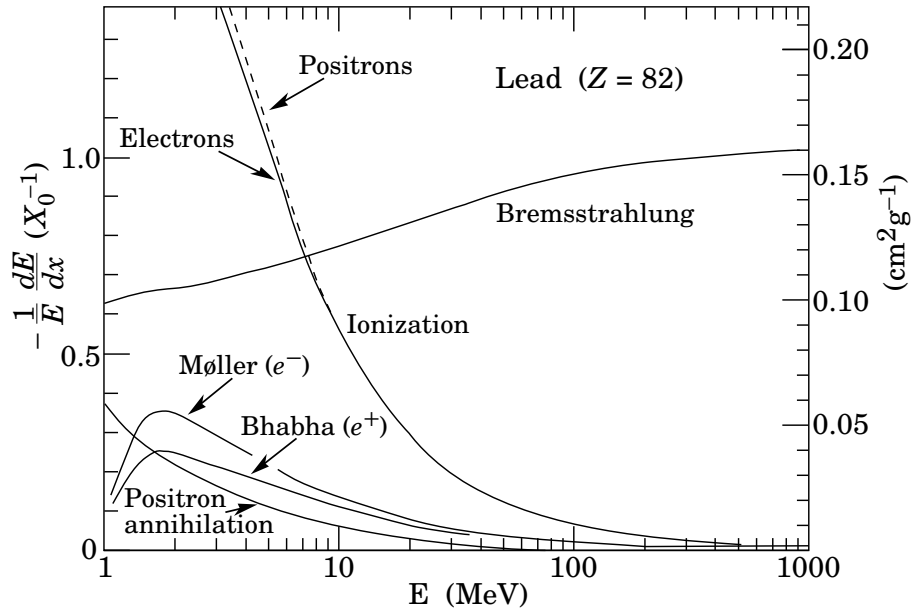}}\\
\subfloat[]{\label{electronEnergyLossFIG2}\includegraphics[height=0.4\textheight]{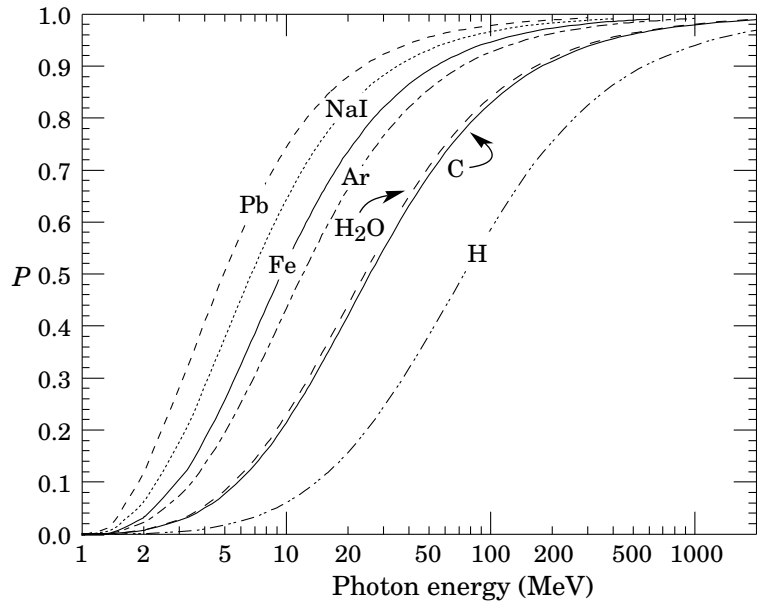}}
\caption{\label{electronEnergyLossFIG}\Subref{electronEnergyLossFIG1} Fractional energy loss per radiation length in lead as a function of electron or positron energy. Electron (positron) scattering is considered as ionization when the energy loss per collision is below 0.255~MeV, and as M\o ller (Bhabha) scattering when it is above. \Subref{electronEnergyLossFIG2} Probability $P$ that a photon interaction will result in conversion to an $e^{+}e^{-}$ pair (the figures are taken from~\cite{introductionBIB20}).}
\end{center}
\end{figure}

At low energies, electrons and positrons primarily loose energy by ionization, although other processes (M\o ller scattering, Bhabha scattering, $e^{+}$ annihilation) contribute as well, as shown in \autoref{electronEnergyLossFIG1}. While ionization loss-rates rise logarithmically with energy, bremsstrahlung losses rise nearly linearly (fractional loss is nearly independent of energy), and dominate above a few tens of MeV in most materials.

At low energies, the photon cross-section is dominated by the photoelectric effect, although Compton scattering, Rayleigh scattering, and photonuclear absorption also contribute. The photoelectric cross-section is characterized by discontinuities (absorption edges) as thresholds for photoionization of various atomic levels are reached. The increasing dominance of pair production as the energy increases is shown in \autoref{electronEnergyLossFIG2}. The cross-section is very closely related to that for bremsstrahlung, since the Feynman diagrams are variants of one another.

\subsection{Electromagnetic Showers}

When a high-energy electron or photon is incident on a thick absorber, it initiates an electromagnetic (EM) shower as pair production and bremsstrahlung generate more electrons and photons with lower energy. The longitudinal development is governed by the high-energy part of the cascade, and therefore scales as the radiation length in the material. Electron energies eventually fall below the critical energy (defined below), and then dissipate their energy by ionization and excitation, rather than by the generation of more shower particles.

The transverse development of electromagnetic showers scales fairly accurately with the Moli\`ere radius, $R_{\mc{M}}$, given by~\cite{introductionBIB20}

\begin{equation}{
R_{\mc{M}} = X_{0} \frac{E_{s}}{E_{c}} \, ,
} \label{molierRadiusEQ} \end{equation}

\noindent where $E_{s} \approx 21$~MeV, and $E_{c}$ is the \textit{critical energy}, which is defined as the energy at which the ionization loss per radiation length is equal to the electron energy~\cite{calorimetryBIB2}. On average, only $10\%$ of the energy of an EM shower lies outside a cylinder with radius $R_{\mc{M}}$ around the shower-center.

\section{Simulation of the Detector Response \label{characteristicsOfShowersSEC}}

The response of LumiCal to the passage of particles was simulated using Mokka, version 06-05-p02 ~\cite{revisedDetectorModelBIB1}. Mokka is an application of a general purpose detector simulation package, GEANT4, of which version 9.0.p01 was used ~\cite{revisedDetectorModelBIB2}. The Mokka model chosen was LDC00\_03Rp, where LumiCal is constructed by the LumiCalX super driver. The output of Mokka is in the LCIO format. Several Marlin  processors were written in order to analyze the LCIO output. Marlin is a C++ software framework for the ILC software~\cite{revisedDetectorModelBIB3}. It uses the LCIO data model and can be used for all tasks that involve processing of LCIO files, e.g. reconstruction and analysis. The idea is that every computing task is implemented as a processor (module) that analyses data in an LCEvent and creates additional output collections that are added to the event. Version 00-09-08 of the program was used.

The geometry of LumiCal which was simulated is that which is described in \autoref{lumiCalDescriptionSEC}.

\begin{figure}[htp]
\begin{center}
\vspace{-20pt}
\subfloat[]{\label{particleCreationSpectrumFIG1}\includegraphics[width=.49\textwidth]{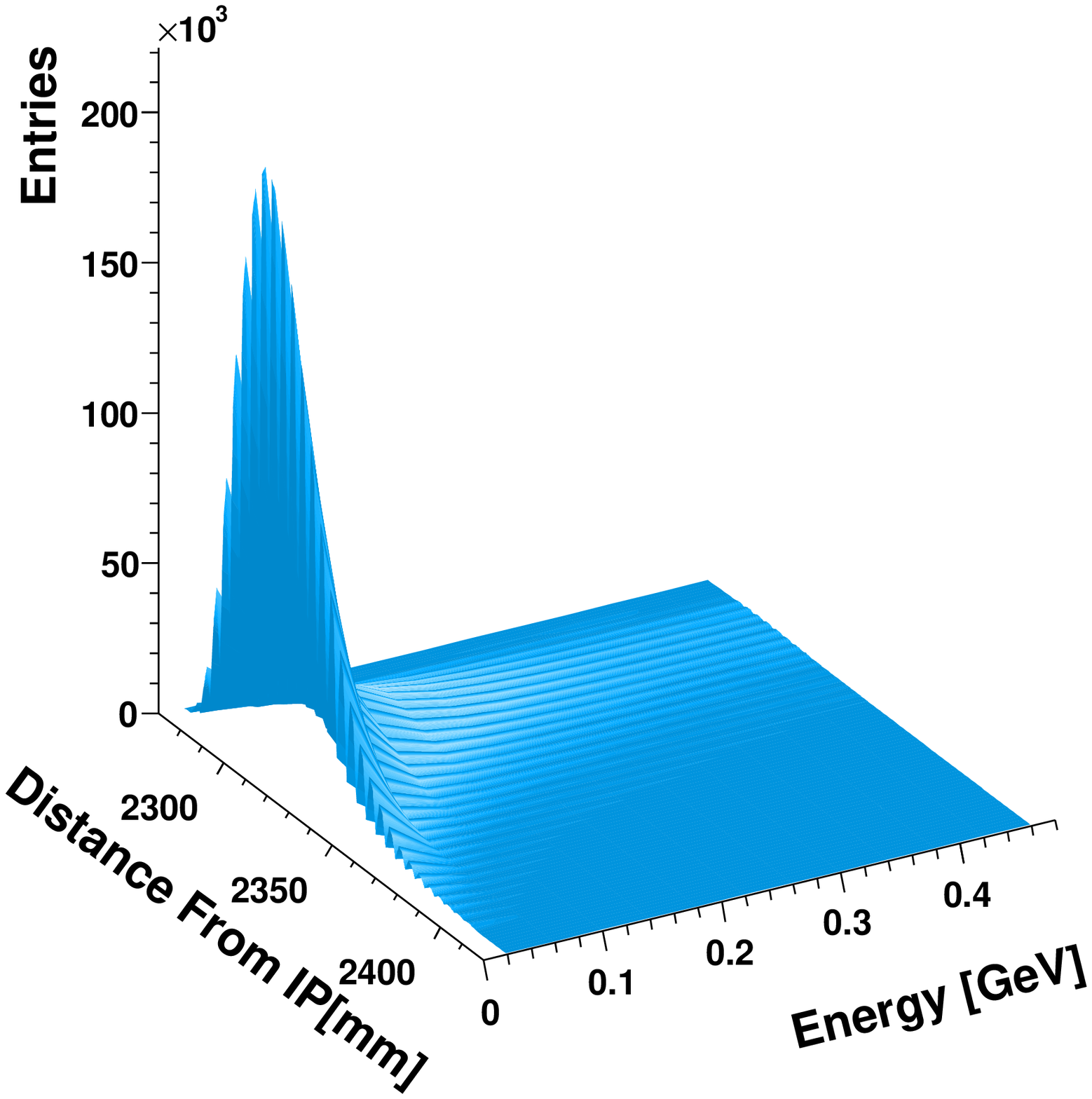}}
\subfloat[]{\label{particleCreationSpectrumFIG2}\includegraphics[width=.49\textwidth]{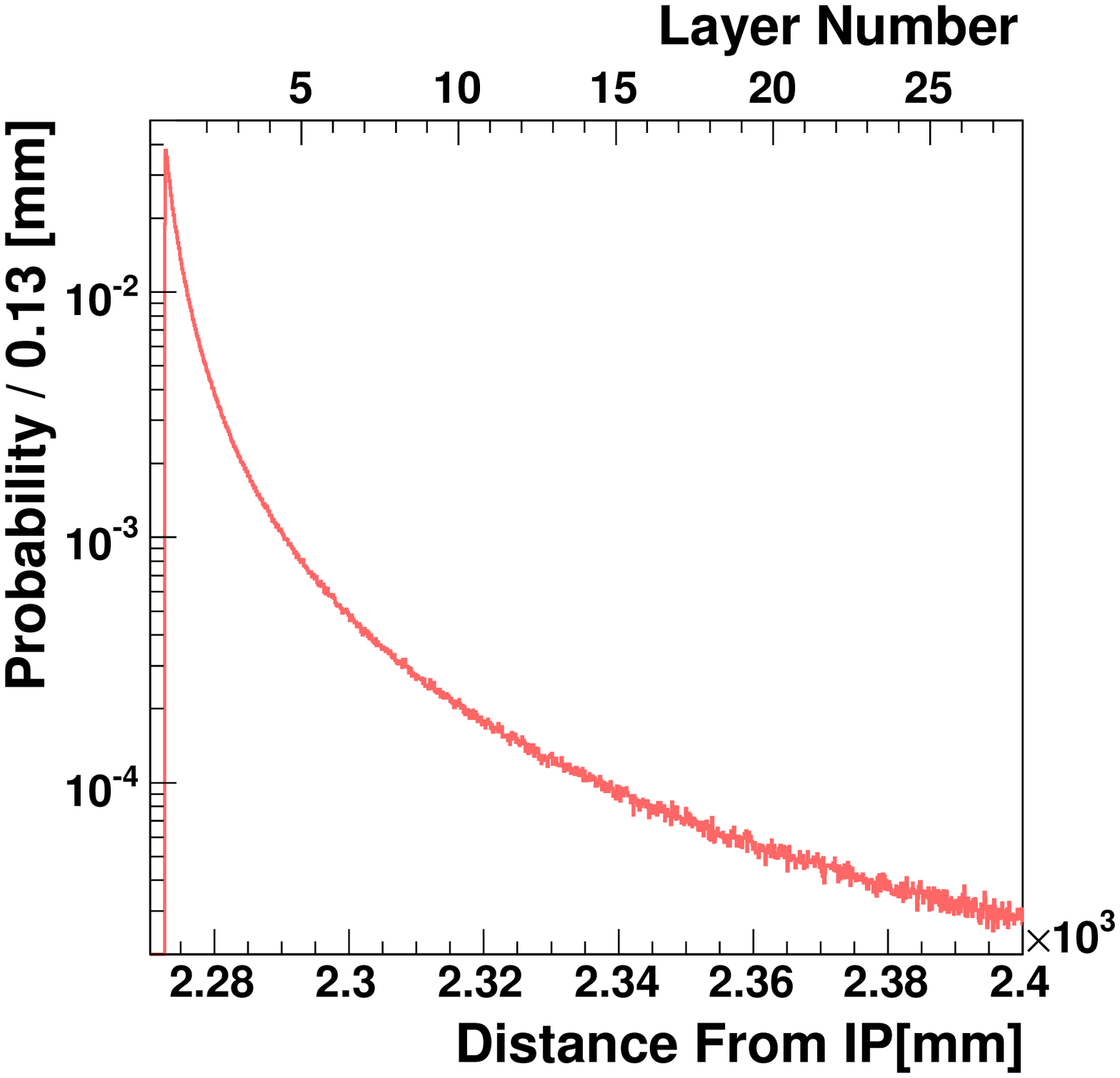}}
\caption{\label{particleCreationSpectrumFIG}\Subref{particleCreationSpectrumFIG1} Distribution of the creation position relative to the IP and of the energy of shower particles in LumiCal for 250~GeV electron showers. \Subref{particleCreationSpectrumFIG2} Normalized distribution of the energy at the creation vertex of shower particles, as a function of the point of creation. The creation point is represented by corresponding scales of the distance from the IP and the LumiCal layer number.}
\subfloat[]{\label{engyDepositionAndParticleNumberFIG1}\includegraphics[width=.49\textwidth]{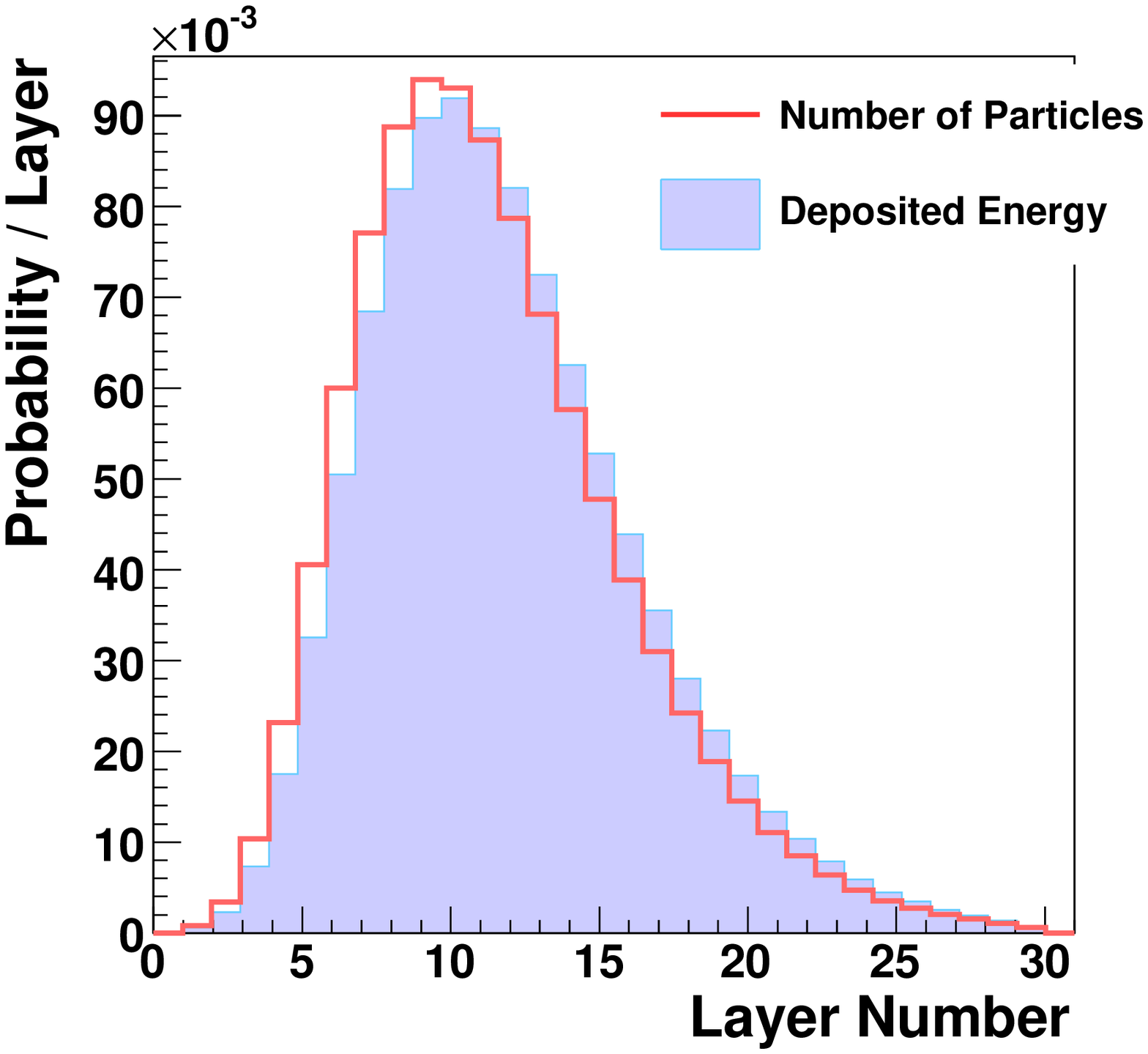}}
\subfloat[]{\label{engyDepositionAndParticleNumberFIG2}\includegraphics[width=.49\textwidth]{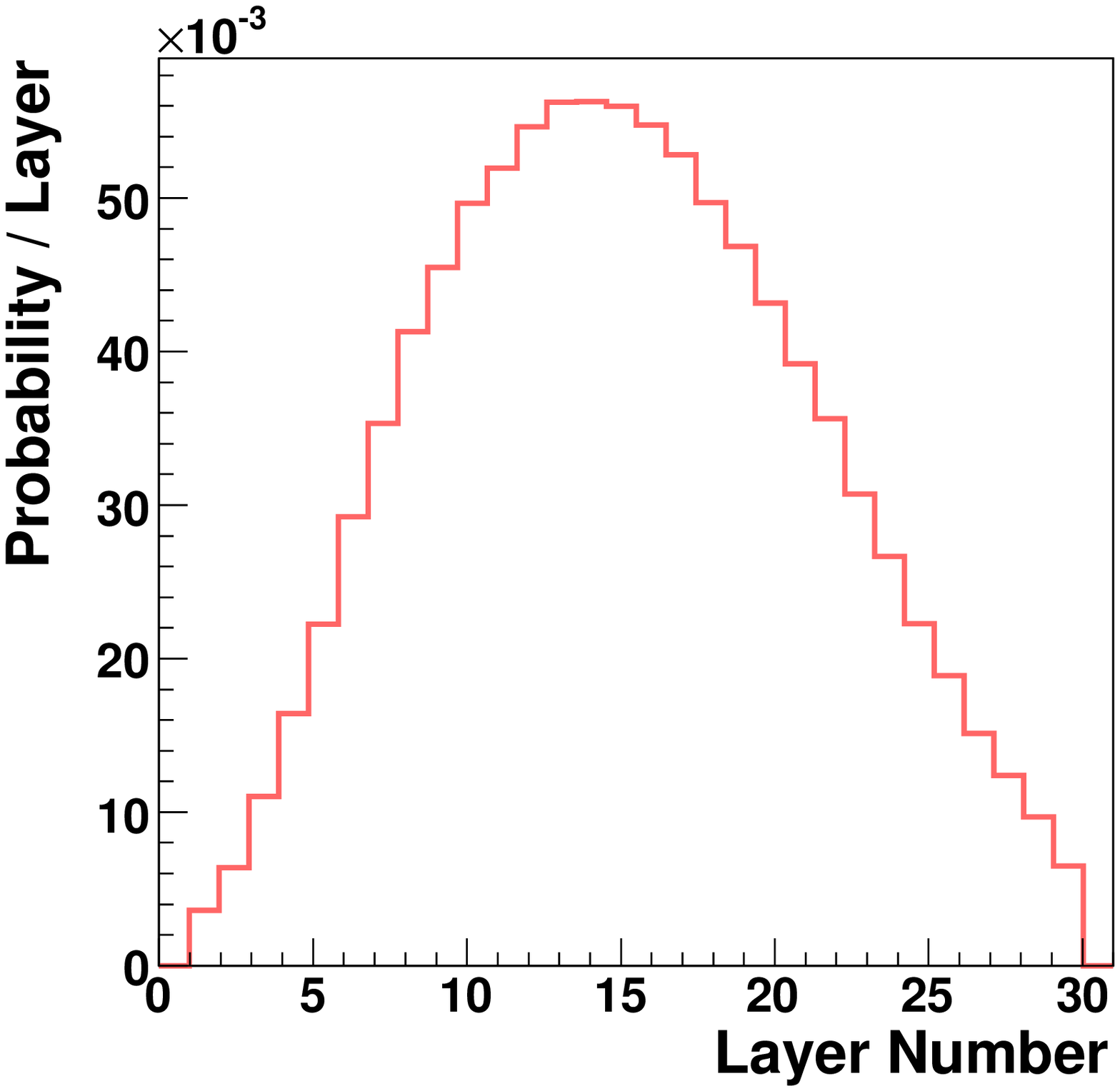}}
\caption{\label{engyDepositionAndParticleNumberFIG}\Subref{engyDepositionAndParticleNumberFIG1} Normalized distributions of the number of shower particles and of the energy deposited in the silicon sensors of LumiCal as a function of the layer, as denoted in the figure. Electron showers of 250~GeV were simulated. \Subref{engyDepositionAndParticleNumberFIG2} Normalized distribution of the number of cell-hits for 250~GeV electrons showers as a function of the layer in LumiCal.}
\end{center}
\end{figure}

\Autoref{particleCreationSpectrumFIG} shows the generation spectrum of electrons, positrons and photons for a 250~GeV EM shower in LumiCal\footnotemark . These particles traverse the layers of tungsten and deposit energy in the silicon sensors mainly through ionization. In \autoref{particleCreationSpectrumFIG1} each entry represents the z-position (relative to the IP) of the creation of a shower particle with a given energy. In \autoref{particleCreationSpectrumFIG2} a normalized profile of the energy as a function of the distance is presented. As the shower develops in depth in the calorimeter, new shower particles are created with less and less energy. Eventually the energy falls off below the threshold of ionization.

\footnotetext{The shower also contains protons and neutrons. The contribution of these particles to the energy deposited in LumiCal is negligible due to their relative low number, and so will not be discussed here.}

Two normalized distributions are overlaid in \autoref{engyDepositionAndParticleNumberFIG1}, the number of shower particles and the deposited energy, both as a function of the layer number in LumiCal. Electron showers of 250~GeV were used. The energy deposited in the silicon sensors is proportional to the number of charged shower particles. This is consistent with the fact that both distributions have a similar shape. However, while the distribution of the number of shower particles peaks at the ninth layer, that of the energy deposition peaks at the tenth. A displacement between the distributions by one layer, which is equivalent to one radiation length, is apparent. This is due to the fact that part of the shower is comprised of photons, which do not deposit energy, but are later converted to electron-positron pairs, which do. In \autoref{engyDepositionAndParticleNumberFIG2} is presented the normalized distribution of the number of cell hits for 250~GeV electron showers as a function of the layer in LumiCal. The number of cells which register a hit peaks around layer number 13.

\begin{figure}[htp]
\begin{center}
\includegraphics[width=.49\textwidth]{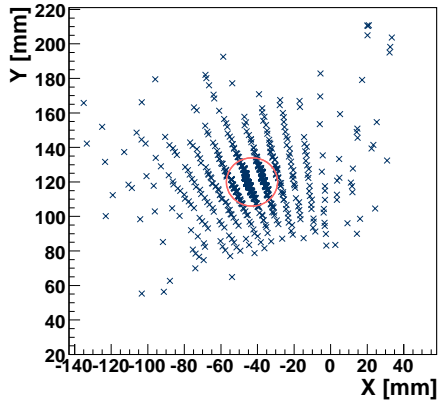}
\caption{\label{showerEnergyProfileFIG}Transverse energy profile for a 250~GeV electron shower in LumiCal. The red circle represents an area bound within one Moli\`ere radius around the shower center.}
\end{center}
\end{figure}

\Autoref{showerEnergyProfileFIG} shows the profile of the energy deposited in LumiCal for a single 250~GeV electron shower. Integration is made along the $z$-direction for equivalent values of the $x$ and $y$ coordinates, which are taken as the centers of the cells of the relevant hits. The \textit{global shower-center} is defined as the center of gravity of the shower profile, using cell energies as weights. The polar symmetry of the segmentation of the detector is also visible in the figure, and it is evident that LumiCal is more finely granulated in the radial direction than in the azimuthal direction.

\begin{figure}[htp]
\begin{center}
\subfloat[]{\label{longitudinalProfileFIG1}\includegraphics[width=.49\textwidth]{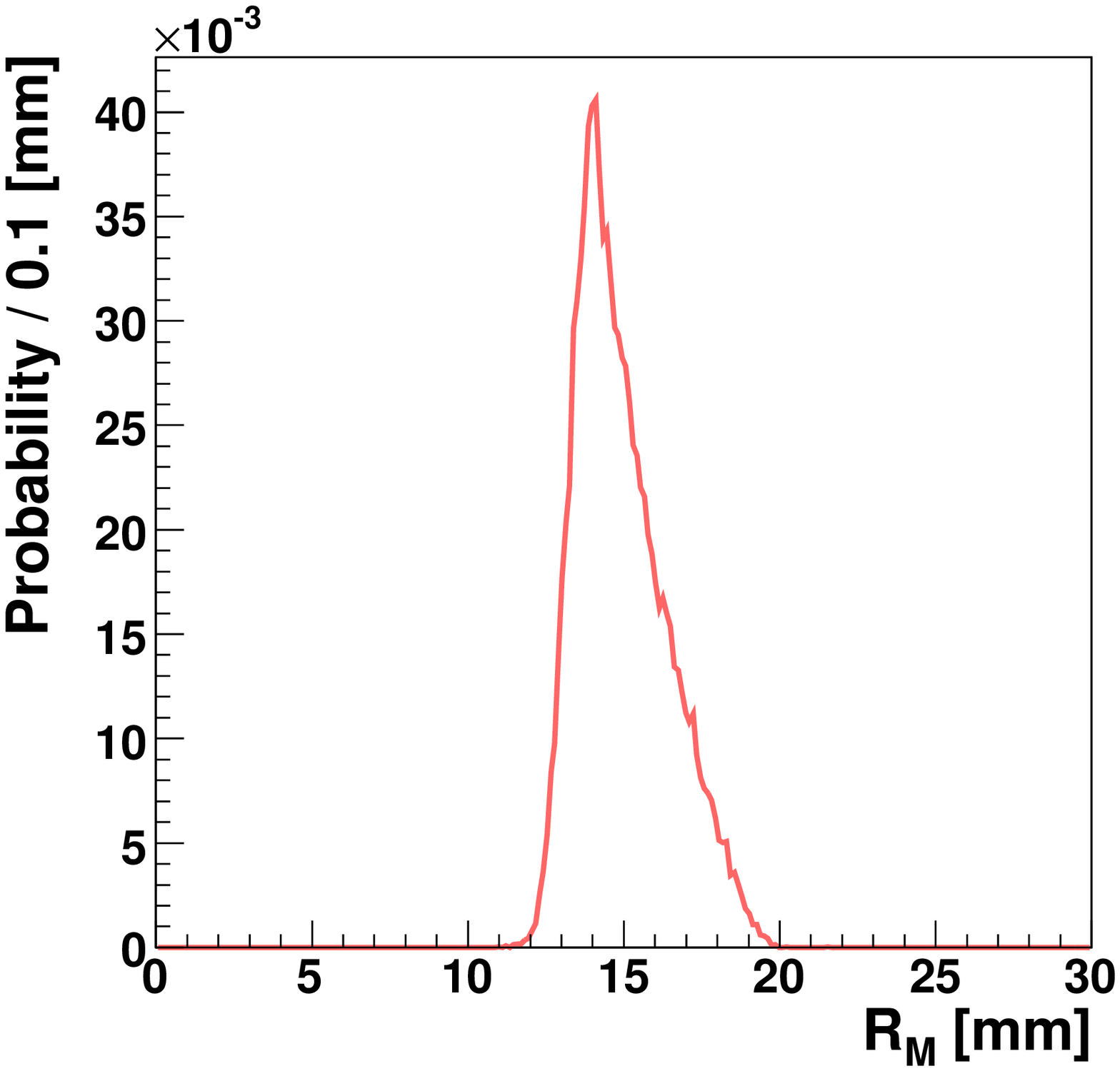}}
\subfloat[]{\label{longitudinalProfileFIG2}\includegraphics[width=.49\textwidth]{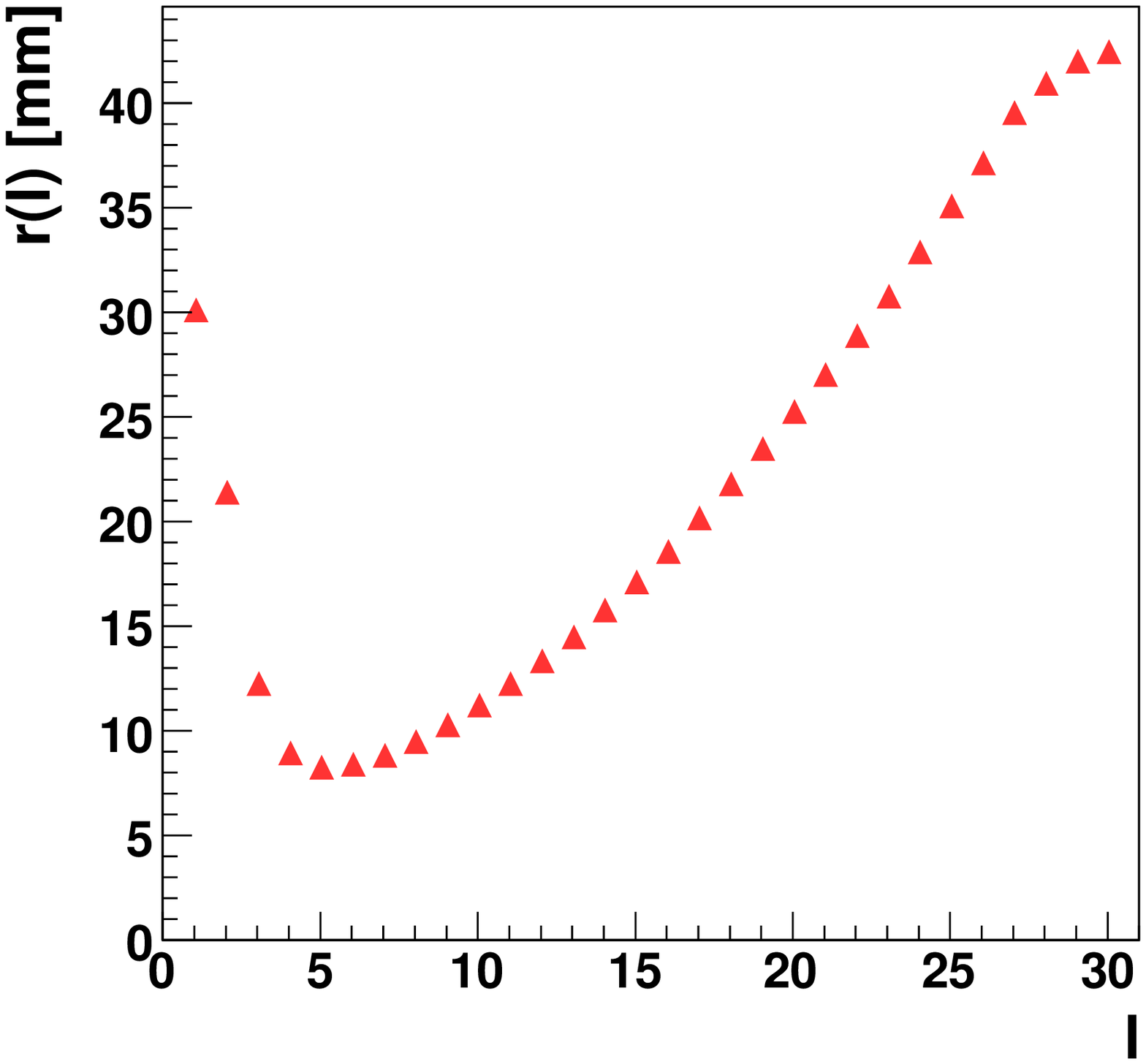}}
\caption{\label{longitudinalProfileFIG}\Subref{longitudinalProfileFIG1} Distribution of $R_{\mc{M}}$, the distance around the global-shower center, in which $90\%$ of the integrated shower energy may be found. \Subref{longitudinalProfileFIG2} Dependence on the layer number, $\ell$, of the layer-radius, $r(\ell)$, which is the distance around the local shower-center in which $90\%$ of the energy of a layer is deposited.}
\end{center}
\end{figure}

\Autoref{longitudinalProfileFIG1} shows the distribution of the distance around the global shower-center, in which $90\%$ of the integrated shower energy may be found. The distribution is centered around 14~mm, which is, by definition, the Moli\`ere radius of LumiCal, $R_{\mc{M}}$. Taking into account all of the hits of the shower, the polar and azimuthal production angles of the initiating particle may be reconstructed. Local shower-centers are defined on a layer-to-layer basis as the extrapolation of the trajectory of the particle according to these angles. We define the distance around the local shower-center of layer $\ell$, in which $90\%$ of the layer's energy is deposited, as its \emph{layer-radius}, $r(\ell)$. \Autoref{longitudinalProfileFIG2} shows the dependence of $r(\ell)$ on the layer number, $\ell$.

According to the distributions in \autoref{engyDepositionAndParticleNumberFIG}, in the first layers there are few cells which register hits. For this reason there is no clear local shower-center, and the area that encompasses $90\%$ of the energy of the layer is large. The information in these layers is, therefore, not sufficient to obtain a clear description of the shower. This effect is lessened as the shower develops in depth and the number of cell-hits increases. Starting at the fifth layer, the shower becomes homogeneous. Beyond this point the shower becomes more and more wide spread in depth, and its diameter may be estimated to good approximation by a power-law. For layer numbers higher than 16, the shower exceeds $150\%$ of $R_{\mc{M}}$ and looses homogeneity  once again. This behavior is supported by \autoref{engyDepositionAndParticleNumberFIG}, which shows that for layers beyond the shower-peak, the number of shower-particles falls off faster than the number of cells which are hit. Since the shower becomes attenuated for high layer numbers, it is difficult to determine the local shower-center with good accuracy. It is, therefore, useful to define an \emph{effective layer-radius}, $r_{eff}(\ell)$, by extrapolation of the behavior of $r(\ell)$ at the middle layers, to the front and back layers,

\begin{equation}{
r_{eff}(\ell) \equiv r(6 \le \ell \le 24) \approx 5 + 0.23 \cdot \ell + 0.04 \cdot \ell ^{2}
} \label{effectiveLayerRadiusEQ} \end{equation}

In summary, the transverse profile of EM showers in LumiCal is characterized by a peak of the shower around the tenth layer. The number of shower-particles before layer six and after layer 16 is small compared to that in the inner layers. The energy of shower particles degrades in depth, and accordingly so do the energy deposits, while the shower becomes more wide spread. The front layers (layers one to five) are, therefore, characterized by a small number of concentrated energy deposits. The middle, so-called, \textit{shower-peak layers} (layers six to 16)\footnotemark register large energy contributions, and the back layers (layers 16 to 30) are characterized by a decreasing number of low-energy shower particles, which deposit little energy in a dispersed manner. The shower has a prominent center, within $R_{\mc{M}}$ (14~mm) of which most of the shower energy is concentrated. On a layer-by-layer basis, most of the energy may be found within an effective layer-radius from the center, which is parameterized by \autoref{effectiveLayerRadiusEQ}.

\footnotetext{On an event-by-event basis the longitudinal profile is not always as smooth as the one represented in \autoref{engyDepositionAndParticleNumberFIG1}. As a result, the shower-peak layers are not necessarily consecutive.}
\chapter{A Clustering Algorithm for LumiCal \label{clusteringCH}}

In the running conditions of the accelerator, the selection of Bhabha candidate events will require pattern recognition in the main detector and in LumiCal. Here the first attempt to perform clustering in LumiCal is presented. The main focus is on clustering optimized for EM showers, with the intent of resolving events in which hard photons were emitted in the final state.

As explained in \autoref{clorimetryCH}, high energy electrons and photons which traverse LumiCal loose energy in the tungsten layers mainly by the creation of electron-positron pairs and of photons, which in turn also loose energy by the same processes. The cascade of particles is propagated until most of the energy of the initial particle has been absorbed in the calorimeter. These secondary particles make up an electromagnetic shower that is sampled in the silicon sensors that make up the back side of each layer. When two (or more) high energy particles enter LumiCal, an EM shower will develop for each particle, and the multiple showers will overlap to some degree, depending on the initial creation conditions of each shower. The ability to separate any pair of showers is subject to the amount of intermixing of the pair.

\section{Outline of the Clustering Algorithm \label{clusteringAlgorithmOutlineSEC}}

The clustering algorithm which was designed for LumiCal was written as a series of Marlin (version 00-09-08) processors~\cite{revisedDetectorModelBIB3}. It operates in three main phases,

\begin{list}{-}{}

\item
selection of shower-peak layers, and two-dimensional clustering therein,

\item
fixing of the number of global (three-dimensional) clusters, and collection of all hits onto these,

\item
characterization of the global-clusters, by means of the evaluation of their energy density.

\end{list}

A short description of each phase will now follow. A complete account can be found in \autoref{clusteringAPP}.

\subsection{Clustering in the Shower-Peak Layers \label{nn2DclustersSPlayersSUBSEC}}

In the first layers of LumiCal, only a few hits from the shower are registered, as was discussed above. In addition to the hits from the main showers, there may also be contributions owing to backscattered particles or background processes. These particles have low energy and do not propagate to the inner layers, but their energy is of the order of the depositions of the showers of interest. In order to make a good estimate of the number of main showers, one must, therefore, begin by considering the information in the shower-peak layers. This process is done in two steps, which are described below, near-neighbor clustering and cluster-merging. 

\subsubsection{Near-Neighbor Clustering}

Initially, clusters are created from groups of closely-connected cells. This is done by means of the method of near-neighbor clustering (NNC), which exploits the gradient of energy around local shower-centers. The assumption is that in first order, the further a hit is relative to the shower center, the lower its energy. By comparing the energy distribution around the center at growing distances, one may check whether the energy is increasing or decreasing. An increase in energy for growing distance from the shower-center would then imply that the hit should be associated with a different shower.

\begin{figure}[htp]
\begin{center}
\subfloat[]{\label{NNclustersFIG1}\includegraphics[height=.37\textheight]{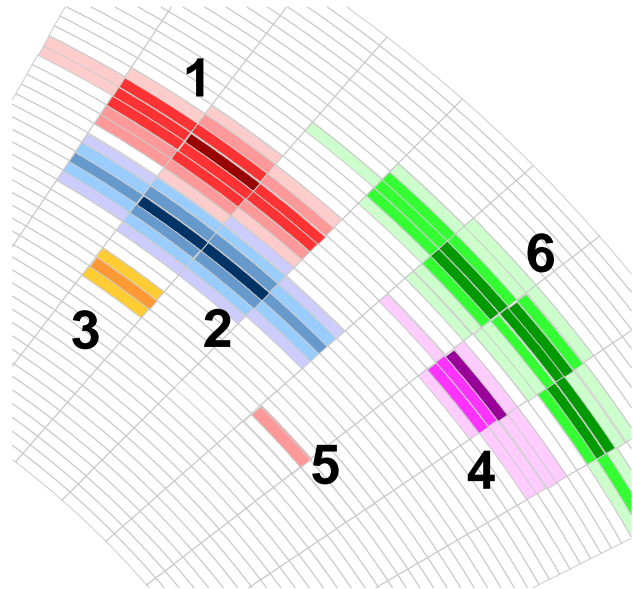}}\\
\subfloat[]{\label{NNclustersFIG2}\includegraphics[height=.37\textheight]{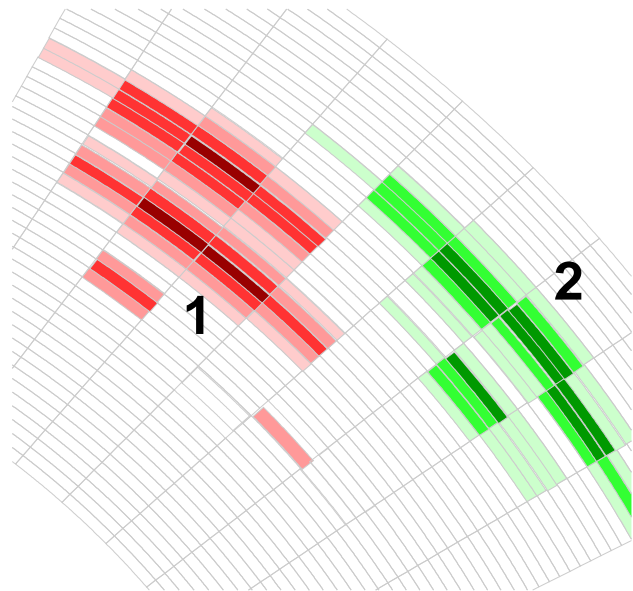}}
\caption{\label{NNclustersFIG}\Subref{NNclustersFIG1} Schematic representation of the results of the NNC phase for a single layer. Six clusters are found by the algorithm. The different clusters are enumerated, as well as distinguished by different color groups. Darker shadings indicate a higher energy content of the cell in question. \Subref{NNclustersFIG2} Evolution of the results from \Subref{NNclustersFIG1} after the cluster-merging phase. Two clusters remain after clusters two, three and five were merged with cluster one, and cluster four was merged with cluster six.}
\end{center}
\end{figure}

For each shower-peak layer separately, the algorithm associates each cell which has an energy deposit with its highest-energy near-neighbor. The result of the NNC phase is a collection of clusters in each layer, centered around local maxima, as illustrated for a single layer in \autoref{NNclustersFIG1}. In this example the algorithm produces six clusters, which are enumerated in the figure. The different clusters are also distinguished by different color groups, where darker shadings indicate a higher energy content of the cell in question.

\subsubsection{Cluster-Merging}

The next step in the algorithm is cluster-merging. The NNC method only connects cells which are relatively close, while showers tend to spread out over a large range of cells, as indicated by \autoref{longitudinalProfileFIG2}. The cluster-merging procedure begins by assigning a center-position to each existing cluster. Weights are then computed for each cluster to merge with the rest of the clusters. In general, the weights are proportional to the energy of the candidate cluster, and inversely proportional to the distance between the pair of clusters. Several variations of the weighting process are tested in consecutive merging attempts. The result of the algorithm after the cluster-merging phase is illustrated for a single layer in \autoref{NNclustersFIG2}.

\subsection{Global Clustering \label{globalClusteringSUBSEC}}

The most important stage of the clustering algorithm is the determination of the number of reconstructed showers. The aftermath of the clustering in the shower-peak layers is several collections of two-dimensional hit aggregates, the number of which varies from layer to layer. The final number of showers is then determined as the most frequent value of the layer-cluster number, derived from the collections in the shower-peak layers.

Once the number of global-showers is fixed, cells from non-shower-peak layers are associated with one of the global-showers. This is done by extrapolating the propagation of each shower through LumiCal, using the information from the shower-peak layers. Following the extrapolation, cells are merged with the extrapolated global-cluster centers in each layer. This process is facilitated by assuming a typical shower-size, defined according to the parameterization of \autoref{effectiveLayerRadiusEQ} (\autoref{characteristicsOfShowersSEC}), which acts as a temporary center-of-gravity. Once the core cells within the assumed shower-radius are associated with the global-centers in these layers, the rest of the cells may also be added. A weighing method, similar to the one used in the shower-peak layers, is used here as well.

\begin{figure}[htp]
\begin{center}
\includegraphics[height=.45\textheight]{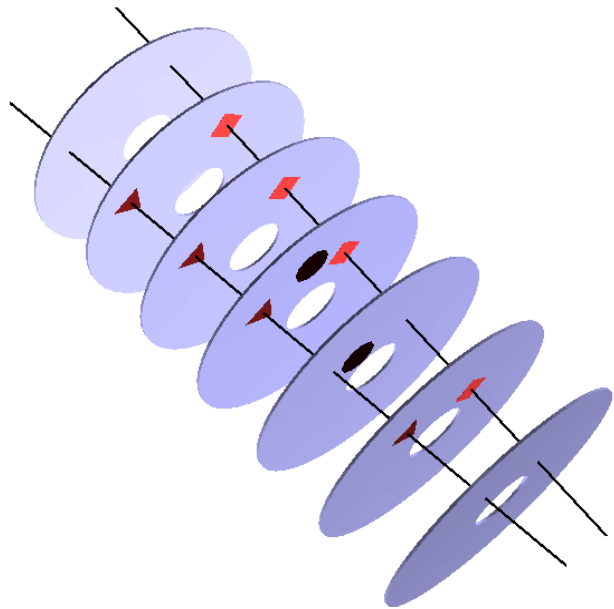}
\caption{\label{globalClusteringFIG}Schematic representation of the global-clustering phase. LumiCal layers are represented by the large blue disks, and layer-clusters are represented by the small triangles, squares and circles. The straight lines show the extrapolated position of the global-clusters in all layers.}
\end{center}
\end{figure}

\Autoref{globalClusteringFIG} shows a schematic representation of the global-clustering phase. LumiCal layers are represented by the large blue disks, and layer-clusters are represented by the small triangles, squares and circles. Three layers have two layer-clusters, one layer has three layer-clusters and one layers has one layer-cluster. The first and last layers have no layer-clusters, since they are not shower-peak layers. The global number of clusters is two, and the layer clusters are associated with each other according to the straight lines. The lines also define the global-cluster positions in the non-shower-peak layers. The cluster represented by a circle in the layer, in which three layer-clusters were found, will be disbanded. Its hits will be associated with either the ``square'' or ``triangle'' global-clusters. The layer-cluster in the layer, in which only one cluster was found, will also be disbanded. The hits will then be clustered around the virtual-cluster positions, represented by the intersection of the straight lines with the layer. A similar procedure will also be performed in the first and last layers, where no layer-clusters were constructed previously.

\subsection{Corrections Based on the Energy Distribution}

At this point all of the hits in the calorimeter have been integrated into one of the global-clusters. Before moving on, it is beneficial to make sure that the clusters have the expected characteristics.

\subsubsection{Energy Density Test}

The EM shower development in LumiCal has been described in \autoref{characteristicsOfShowersSEC}. Accordingly, one would expect that $90\%$ of a cluster's energy would be found within one Moli\`ere radius, $R_{\mc{M}}$, of its center. While statistically this is true, on a case-by-case basis fluctuations may occur, and thus it should not be set as a hard rule. It is possible to define, instead, a set of tests based on the amount of energy which is located in proximity to each cluster-center. Both the energy-density of individual clusters and that of all the clusters together is evaluated. When the global-clusters fall short, a quick re-clustering is possible. The first step is to profile the energy of all cells in the longitudinal direction (see \autoref{showerEnergyProfileFIG}), and then strip away low energy cell contributions and perform the profiling procedure again in successive  iterations. This way, it is sometimes possible to reliably locate the high density shower centers. Global clusters are then constructed around these centers. The energy density of the new clusters is compared to that of the original clusters, and the best set is finally kept.

\subsubsection{Unfolding of Mixed Clusters}

Another modification that can be made in the aftermath of the clustering procedure, is allocation of hits for mixed cluster pairs. When a pair of showers develop in close proximity to each other (in terms of their Moli\`ere radius), some cells receive energy depositions from both showers. The problem is, that the clustering procedure associates each cell with only one cluster. This biases the energy content, especially of low-energy clusters, due to the fact that their energy tends to be greatly over-estimated  by contributions from high-energy clusters. High-energy clusters are less affected, because percentage-wise, the variance in energy caused by low-energy clusters is insignificant. A way to correct for this effect is to evaluate the energy distribution of each cluster in the region furthest away from the position of its counterpart. If one assumes that the shape of each shower is smooth\footnotemark, the distribution of hits in the area where there is no mixing can be used to predict the distribution in the mixed area. Correction factors are then derived on a cell-by-cell basis, and the energy is split between the pair of clusters accordingly.

\footnotetext{In fact, the assumption of smoothness is not always correct. This is due to statistical fluctuations in the shower development, and also to the fact that the difference of the cell sizes in play are not taken into account. Despite this, the method does improve the estimation of cluster energy.}

\section{Physics Sample \label{clusteringPhysicsSample}}

The physics sample which was investigated consisted of $3 \cdot 10^{4}$ Bhabha scattering events with center-of-mass energy $\sqrt{s} = 500$~GeV. The events were generated using BHWIDE, version 1.04~\cite{revisedDetectorModelBIB6}. BHWIDE is a wide angle Bhabha MC, which contains the electro-weak contributions, which are important for the high energy $e^{+}e^{-}$ interactions considered here. The sample contains only events in which the leptons are scattered within the polar angular range $35<\theta<153$~mrad. While all of these events were processed by the clustering algorithm, some were eventually discarded. Only events in which the reconstructed cluster with the highest energy content was found within the fiducial volume (acceptance range) of LumiCal, $41.5<\theta<131$~mrad, were kept. Individual clusters were constrained in the same way. The reason for this is that showers whose position is reconstructed outside the fiducial volume are not fully contained, i.e. a significant amount of their energy leaks out of the detector, making the reconstruction process unreliable.

\begin{figure}[htp]
\begin{center}
\includegraphics[width=.49\textwidth]{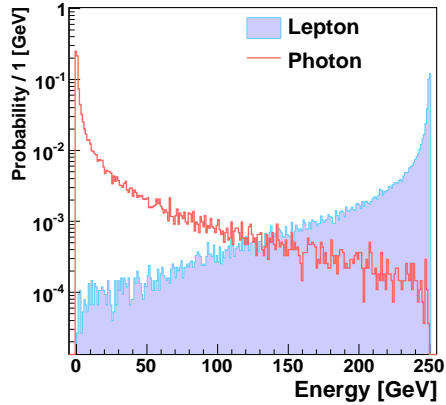}
\caption{\label{bhabhaProduction2FIG}Distribution of the production energy of scattered leptons and radiative photons for Bhabha scattering events with center-of-mass energy $\sqrt{s} = 500$~GeV in the LumiCal fiducial volume.}
\end{center}
\end{figure}

\Autoref{bhabhaProduction2FIG} shows the energy spectrum of the scattered leptons and radiative photons. The lepton distribution peaks at 250~GeV, as expected, and has a long tail of lower energies, accounting for the energy which was carried away by the photons.

\begin{figure}[htp]
\begin{center}
\subfloat[]{\label{bhabhaProduction1FIG1}\includegraphics[width=.49\textwidth]{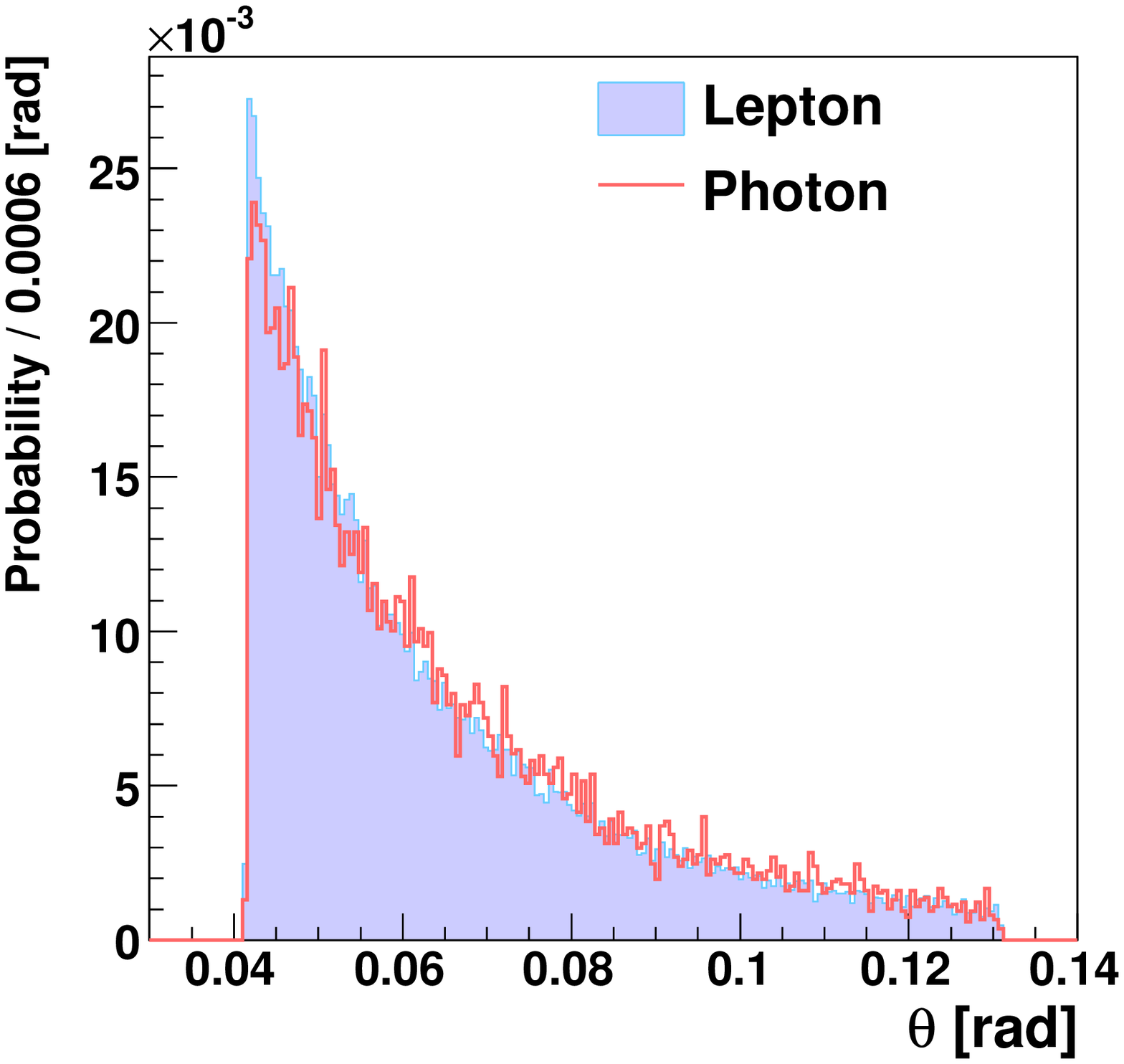}}
\subfloat[]{\label{bhabhaProduction1FIG2}\includegraphics[width=.49\textwidth]{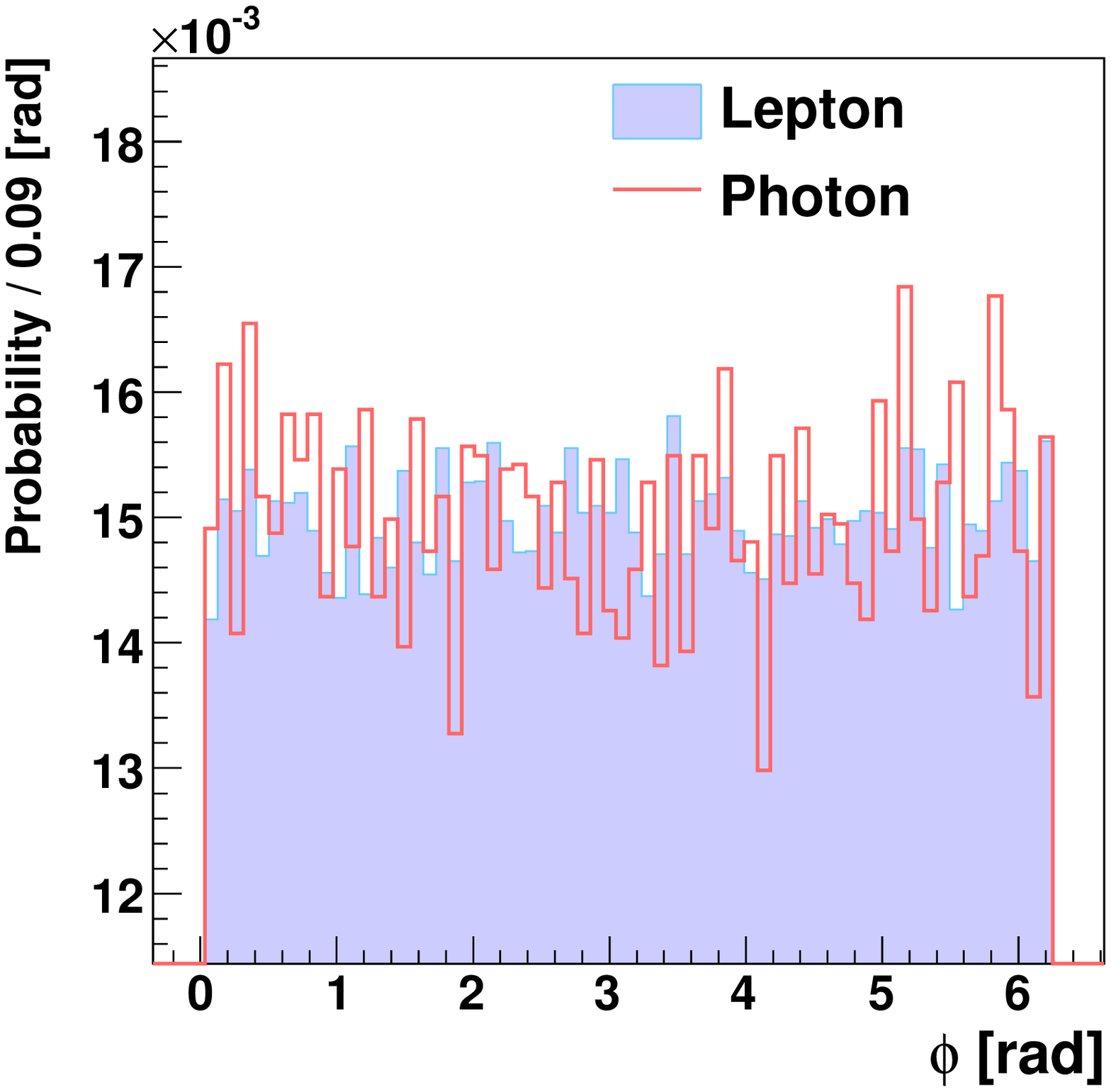}}
\caption{\label{bhabhaProduction1FIG}Distributions of the polar \Subref{bhabhaProduction1FIG1} and azimuthal \Subref{bhabhaProduction1FIG2} production angles, $\theta$ and $\phi$, of leptons and photons, as denoted in the figures. The Bhabha scattering events were simulated with center-of-mass energy $\sqrt{s} = 500$~GeV in the LumiCal fiducial volume.}
\end{center}
\end{figure}

\Autoref{bhabhaProduction1FIG} shows the polar\footnotemark and azimuthal production angles, $\theta$ and $\phi$, of scattered leptons and radiative photons. The distribution of the polar angle is cut according to the fiducial volume of LumiCal. As expected in light of \autoref{bhabhaXs2EQ}, the distribution of the polar angle falls off rapidly with $\theta$, and the distribution of the azimuthal angle is flat.

\footnotetext{Naturally the electron and the positron have polar angles of opposite signs, but as the distributions of the production angles are equivalent for either one, this sign will be ignored throughout the following.}

\begin{figure}[htp]
\begin{center}
\subfloat[]{\label{bhabhaProduction3FIG1}\includegraphics[width=.49\textwidth]{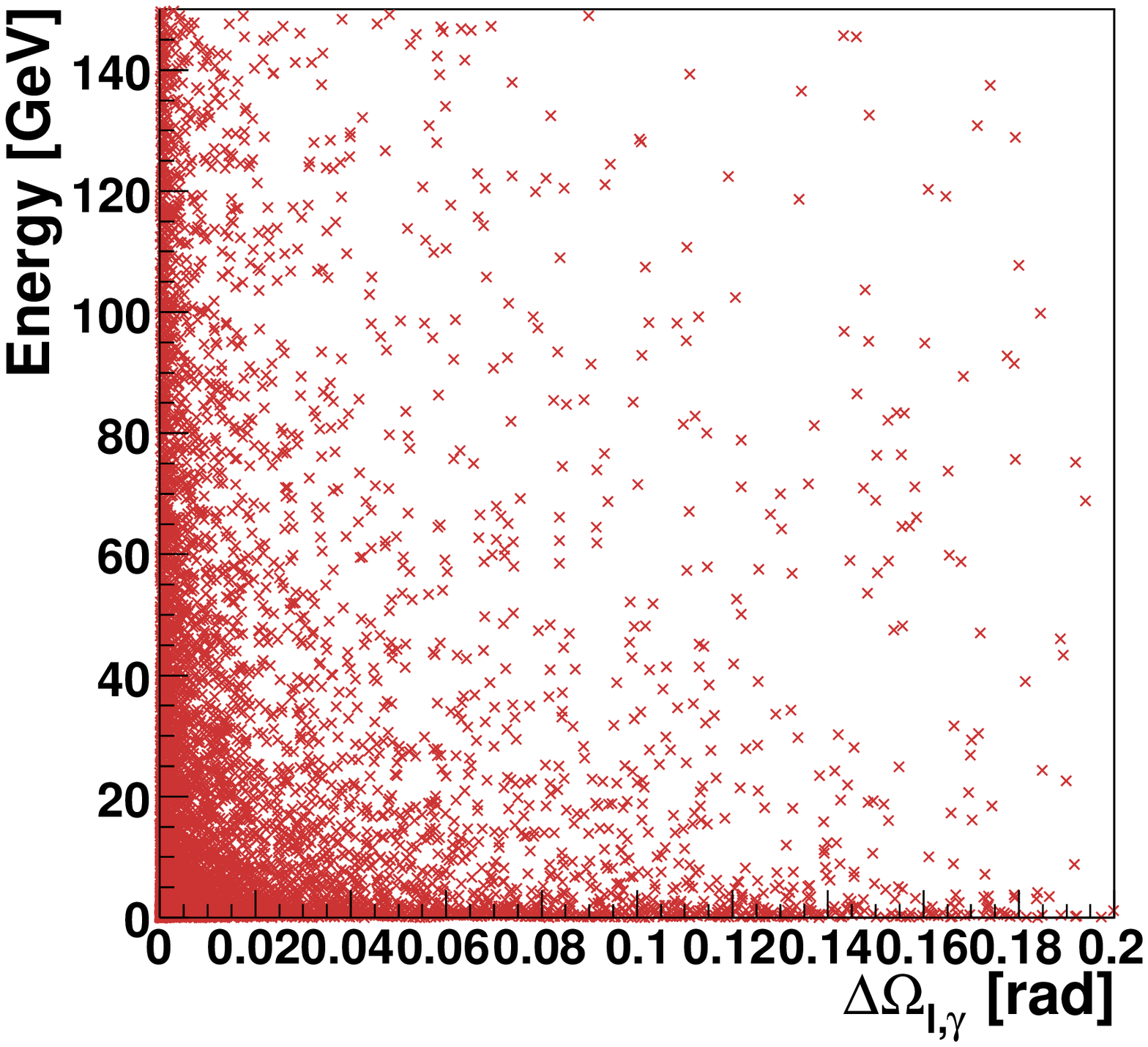}}
\subfloat[]{\label{bhabhaProduction3FIG2}\includegraphics[width=.49\textwidth]{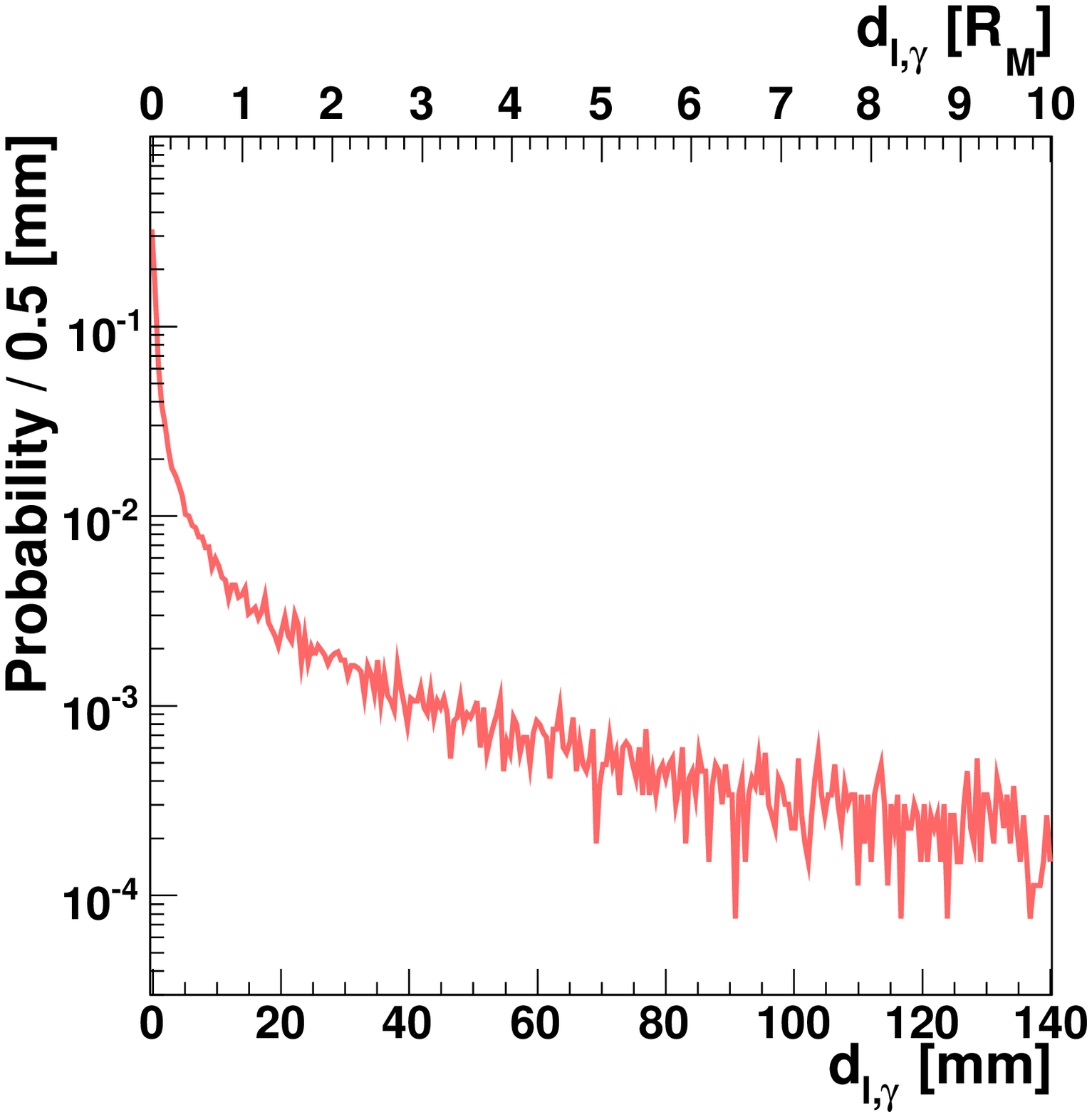}}
\caption{\label{bhabhaProduction3FIG}\Subref{bhabhaProduction3FIG1}  Correlation between the angular separation between leptons and radiative photons, $\Delta\Omega_{\ell,\gamma}$, and the photon energy. The spectrum of particles is generated for Bhabha scattering with center-of-mass energy $\sqrt{s} = 500$~GeV in the LumiCal fiducial volume. \Subref{bhabhaProduction3FIG2} Energy profile of the distribution in \Subref{bhabhaProduction3FIG1}. The distance in this case is expressed in units of mm, as the separation between the pair of particles on the face of LumiCal, $d_{\ell,\gamma}$. An equivalent scale is also shown in units of Moli\`ere radius, $R_{\mc{M}}$.}
\end{center}
\end{figure}

Since most initial state radiative photons travel through the beampipe and are undetected (see \autoref{bhabhaScatteringIntroSEC}), only final state photons are considered. Conservation of momentum dictates that the more energy these photons take from the lepton, the smaller the angular separation between the two. This is confirmed by \autoref{bhabhaProduction3FIG1}, which shows the correlation between the photon energy and its angular separation from the accompanying lepton, $\Delta\Omega_{\ell,\gamma}$. In \autoref{bhabhaProduction3FIG2} the energy dependence of \autoref{bhabhaProduction3FIG1} is integrated and normalized, showing the event rate for the photon emission. The distance in this case is expressed  as the separation between the pair of particles on the face of LumiCal in units of mm and of Moli\`ere radius. It is apparent from the distributions that the vast majority of radiative photons is of low energy, and enter LumiCal in close proximity to the lepton.

\section{Performance of the Clustering Algorithm}

The distributions for the position and energy presented in the previous section were drawn from the raw output of the BHWIDE event generator. As such, they represent an ideal description of Bhabha scattering. In reality, observables are distorted by the inherent resolution of the measuring device. The energy resolution, which is determined by the amount of leakage, and by the sampling rate of the calorimeter (see \autoref{revisedDetectorModelCH}), incurs an error on the signal-to-energy calibration of LumiCal. Similarly, the polar and azimuthal reconstructed angles have a resolution, and also a bias, of their own. In order to analyze the output of the clustering algorithm it is necessary to isolate the errors in reconstruction resulting from the clustering, from the other systematic uncertainties of LumiCal.

To this effect, two classes of objects may be defined. The basic simulation-truth data will be represented by showers, which contain all of the hits which belong to an EM shower initiated by a single particle. These will be referred to as \textit{generated showers}. Since a single detector cell may contain contributions from more than one EM shower, generated showers may share cells. Hit collections, built by the clustering algorithm, will be referred to as \textit{reconstructed clusters}. In order to remove the systematic uncertainties, the properties of both the showers and the clusters are reconstructed in the same manner, using information from the detector cells.

Since there is no way to distinguish in practice between EM showers initiated by leptons and those started by photons, reconstructed clusters and generated showers will be referred to as having either \textit{high-energy}, or \textit{low-energy}, which correspond to \textit{effective leptons}, and \textit{effective photons} respectively. High-energy clusters (showers) are identified as those that have the highest integrated energy content among the set of all reconstructed clusters (generated showers). The rest of the clusters (showers) are identified as low-energy clusters (showers).

\subsection{Event Selection \label{clusteringEventSelectionSEC}}

\begin{figure}[htp]
\begin{center}
\subfloat[]{\label{missClusteredSeparatedFIG1}\includegraphics[height=0.37\textheight]{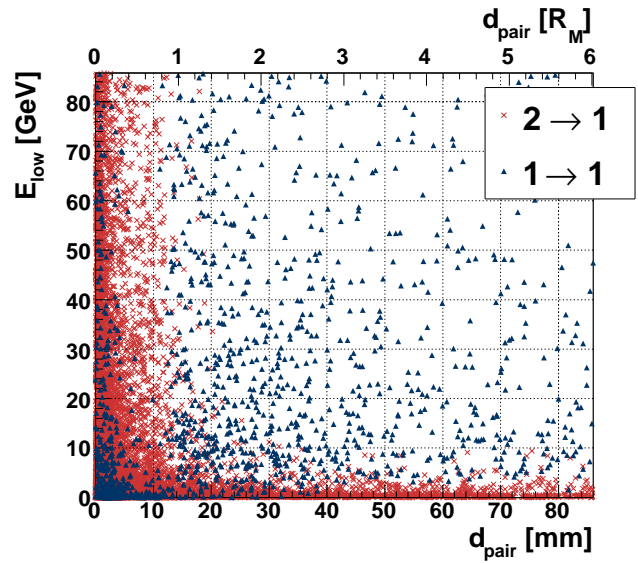}} \\
\subfloat[]{\label{missClusteredSeparatedFIG2}\includegraphics[height=0.37\textheight]{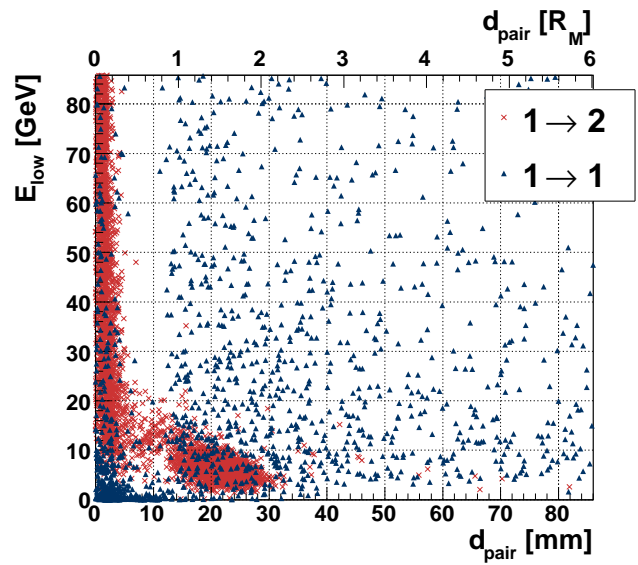}}
\caption{\label{missClusteredSeparatedFIG}Success and failure of the clustering algorithm in distinguishing between a pair of generated showers as a function of the separation distance between the pair, $d_{pair}$, and of the energy of the low-energy shower, $E_{low}$. The distance $d_{pair}$ is expressed in units of mm and of Moli\`ere radius, $R_{\mc{M}}$. Both figures show success $(1 \rightarrow 1)$ of the algorithm; \Subref{missClusteredSeparatedFIG1} also shows cases where a pair of generated showers are merged into one reconstructed cluster $(2 \rightarrow 1)$, and \Subref{missClusteredSeparatedFIG2} also shows cases where a single shower is split into two clusters $(1 \rightarrow 2)$.}
\end{center}
\end{figure}

\Autoref{missClusteredSeparatedFIG} shows the success and failure of the clustering algorithm in distinguishing between a pair of generated showers as a function of the separation distance between the pair, $d_{pair}$, and of the energy of the low-energy shower, $E_{low}$. Failure of the algorithm may take two forms. A pair of generated showers may be merged into one reconstructed cluster (\autoref{missClusteredSeparatedFIG1}), or one shower may be split into two clusters (\autoref{missClusteredSeparatedFIG2}). As expected, since the great majority of radiative photons enter LumiCal within a small distance from the leptons, separation between the showers of the two particles is not trivial. The difficulty is enhanced due to the increasing size of showers, as they develop in depth in LumiCal. Distinguishing between pairs of showers becomes easier when either $E_{low}$ or $d_{pair}$ increase in value. 

It is, therefore, required to set low bounds on the energy of a cluster, and on the separation between any pair of clusters. When the algorithm produces results that do not pass the cuts, the two clusters are integrated into one. In order to compare with theory the distribution of clusters after making this \textit{merging-cut} on $E_{low}$ and $d_{pair}$, one must also apply the same restrictions on the generated showers. The generated showers follow a distribution complying with an \textit{effective Bhabha cross-section}.

The distinction between the original and the effective cross-sections is important, and it must be noted that the effective cross-section can only be computed by simulating the detector response. The position of a cluster is reconstructed by making a cut on cell energy, relative to the entire cluster energy (see \autoref{logWeighClustering2EQ}). As a result, an integration of a pair of clusters into one, sets the position of the merged cluster to an a-priori unpredictable value\footnotemark. The momentum of the initiating particles will, in some cases, not balance with that of the effective (merged) particle. Summing up deposits from multiple showers in LumiCal is, therefore, not equivalent to any summation procedure that might be done on the cross-section, at the generated-particle level.

\footnotetext{For instance, if a cluster is of much higher energy than its counterpart, then the energy contributions of the low-energy cluster will not be taken into account in the position reconstruction at all. For the reconstruction of single showers this is not a problem, since the shower is of homogeneous shape around a defined center. For the case of two showers, which are far apart, this is no longer the case.}

\subsection{Observables \label{clusteringObservablesSEC}}

\subsubsection{Quantification of the Performance}

The error on the effective cross-section will depend on the number of miscounted showers. In order to judge the success of the algorithm, one may evaluate its acceptance, $\mc{A}$, purity, $\mc{P}$, and efficiency, $\mc{E}$, which are defined as

\begin{equation}{
\mc{A} = \frac{N_{1 \rightarrow 1}}{N_{1 \rightarrow 1} + N_{2 \rightarrow 1}} \, , \quad
\mc{P} = \frac{N_{1 \rightarrow 1}}{N_{1 \rightarrow 1} + N_{1 \rightarrow 2}} \, , \quad \mathrm{and}  \quad
\mc{E} = \frac{\mc{A}}{\mc{P}} \, ,  
} \label{apeEQ} \end{equation}

\noindent where $N_{1 \rightarrow 1}$ is the number of generated showers which were reconstructed as one cluster by the algorithm, $N_{2 \rightarrow 1}$ is the number of pairs of showers which were reconstructed as one cluster, and $N_{1 \rightarrow 2}$ is the number of single showers which were separated into two reconstructed clusters.

The values of the acceptance, purity and efficiency are presented in \autoref{apeTABLE} for several pairs of merging-cuts on the minimal energy and the separation distance between a pair of clusters. Also shown is the fraction of radiative photons which are available for reconstruction after applying the merging-cuts, 

\begin{equation}{
\wp_{\gamma} = \frac{N_{\gamma}(\mathrm{cut})}{N_{\gamma}(\mathrm{all})} \, ,
} \label{photonFracEQ} \end{equation}

\noindent where $N_{\gamma}(\mathrm{all})$ is the total number of radiative photons in the fiducial volume of LumiCal, and $N_{\gamma}(\mathrm{cut})$ is the number of photons in LumiCal which also pass the merging-cuts on $E_{low}$ and $d_{pair}$.

\begin{table}[htp]
\begin{center} \begin{tabular}{ |c|c|c|c|c|c| }
\hline
\multicolumn {2}{|c|}{Cuts}  &
\multirow {2}{*}{\; $\wp_{\gamma} ~[\%]$ \;}  &
\multirow {2}{*}{\; $\mc{A} ~[\%]$ \;}  &
\multirow {2}{*}{\; $\mc{P} ~[\%]$ \;}  &
\multirow {2}{*}{\; $\mc{E} ~[\%]$ \;}  \\  \cline{1-2}
\; $d_{pair} ~ [R_{\mc{M}}]$ \; & $E_{min}$ [GeV] & & & & \\ [2pt]
\hline \hline
0.5  &  25  & 6.6  & 69  & 96  &  71 \\ \hline
0.75 &  20  & 5.9  & 85  & 95  & 90  \\ \hline
0.75 &  25  & 5.2  & 58  & 96  & 89  \\ \hline
1    &  15  & 6    & 94  & 93  & 100 \\ \hline
1    &  20  & 5.2  & 95  & 95  & 100 \\ \hline
1    &  25  & 4.6  & 95  & 96  & 98  \\ \hline
1.5  &  20  & 4.3  & 99  & 98  & 100 \\ \hline
\end{tabular} \end{center}
\caption{\label{apeTABLE}The values of the percentage of photon showers, which are available for reconstruction, $\wp_{\gamma}$, and of the acceptance, $\mc{A}$, purity, $\mc{P}$, and efficiency, $\mc{E}$, of the algorithm, for several pairs of merging-cuts on the minimal energy of a cluster, $E_{min}$, and on the separation distance between a pair of clusters, $d_{pair}$. The merging-cut $d_{pair}$ is expressed in units of the Moli\`ere radius, $R_{\mc{M}}$.}
\end{table}

%
%
%
%
%

The relative error of the effective cross-section as a result of miscounting depends on the observed number of effective leptons and photons, and on the fractions of miscounted events out of the relevant event population. The probability of finding a given value for $N_{1 \rightarrow 2}$ or for $N_{2 \rightarrow 1}$ is given by the binomial distribution, and so the relative error on either one is

\begin{equation}
\frac{\Delta N_{\ell,\gamma}}{N_{\ell,\gamma }} = \frac{\sqrt{N_{\ell,\gamma } p q}}{N_{\ell,\gamma }} = \sqrt{\frac{p q}{N_{\ell,\gamma }}} \, ,
\label{missCountingRelErr1EQ} \end{equation}

\noindent where $p$ is the probability to miscount in a given event, $q = 1 - p$, and $N_{\ell,\gamma }$ is either the number of effective leptons, $N_{\ell}$, or the number of effective photons, $N_{\gamma}$, depending on the type of miscounting\footnotemark. Values for $p$ and $q$ were derived from running the clustering algorithm on the sample of Bhabha events with different sets of merging-cuts on $E_{low}$ and $d_{pair}$. The corresponding relative errors are shown in \autoref{clusteringErrorTABLE}. Also shown there is the relative error

\footnotetext{When computing the number of single showers which were split into two clusters, $N_{\ell,\gamma } \rightarrow N_{\ell}$, since the candidate for false splitting will come from the population of single (lepton) showers. On the other hand, merging of two showers into one may only happen when a photon shower exists, so that in this case $N_{\ell,\gamma } \rightarrow N_{\gamma}$. This distinction is important, as the number of effective leptons far outweighs the number of effective photons.}

\begin{equation}
\frac{\Delta N_{tot}}{N_{tot}} = \left( \frac{\Delta N}{N} \right)_{1 \rightarrow 2} \oplus \; \left( \frac{\Delta N}{N} \right)_{2 \rightarrow 1} \quad ,
\label{missCountingRelErr2EQ} \end{equation}

\noindent which corresponds to the total error resulting from both types of miscounting, rescaled for an integrated luminosity of $500~\mathrm{fb}^{-1}$ .

\begin{table}[htp]
\begin{center} \begin{tabular}{ |c|c|c|c|c| }
\hline
\multicolumn {2}{|c|}{Cuts}  &
\multirow {2}{*}{\; \begin{large}$ \frac{\Delta N_{1 \rightarrow 2}}{N_{1 \rightarrow 2}}$ \end{large}} &  
\multirow {2}{*}{\; \begin{large}$ \frac{\Delta N_{2 \rightarrow 1}}{N_{2 \rightarrow 1}}$ \end{large}} &  
\multirow {2}{*}{\; \begin{large} $\frac{\Delta N_{tot}}{N_{tot}}$ \end{large} \;} \\  \cline{1-2}
\; $d_{pair} ~ [R_{\mc{M}}]$ \; & $E_{min}$ [GeV] & & & \\ [2pt]
\hline \hline
0.5   &  25  & $4.2\cdot10^{-4}$   & $31.5\cdot10^{-2}$  & $10.3\cdot10^{-5}$ \\ \hline
0.75  &  20  & $7.6\cdot10^{-4}$   & $14.6\cdot10^{-2}$  & $7.5\cdot10^{-5}$  \\ \hline
0.75  &  25  & $5.4\cdot10^{-4}$   & $14.6\cdot10^{-2}$  & $8\cdot10^{-5}$    \\ \hline
1     &  15  & $12.9\cdot10^{-4}$  & $6.3\cdot10^{-2}$   & $5.1\cdot10^{-5}$  \\ \hline
1     &  20  & $7\cdot10^{-4}$     & $4.6\cdot10^{-2}$   & $4.9\cdot10^{-5}$  \\ \hline
1     &  25  & $5.1\cdot10^{-4}$   & $5.2\cdot10^{-2}$   & $5.3\cdot10^{-5}$  \\ \hline
1.5   &  20  & $3.2\cdot10^{-4}$   & $0.5\cdot10^{-2}$   & $1.8\cdot10^{-5}$  \\ \hline
\end{tabular} \end{center}
\caption{\label{clusteringErrorTABLE}The relative errors on the miscounting of clusters (using $N_{1 \rightarrow 2}$ and $N_{2 \rightarrow 1}$), and the total relative error on the measurement of the effective Bhabha cross-section (using $N_{tot}$), for several pairs of merging-cuts on the minimal energy of a cluster, $E_{min}$, and on the separation distance between a pair of clusters, $d_{pair}$. The merging-cut $d_{pair}$ is expressed in units of the Moli\`ere radius, $R_{\mc{M}}$. The relative errors of the numbers for miscounted showers, $N_{1 \rightarrow 2}$ and $N_{2 \rightarrow 1}$, are computed for a sample of $3 \cdot 10^{4}$ Bhabha events. The number $N_{tot}$ takes into account both of the miscounting errors, and is computed for an integrated luminosity of $500~\mathrm{fb}^{-1}$.}
\end{table}

It is apparent from \autorefs{apeTABLE} and \ref{clusteringErrorTABLE} that achieving a minimum of error in counting the number of effective photons, depends both on the size of the sample of available photons, and on the sensitivity of the algorithm to miscounting. For merging-cuts in energy~$\ge 20$~GeV and distance~$\ge R_{\mc{M}}$, the algorithm makes relatively few mistakes. The decision on where exactly to set the merging-cuts reduces to the choice of maximizing the measurable amount of statistics.

\subsubsection{Event-by-Event Comparison of Observables}

Other than counting the number of low and high-energy clusters and comparing the results to the expected numbers, deduced from the effective Bhabha cross-section, the properties of the clusters may also be evaluated. For this purpose, one may produce such distributions as the production angles of clusters, the angular separation between pairs of clusters, and the value of cluster-energy. A first step in this process is to look at the shower/cluster differences on an event-by-event basis.

\begin{figure}[htp]
\begin{center}
\label{engyRecGenDifferanceEvtByEvtFIG1}\includegraphics[width=.49\textwidth]{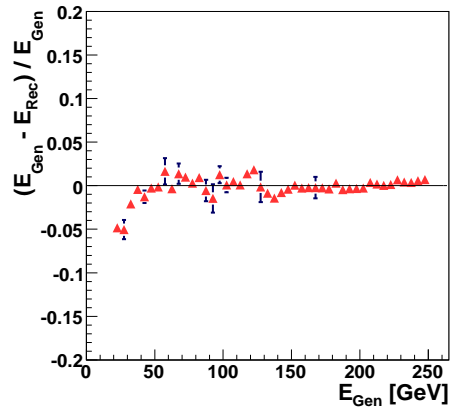}
\caption{\label{engyRecGenDifferanceEvtByEvtFIG}The normalized difference between the energy of generated showers, $E_{gen}$, and their respective reconstructed clusters, $E_{rec}$, as a function of the energy of the generated shower.}
\end{center}
\end{figure}

The energy of the particle which initiated a generated shower (reconstructed cluster) is determined by integrating all the contributions of the shower (cluster) and multiplying by a calibration constant. The constant transforms between the values of the detector signal and the particle energy, and is determined by a calibration curve, such as the one shown in \autoref{engyCalibrationFIG} (\autoref{digitizationSEC}). \Autoref{engyRecGenDifferanceEvtByEvtFIG} shows the normalized difference between the energy of reconstructed clusters and their respective generated showers, as a function of the energy of the generated shower. The fluctuations are of $\mc{O}(2~\mathrm{GeV})$ or lower. The reconstructed clusters, which are taken into account here, belong to the effective Bhabha cross-section for which $E_{low} \ge 20$~GeV and $d_{pair} \ge R_{\mc{M}}$. Increasing the merging-cut on separation distance reduces the fluctuations significantly, due to reduced cluster-mixing (see \autoref{clusterMixingSEC}), but also reduces the available statistics.

\begin{figure}[htp]
\begin{center}
\subfloat[]{\label{posRecGenDifferanceEvtByEvtFIG1}\includegraphics[width=.49\textwidth]{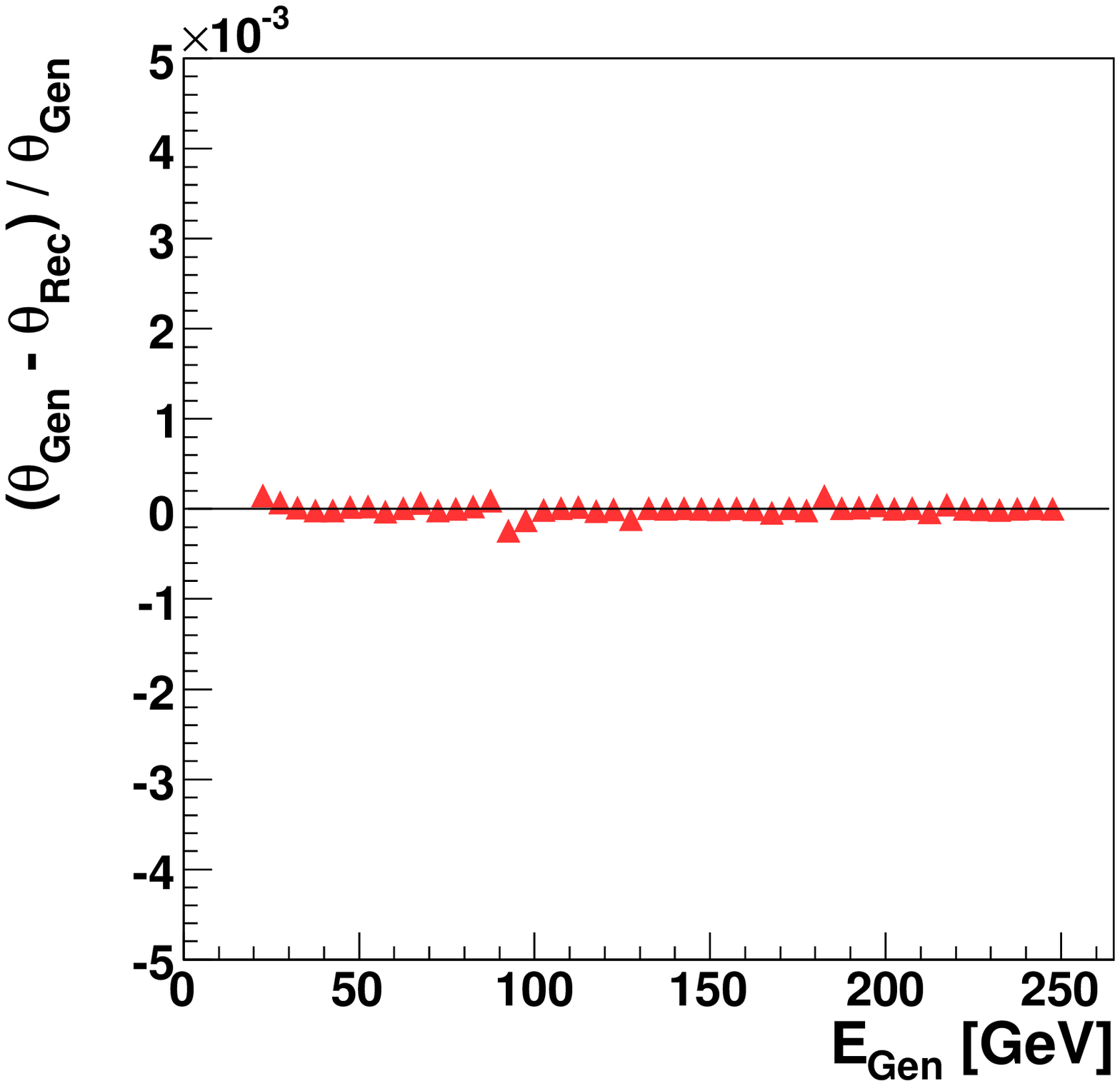}}
\subfloat[]{\label{posRecGenDifferanceEvtByEvtFIG2}\includegraphics[width=.49\textwidth]{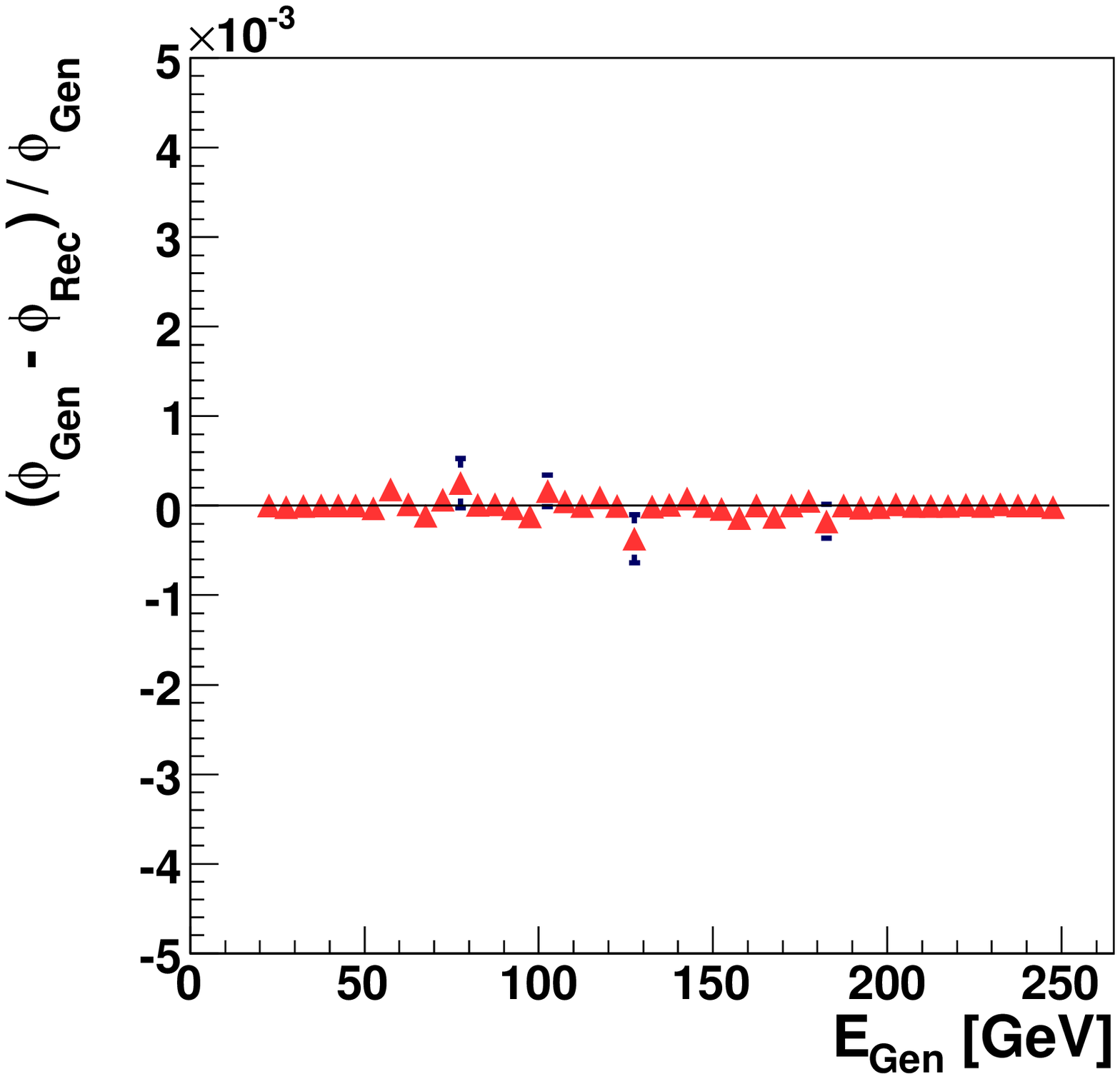}}
\caption{\label{posRecGenDifferanceEvtByEvtFIG}The normalized difference between the position of generated showers and their respective reconstructed clusters. The position is parameterized by the reconstructed and the generated polar angles, $\theta_{Rec}$ and $\theta_{Gen}$, \Subref{posRecGenDifferanceEvtByEvtFIG1} and by the reconstructed and the generated azimuthal angles, $\phi_{Rec}$ and $\phi_{Gen}$, \Subref{posRecGenDifferanceEvtByEvtFIG2} and are presented as a function of the energy of the generated shower, $E_{Gen}$.}
\end{center}
\end{figure}

\Autoref{posRecGenDifferanceEvtByEvtFIG} shows the normalized difference between the position of reconstructed clusters and their respective generated showers. The position is parameterized by the polar angle, $\theta$, and the azimuthal angle, $\phi$. The difference is presented as a function of the energy of the generated shower. The angles $\theta$ and $\phi$ are reconstructed according to \autorefs{logWeighClustering1EQ} and \ref{logWeighClustering2EQ}. Since the fluctuations in all cases are of $\mc{O}(10^{-5})$ or lower, it is concluded that the position reconstruction is performed well. This makes sense in light of \autoref{logWeighClustering1EQ}, since only the core of high energy cells, which are in close proximity to the cluster center, contribute to the position reconstruction. Low-energy cells which are miss-assigned between clusters, therefore, do not degrade the reconstruction.

\subsubsection{Measurable Distributions}

\begin{figure}[htp]
\begin{center}
\subfloat[]{\label{engyRecGenDistributionFIG1}\includegraphics[width=.49\textwidth]{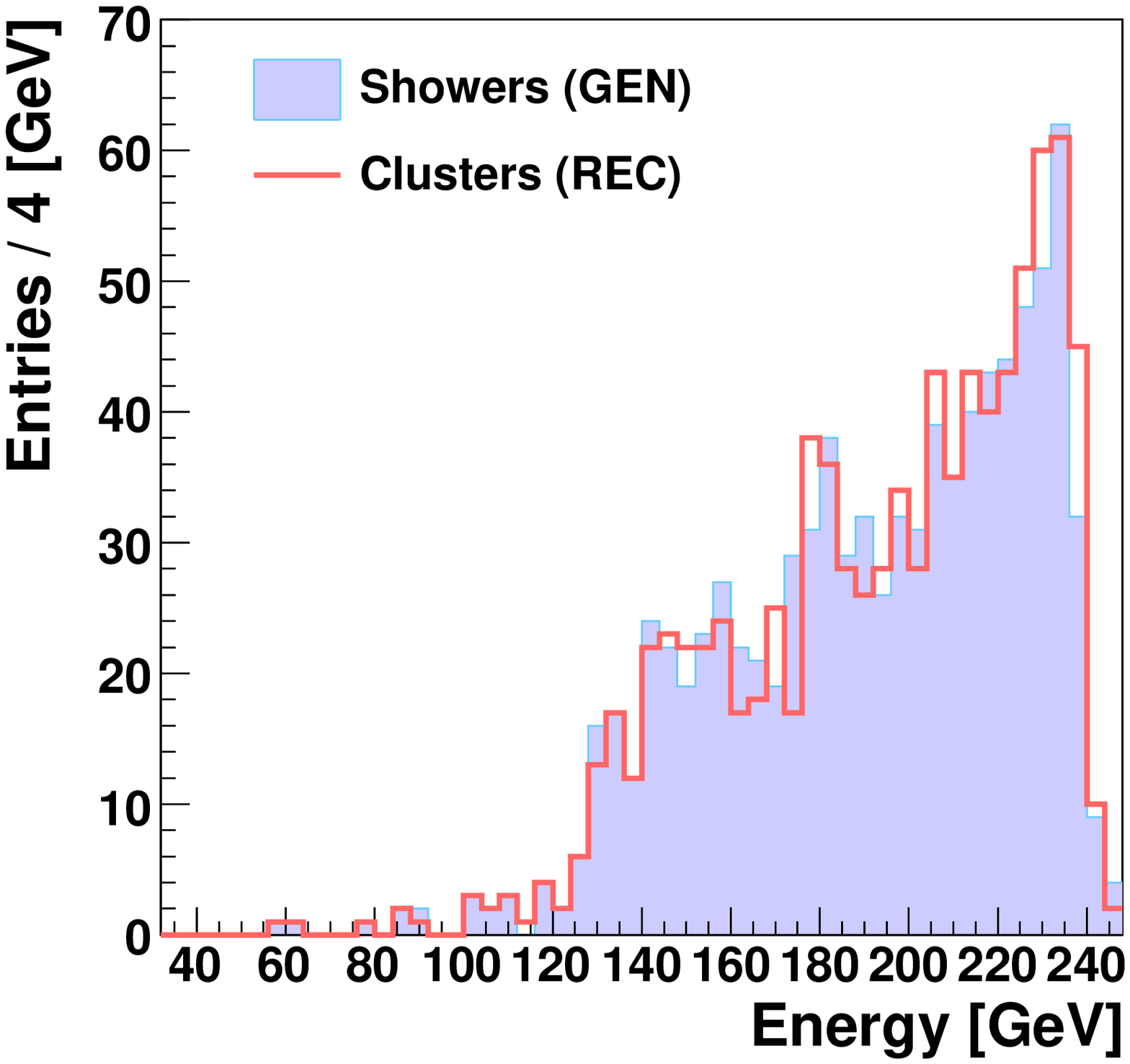}}
\subfloat[]{\label{engyRecGenDistributionFIG2}\includegraphics[width=.49\textwidth]{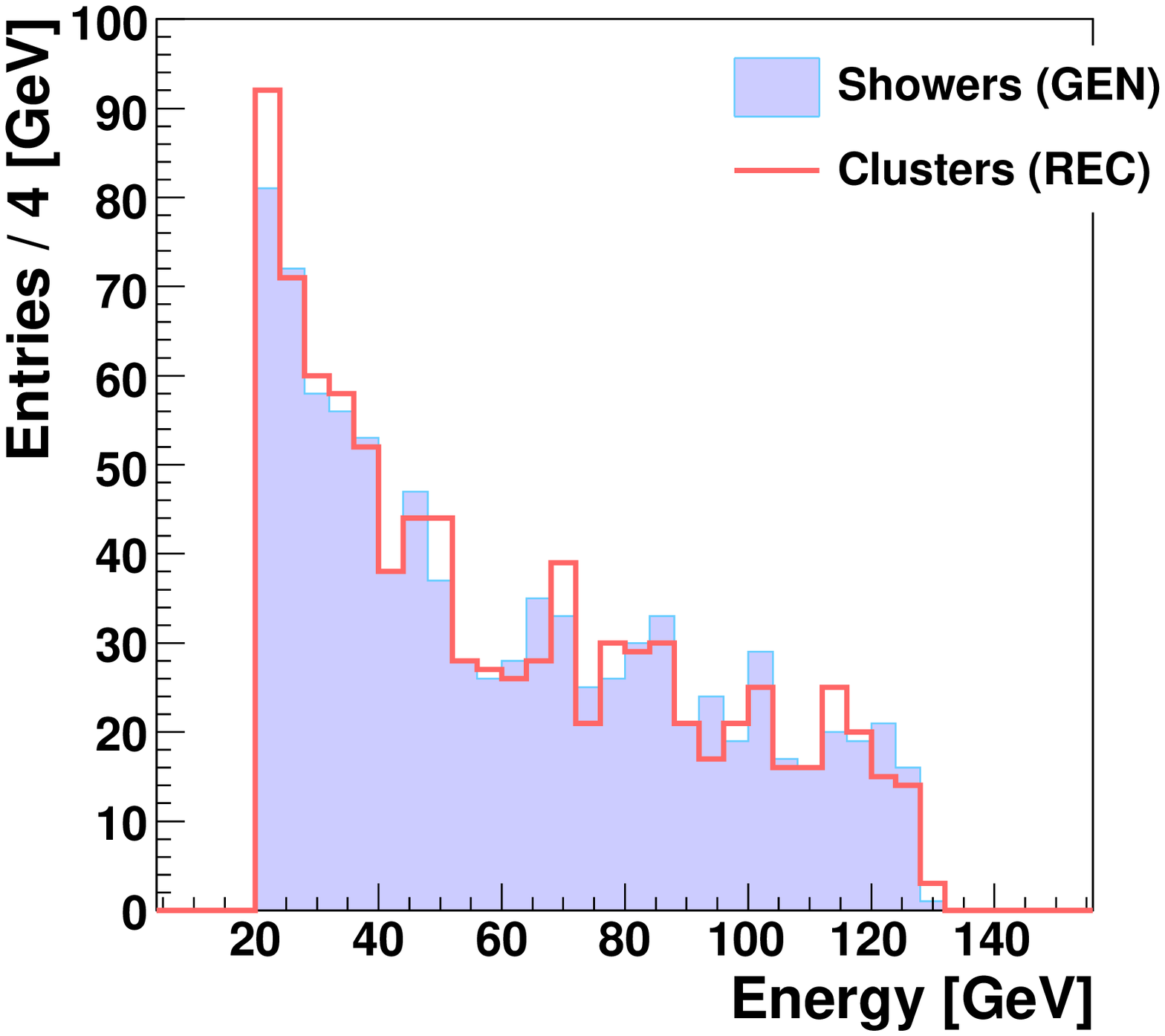}}
\caption{\label{engyRecGenDistributionFIG}Distributions of the energy of reconstructed clusters (REC) and their respective generated showers (GEN), as denoted in the figures. The sample is divided into high \Subref{engyRecGenDistributionFIG1} and low-energy \Subref{engyRecGenDistributionFIG2}  clusters (showers).}
\end{center}
\end{figure}

The distribution of  the energy of reconstructed clusters and their respective generated showers for high\footnotemark and low-energy clusters (showers) is shown in \autorefs{engyRecGenDistributionFIG1} and \ref{engyRecGenDistributionFIG2}, respectively.

\footnotetext{It may be noticed that the distribution of high-energy clusters has values smaller than 125~GeV (less than half the maximal possible energy). This is due to the fact that occasionally part of the energy is not included in the reconstruction. This may occur when some of the particles do not enter LumiCal, or when the position of a shower is reconstructed outside the fiducial volume, and so all cell-information is discarded.}

\begin{figure}[htp]
\begin{center}
\subfloat[]{\label{posRecGenDistributionFIG1}\includegraphics[width=.49\textwidth]{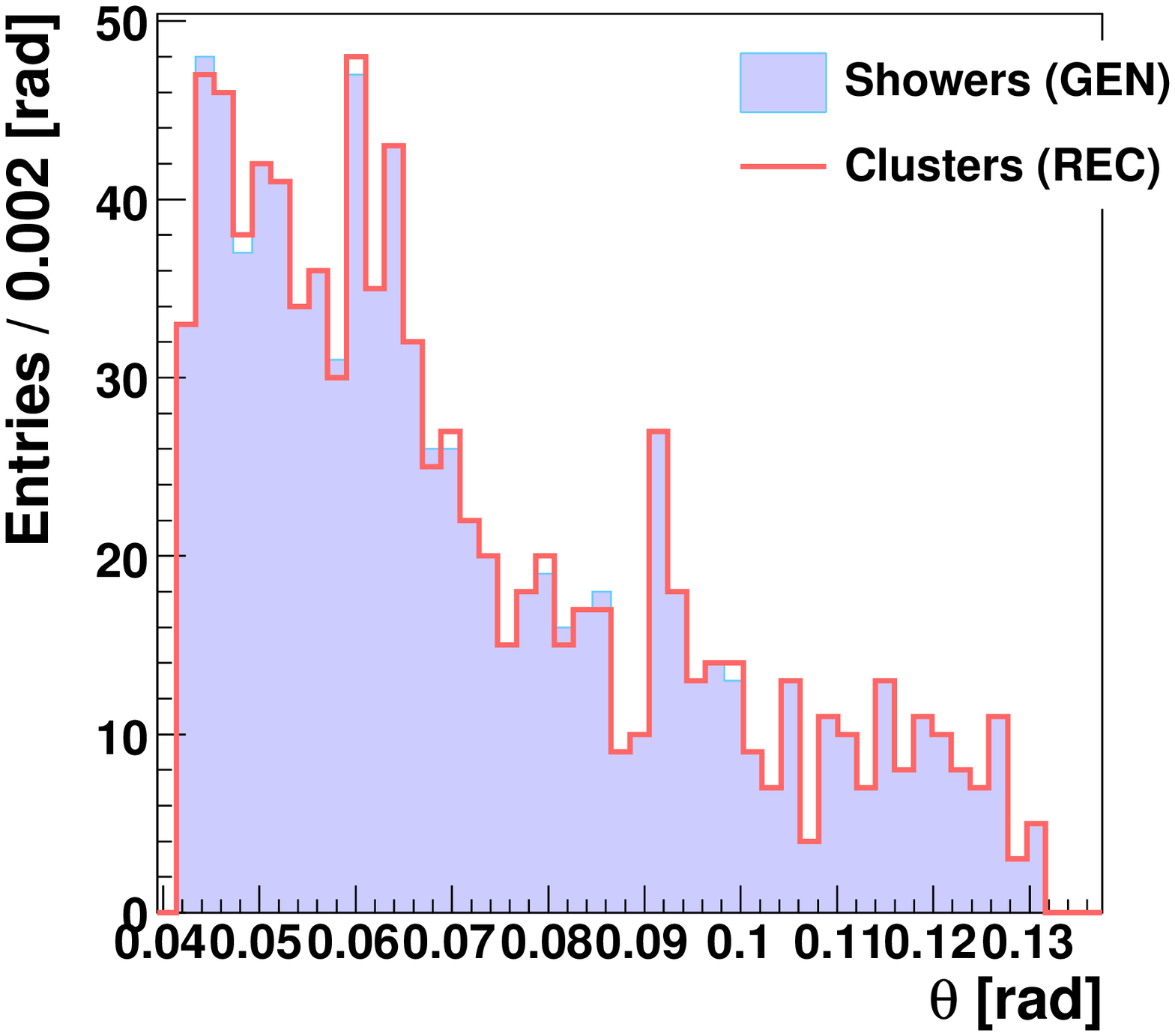}}
\subfloat[]{\label{posRecGenDistributionFIG2}\includegraphics[width=.49\textwidth]{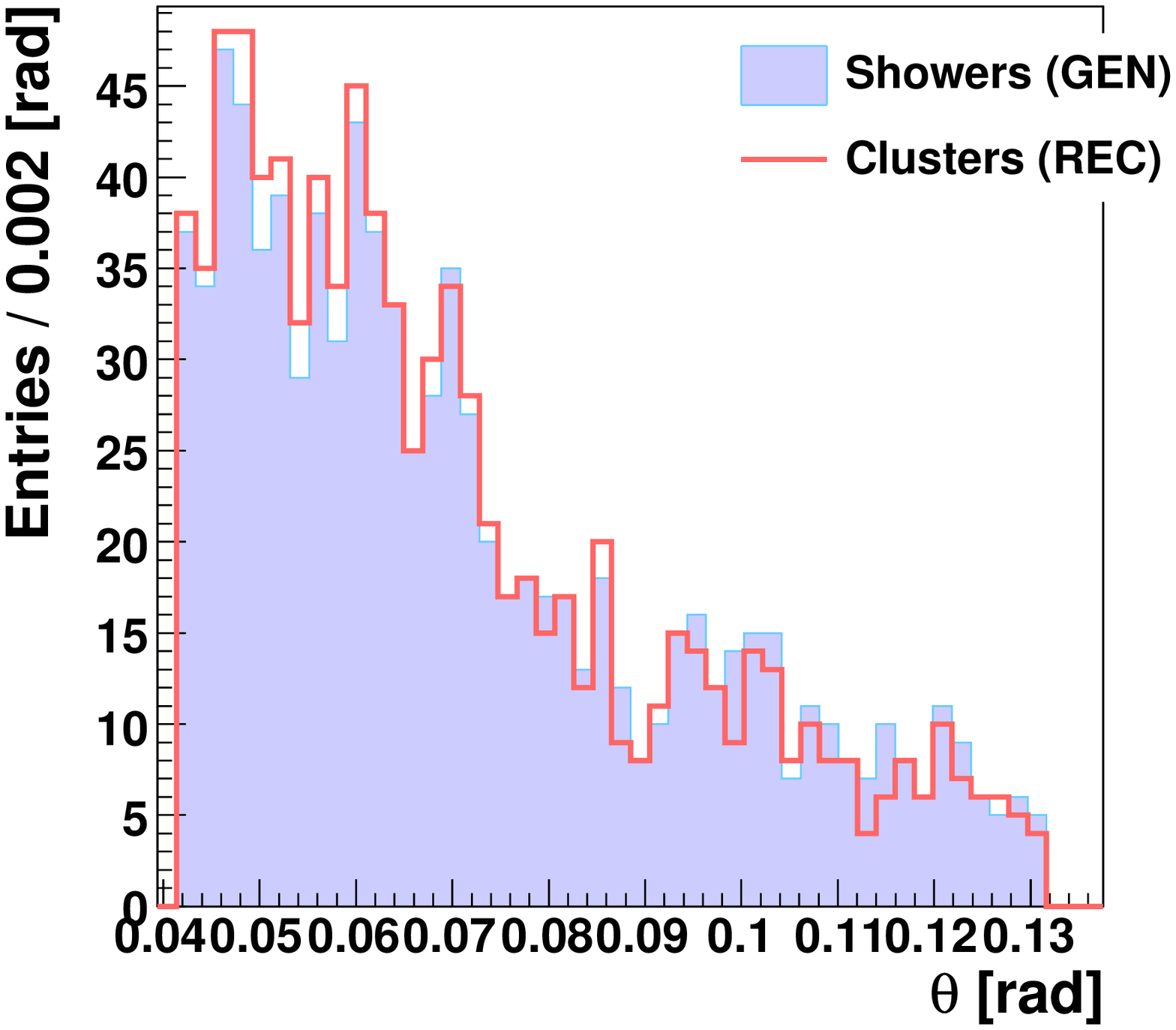}}
\caption{\label{posRecGenDistributionFIG}Distributions of the polar angle, $\theta$, of reconstructed clusters (REC) and their respective generated showers (GEN), as denoted in the figures. The sample is divided into high \Subref{posRecGenDistributionFIG1} and low-energy \Subref{posRecGenDistributionFIG2}  clusters (showers).}
\end{center}
\end{figure}

\Autoref{posRecGenDistributionFIG} shows the distributions of the polar angle, $\theta$, of reconstructed clusters and their respective generated showers. The sample is divided into high and low-energy clusters (showers). In \autoref{thetaHighMinLowDistributionFIG} is presented the distribution of the difference in polar angle, $\Delta \theta_{high,low} \equiv \theta_{high} - \theta_{low}$, between the high and low reconstructed clusters and their respective generated showers.

\begin{figure}[htp]
\begin{center}
\label{thetaHighMinLowDistributionFIG}\includegraphics[width=.49\textwidth]{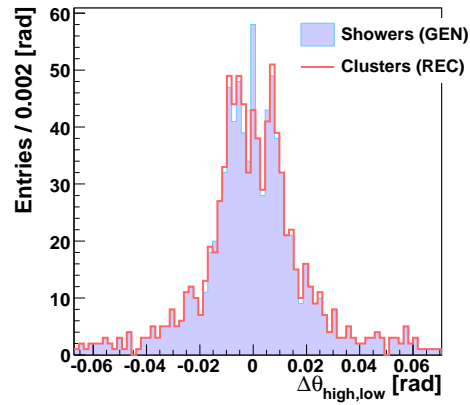}
\caption{\label{thetaHighMinLowDistributionFIG}Distributions of the difference in polar angle, $\Delta \theta_{high,low}$, between the high and low reconstructed clusters (REC) and their respective generated showers (GEN), as denoted in the figure.}
\end{center}
\end{figure}

\begin{figure}[htp]
\begin{center}
\label{thetaHighMinLowDistributionFIG}\includegraphics[height=0.37\textheight]{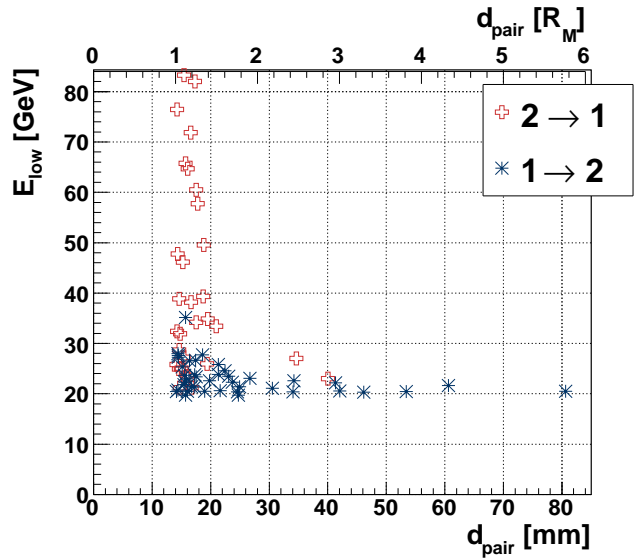}
\caption{\label{missClusteredSeparatedAfterSumFIG}Instances of failure of the clustering algorithm in distinguishing between a pair of generated showers, as a function of the separation distance between the pair, $d_{pair}$, and of the energy of the low-energy shower, $E_{low}$. The distance $d_{pair}$ is expressed in units of mm and of Moli\`ere radius, $R_{\mc{M}}$. Two cases are possible, a pair of generated showers may be merged into one reconstructed cluster $(2 \rightarrow 1)$, or one shower may be separated into two clusters $(1 \rightarrow 2)$. The event sample considered complies with the merging-cuts $E_{low} \ge 20$~GeV and $d_{pair} \ge R_{\mc{M}}$.}
\end{center}
\end{figure}

In light of the relations shown in \autoref{posRecGenDifferanceEvtByEvtFIG} one might naively expect that the match between the distributions of cluster and shower positions would be better. The small noticeable discrepancies originate from miscounted events. On a case-by-case basis the difference in reconstruction of the polar and azimuthal angles is usually below the resolving power of LumiCal. However, single showers which are reconstructed as two clusters, and shower pairs that are reconstructed as single clusters, must also be taken into account. The distributions in \autoref{missClusteredSeparatedFIG} indicate that the sources of the discrepancies are showers with small angular separation. This is indeed the case, as can be deduced from \autoref{missClusteredSeparatedAfterSumFIG}, where the instances of failure of the algorithm are shown as a function of the separation distance between pairs of showers, $d_{pair}$, and of the energy of the low-energy shower, $E_{low}$. The merging-cuts $E_{low} \ge 20$~GeV and $d_{pair} \ge R_{\mc{M}}$ have been used for selection of reconstructed clusters. The algorithm tends to produce mistakes when merging showers for which $E_{low}$ and $d_{pair}$ are close to the merging-cut values, which is due to errors in either the position or the energy reconstruction. Thus, it is possible to improve the results of the comparison between generated showers to reconstructed clusters, by making a selection-cut on events with low-cluster energy and cluster-pair distance, which are close to the merging-cut values.

\section{Conclusions on Clustering}

It has been shown that clustering of EM showers in LumiCal is possible. In order to achieve results of high acceptance and purity, a merging-cut on minimal energy for each cluster, and on the separation distance between any pair of clusters, needs to be made. The merging leads to a measurement of an effective Bhabha cross-section. The number of effective photons may then be counted with an uncertainty that corresponds to the required precision for the measurement of the luminosity spectrum. The distributions of the position and energy of the effective leptons and photons may also be measured and compared to the expected results. A merging-cut should be used in this case for summation of clusters, as is done for the counting of effective photons. Imposing an additional selection-cut on events can improve the results, by restricting even further the separation distance between the pair of clusters and the energy of the low-energy cluster, and thus effectively discarding most of the miscounted clusters.

\chapter{The Performance of LumiCal \label{revisedDetectorModelCH}}

The performance of LumiCal may be evaluated using several parameters; the precision with which luminosity is measured, the energy resolution, the ability to separate multiple showers, viability of the electronics readout, and finally, the integration of LumiCal in the detector. In the following, each of these criteria will be discussed. The baseline geometrical parameters of LumiCal are presented in \autoref{baselineGeometryTABLE}. It will be shown that for the present detector concept, these parameters fulfill the requirement of best performance of LumiCal. Finally, the influence of making changes to the different parameters on the calorimeter performance will be summarized. This is necessary in order to facilitate setting an optimization procedure for future changes in the design of LumiCal.

\begin{table}[htp]
\begin{center} \begin{tabular}{ |c|c| }
\hline
Parameter & Value  \\ [2pt]
\hline \hline
Distance from the IP & 2270~mm  \\ \hline
Number of Radial divisions & ~~~ 64 (0.75~mrad pitch) ~~~ \\
Number of Azimuthal divisions & 48 (131~rad pitch) \\ \hline
Number of Layers & 30 \\
Tungsten thickness & 3.5~mm \\
Silicon thickness & 0.3~mm \\
Support thickness & 0.6~mm \\
Layer gap & 0.1~mm  \\ \hline
Inner radius & 80~mm \\
Outer radius & 190~mm \\ \hline
\end{tabular} \end{center}
\caption{\label{baselineGeometryTABLE}Baseline properties of LumiCal.}
\end{table}

\section{Intrinsic Parameters \label{fiducialVolumeSEC}}

\subsection{Energy Resolution \label{fiducialVolumeSEC}}

LumiCal is designed in such a way that incident high energy electrons and photons deposit practically all of their energy in the detector. Energy degradation is achieved by the creation of EM showers, due to the passage of particles in the layers of tungsten (see \autoref{calorimetryConceptsSEC}).

Prevention of leakage through the edges of LumiCal is possible by defining fiducial cuts on the minimal and on the maximal reconstructed polar angle, $\theta$, of the particle showering in LumiCal. Stable energy resolution is the hallmark of well-contained showers. The relative energy resolution, $\sigma_{E} / E$, is usually parameterized as

\begin{equation}{
\frac{\sigma_{E}}{E} = \frac{E_{res}}{\sqrt{E_{beam}~\mathrm{(GeV)}}},
}\label{engyResEQ} \end{equation}

\noindent where $E$ and $\sigma_{E}$ are the most probable value, and the root-mean-square of the signal distribution for a beam of electrons of energy $E_{beam}$. Very often the parameter $E_{res}$ is quoted as resolution, a convention which will be followed in the analysis presented here.

\begin{figure}[htp]
\begin{center}
\subfloat[]{\label{thetaCutEngyResFIG1}\includegraphics[width=.49\textwidth]{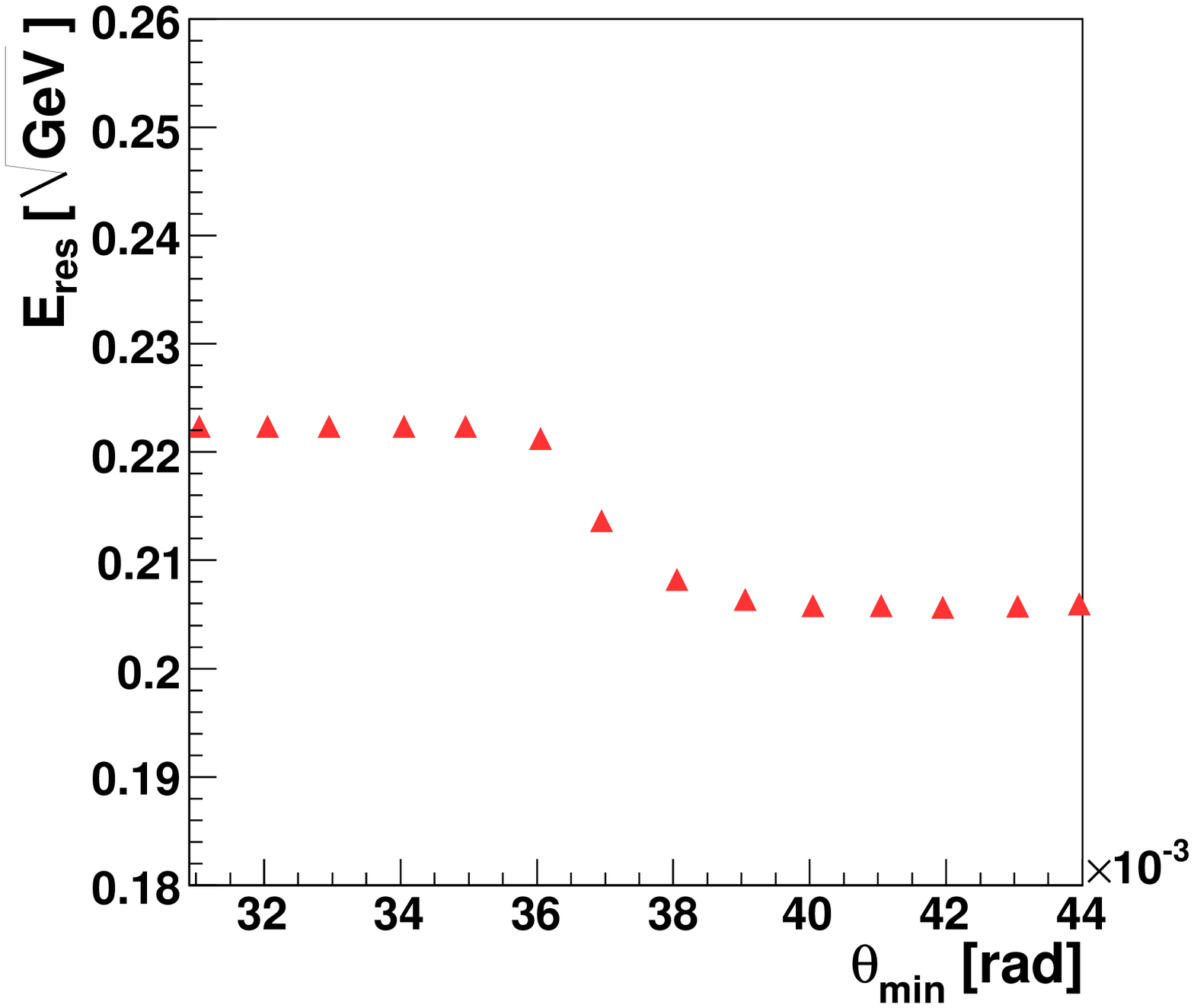}}
\subfloat[]{\label{thetaCutEngyResFIG2}\includegraphics[width=.49\textwidth]{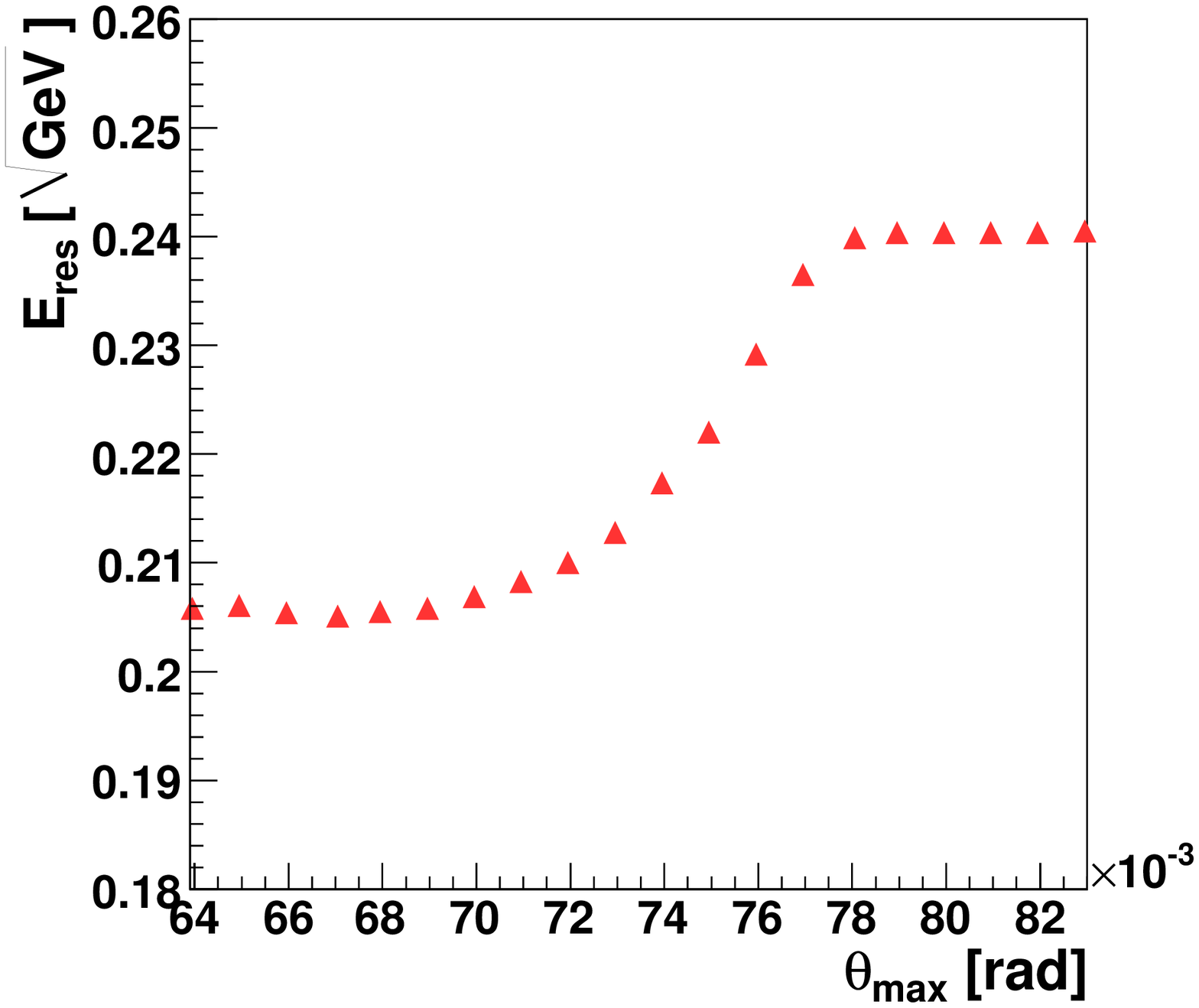}}
\caption{\label{thetaCutEngyResFIG}Energy resolution for 250~GeV electrons as a function of the minimal polar angle, $\theta_{min}$, \Subref{thetaCutEngyResFIG1} and as a function of the maximal polar angle, $\theta_{max}$, \Subref{thetaCutEngyResFIG2}.}
\end{center}
\end{figure}

\Autoref{thetaCutEngyResFIG1} shows the energy resolution as a function of $\theta_{min}$. The maximal angle is kept constant. The best energy resolution is achieved for $\theta_{min}$ = 41~mrad. A similar evaluation was done for a constant $\theta_{min}$ and a changing $\theta_{max}$, resulting in an optimal cut at $\theta_{max}$ = 69~mrad, as shown in \autoref{thetaCutEngyResFIG2}. The fiducial volume of LumiCal is thus defined to be within the polar angular range: $41<\theta<69$~mrad. For this fiducial volume $E_{res} = (20.50 \pm 0.05) \cdot 10^{-2} ~ \sqrt{\mathrm{(GeV)}}$.

\subsection{Polar Angle Resolution and Bias \label{polarAngleRecSEC}}

The polar angle is reconstructed by averaging over the individual cells hit in the detector, using the cell centers and a weight function, $\mc{W}_{i}$, such that

\begin{equation}{
<\theta> = \frac{\sum_{i}\theta_{i} \cdot \mc{W}_{i}}{\sum_{i} \mc{W}_{i}}.
}\label{logWeigh1EQ} \end{equation}

\begin{figure}[htp]
\begin{center}
\subfloat[]{\label{thetaRecFIG1}\includegraphics[width=.49\textwidth]{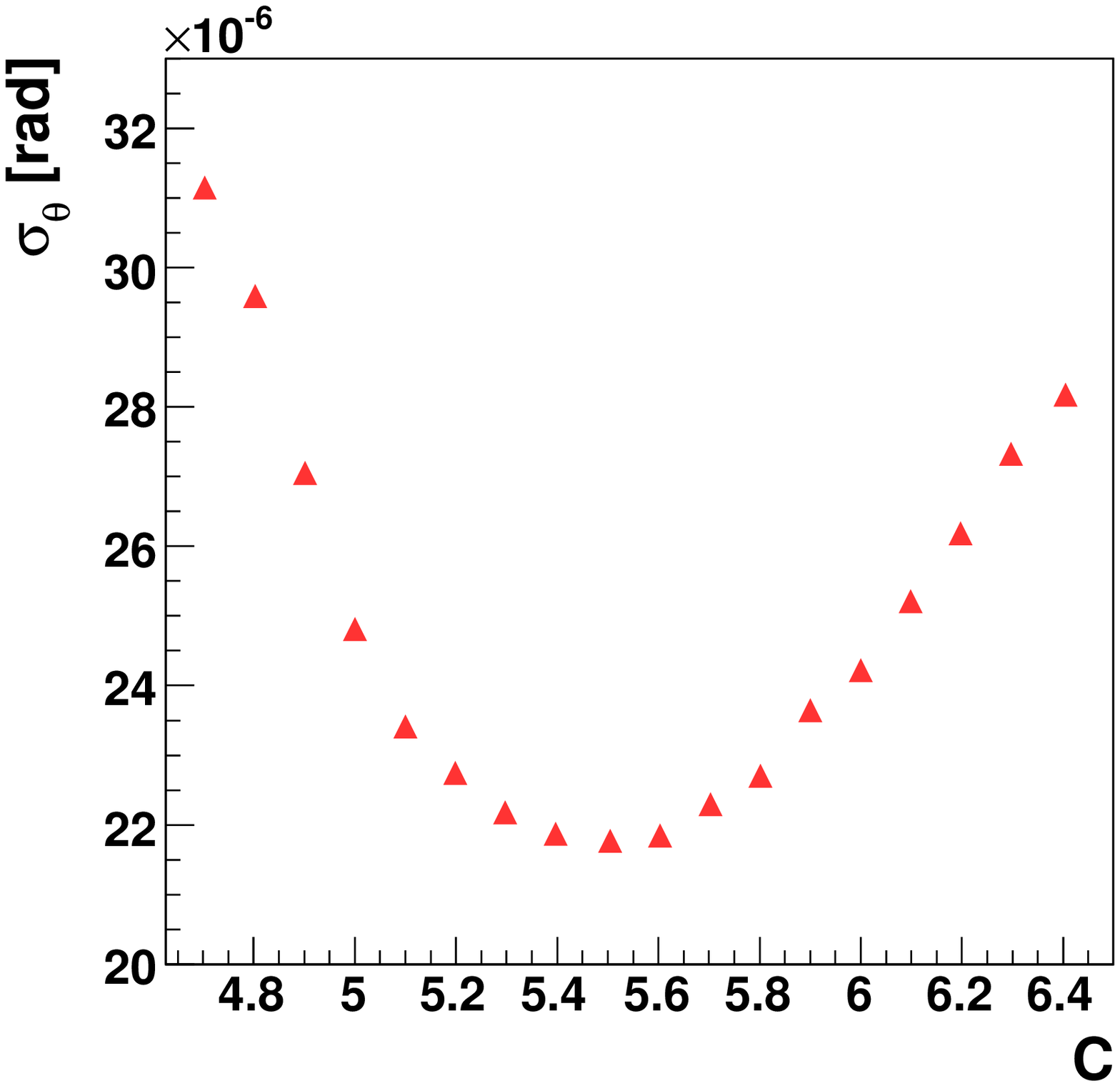}}
\subfloat[]{\label{thetaRecFIG2}\includegraphics[width=.49\textwidth]{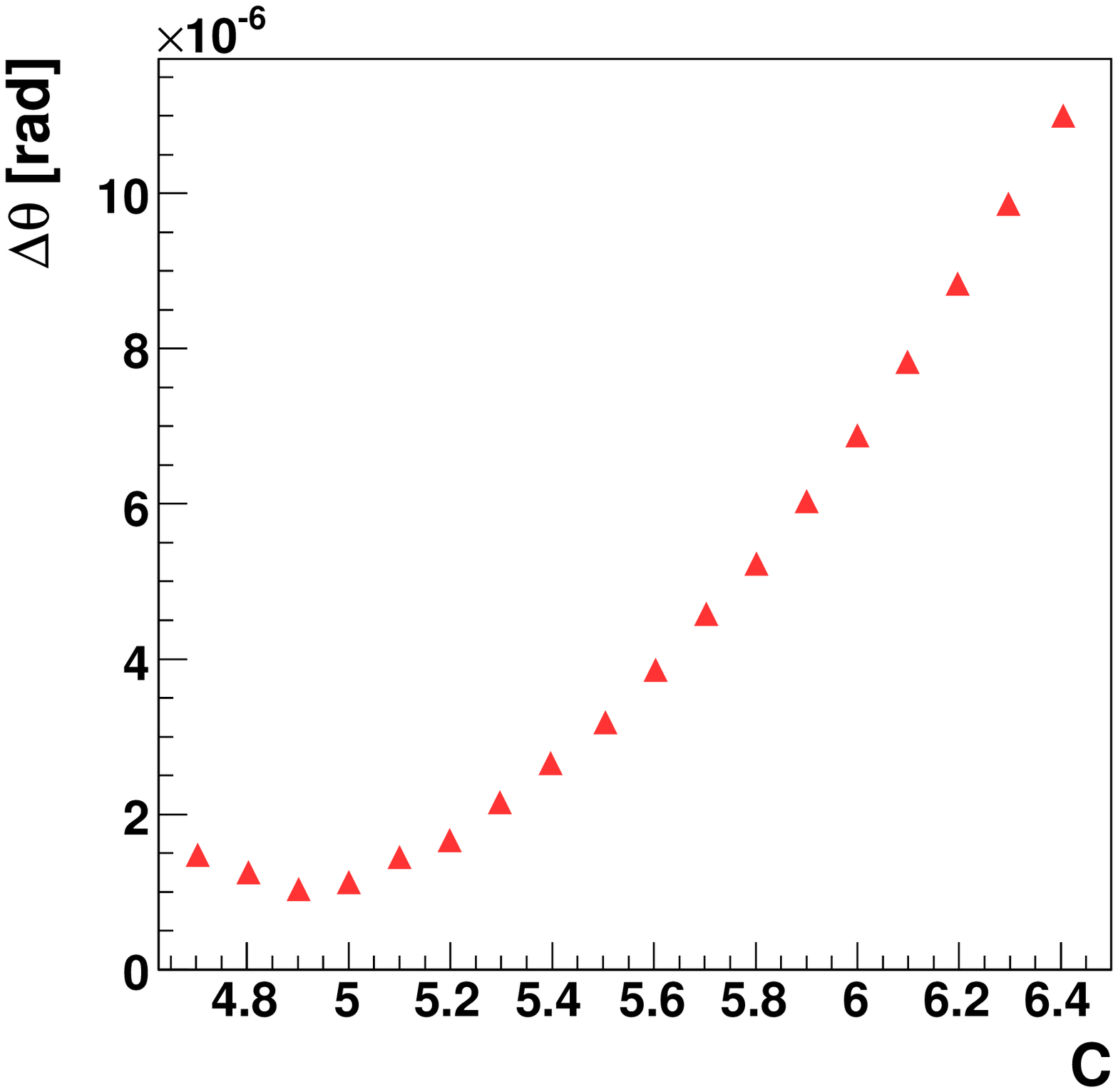}}
\caption{\label{thetaRecFIG}The polar resolution, $\sigma_{\theta}$, \Subref{thetaRecFIG1} and the polar bias, $\Delta \theta$, \Subref{thetaRecFIG2} as a function of the logarithmic weighing constant, $\mc{C}$, using 250~GeV electron showers.}
\end{center}
\end{figure}

Weights are determined by the so-called logarithmic weighting~\cite{revisedDetectorModelBIB4}, for which

\begin{equation}{
\mc{W}_{i} = \mathrm{max} \{~ 0 ~,~ \mc{C} + \mathrm{ln} \frac{E_{i}}{E_{tot} ~} \},
}\label{logWeigh2EQ} \end{equation}

\noindent where $E_{i}$ is the individual cell energy, $E_{tot}$ is the total energy in all cells, and $\mc{C}$ is a constant. In this way, an effective cutoff is introduced on individual hits, and only cells which contain a high percentage of the event energy contribute to the reconstruction. This cut, which depends on the size of the different cells, and on the total absorbed energy, is determined by $\mc{C}$. There is an optimal value for $\mc{C}$, for which the polar resolution, $\sigma_{\theta}$, is minimal. This is shown in \autoref{thetaRecFIG1} using 250~GeV electron showers. The corresponding polar bias, $\Delta \theta$, is presented in \autoref{thetaRecFIG2}. Accordingly, the polar resolution and bias of LumiCal are $\sigma_{\theta} = (2.18\, \pm\, 0.01) \cdot 10^{-2}$~mrad and $\Delta \theta = (3.2\, \pm\, 0.1) \cdot 10^{-3}$~mrad.

\section{Readout Scheme\label{constraintsElectronicsSEC}}

Upon deciding on the granularity of LumiCal, it is necessary to define the dynamical range of the electronics required to process the signal from the detector. Once the dynamical range is set, the digitization scheme depends on the ADC precision. The energy resolution and the polar-angle reconstruction depend on the digitization scheme. For the present study, it is assumed that the dynamical range of the electronics has to be such, that it enables to measure signals from minimum ionizing particles (MIP) up to the highest-energy EM showers, which are allowed by kinematics.

In order to determine the lower bound on the signal in LumiCal, the passage of muons through the detector was simulated. Muons do not shower, and are, therefore, MIPs. In the present conceptual approach, muons will be used to inter-calibrate the cells of the detector, and may also be used to check \textit{in-situ} the alignment of the detector. The detection of muons in the forward region also has significance for many searches for physics beyond the Standard Model, such as implied by certain supersymmetry models, or by theories with universal extra dimensions~\cite{constraintsElectronicsBIB1}.

In order to measure the signals of both MIPs and high energy electrons in LumiCal, the detector would have to operate in two different modes. In the \textit{calibration} (high gain) mode the electronics will be sensitive to MIP signals. In the \textit{physics} (low gain) mode the signals of high energy showers will be processed. The signature of a Bhabha event is an $e^{+} e^{-}$ pair, where the leptons are back to back and carry almost all of the initial energy. For the case of a nominal center of mass energy of 500~GeV, the maximal energy to be absorbed in LumiCal is, therefore, 250~GeV, and so 250~GeV electrons were used in order to find the upper bound on the detector signal. The low limit on the signal will have to be of the order of a single MIP, and will be precisely determined according to the restrictions imposed by the energy resolution.

The output of Mokka is in terms of energy lost in the active material, silicon in the case of LumiCal. In order to translate the energy signal into units of charge, the following formula was used:

\begin{equation}{
\mc{S_{Q}} [\mathrm{fC}] = \frac{1.6 \cdot 10^{-4}}{3.67} \; \mc{S_{E}} [\mathrm{eV}]
}\label{engySignalToChargeEQ} \end{equation}

\noindent where $\mc{S_{E}}$ denotes the signal in units of eV, and $\mc{S_{Q}}$ the signal in units of fC. The value  3.67~eV is the energy to create an electron-hole pair in silicon. The number $1.6 \cdot 10^{-4}$~fC is the charge of an electron.

The detector model described in \autoref{baselineGeometryTABLE} was simulated.

\subsection{Dynamical Range of the Signal \label{muonElectronSignalsSEC}}

\begin{figure}[htp]
\begin{center}
\includegraphics[width=.49\textwidth]{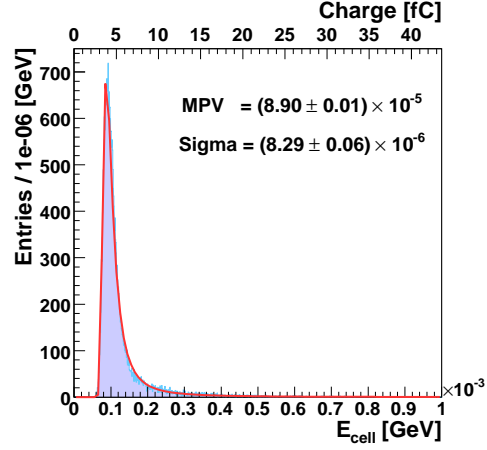}
\caption{\label{mipSignalFIG}Distribution of the energy deposited in a detector cell, $E_{cell}$, by 250~GeV muons. A corresponding scale in units of charge is also shown.}
\end{center}
\end{figure}

The distribution of the energy deposited in a detector cell by 250~GeV muons is presented in \autoref{mipSignalFIG}. According to this, the most probable value of induced charge for a muon traversing $300~\mu\mathrm{m}$ of silicon is 89~keV, which is equivalent to 3.9~fC. 

\begin{figure}[htp]
\begin{center}
\subfloat[]{\label{maxChargeCell250GeVFIG1}\includegraphics[width=.49\textwidth]{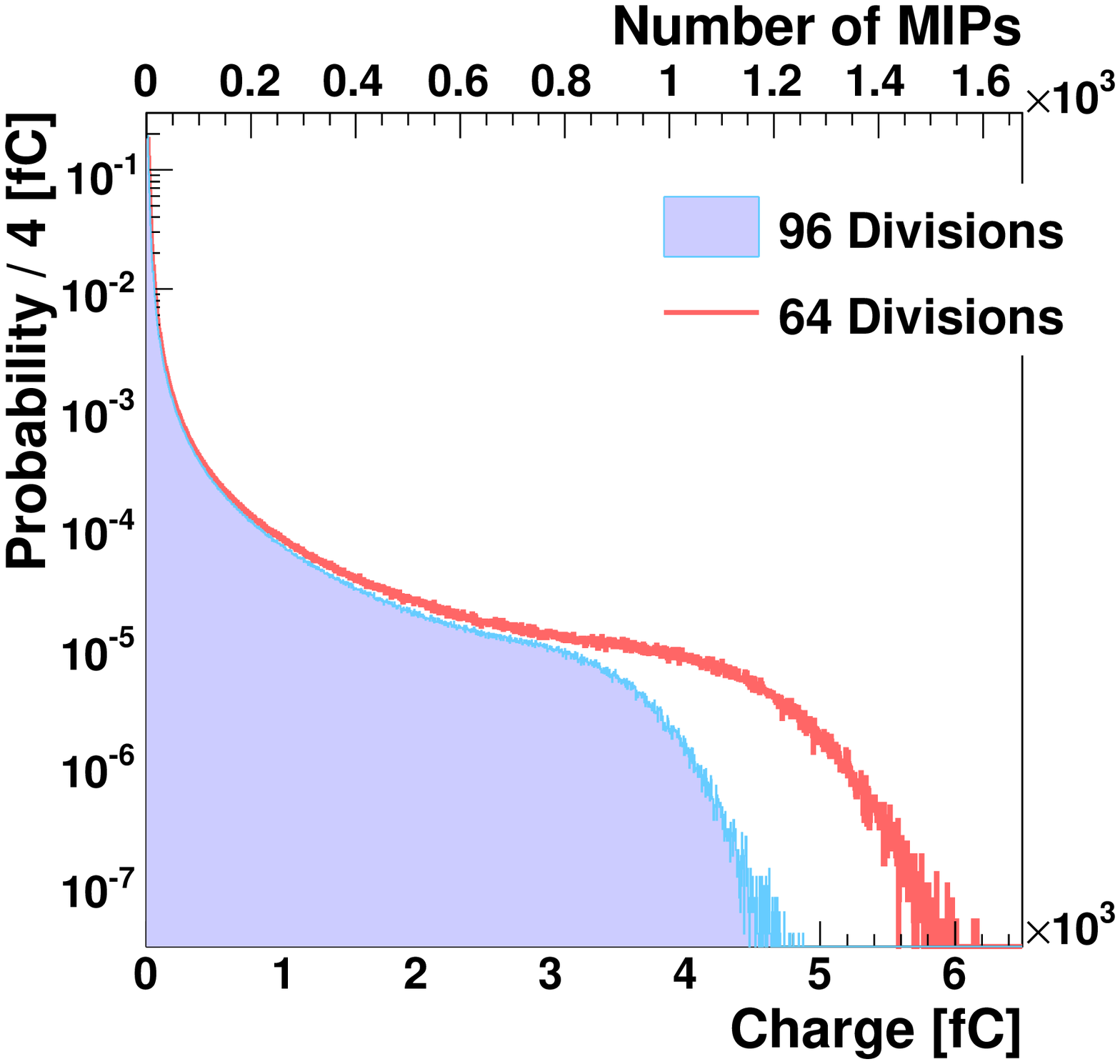}}
\subfloat[]{\label{maxChargeCell250GeVFIG2}\includegraphics[width=.49\textwidth]{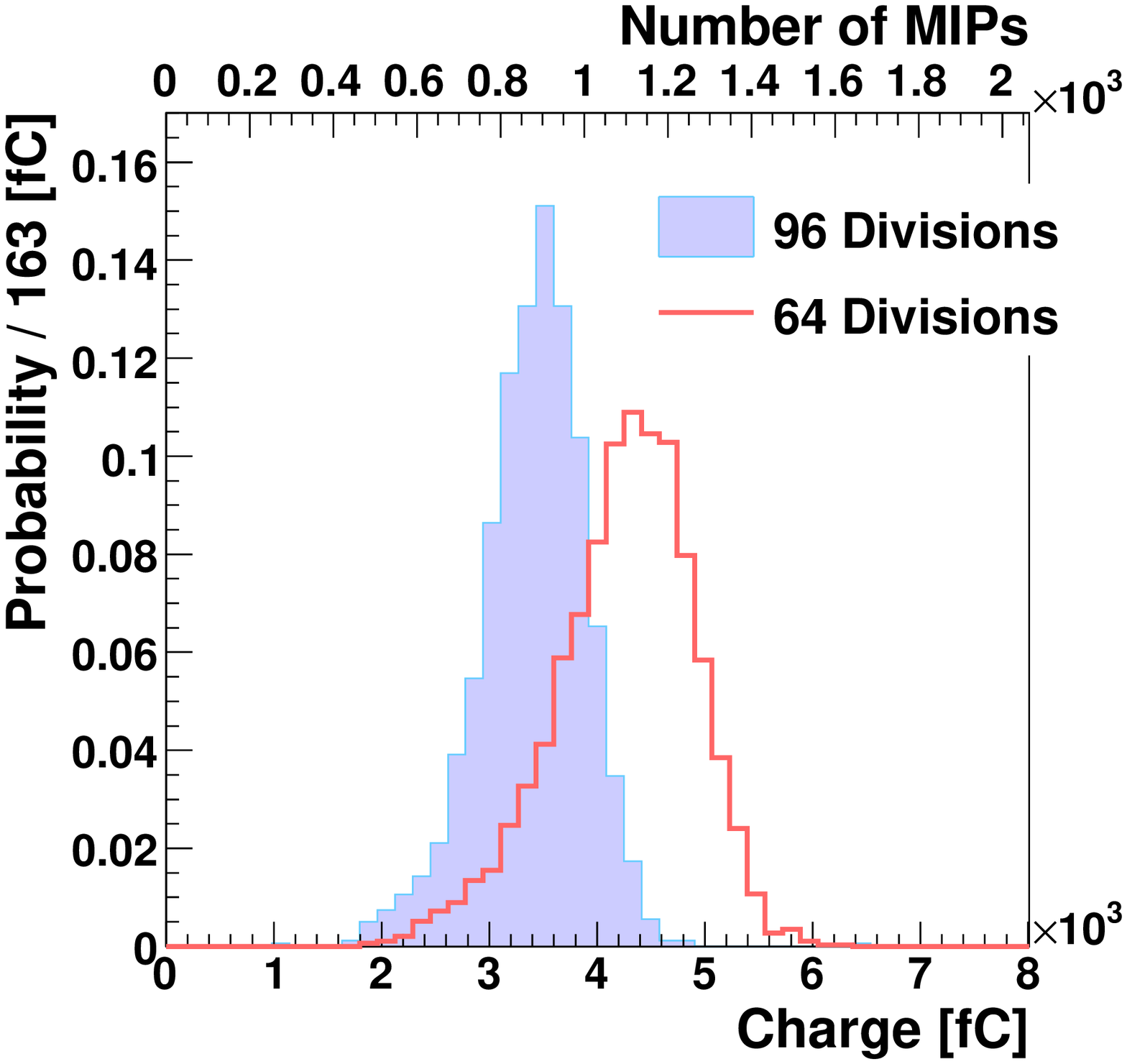}}
\caption{\label{maxChargeCell250GeVFIG}\Subref{maxChargeCell250GeVFIG1} Normalized distribution of the charge deposited in a detector cell by 250~GeV electron showers for a LumiCal with 96 or 64 radial divisions, as denoted in the figure. \Subref{maxChargeCell250GeVFIG2} Normalized distribution of the maximal charge collected in a single cell per shower for 250~GeV electron showers for a LumiCal with 96 or 64 radial divisions, as denoted in the figure. In both figures a corresponding scale in units of MIPs is also shown.}
\end{center}
\end{figure}

The distribution of collected charge per cell for 250~GeV electron showers is presented in \autoref{maxChargeCell250GeVFIG1} for a LumiCal with 96 or 64 radial divisions, which correspond to angular cell sizes of 0.5 and 0.8~mrad respectively. For the baseline case of 64 radial divisions, the value of the collected charge extends up to 6~pC, which is equivalent to $\sim$~1540~MIPs. The distribution of the maximal charge collected in a single cell per shower for 250~GeV electrons is shown in \autoref{maxChargeCell250GeVFIG2} for the two granularity options. As expected, for the case of smaller cell sizes the signal per cell is lower.

\subsection{Digitization \label{digitizationSEC}}

Once a low or high bound on the dynamical range for each mode of operation is set, it is necessary to digitize the signal. For each mode of operation separately

\begin{equation}{
\sigma_{\mathrm{ADC}} \equiv q_{min} = \frac{q_{max}}{2^{\mc{B}_{digi}}},
}\label{digitizationResEQ} \end{equation}

\noindent where $\sigma_{\mathrm{ADC}}$ is the ADC channel resolution (bin size), $q_{min}$ and $q_{max}$ are, respectively, the low and high charge bounds and $2^{\mc{B}_{digi}}$ is the number of channels for a given number of available ADC bits,  $\mc{B}_{digi}$. Each cell of deposited charge, $q_{dep}$, is read-out as having a charge, $q_{digi}$, (rounding error) where

\begin{equation}{
q_{digi} = ( \mc{Q}(q_{dep},\sigma_{\mathrm{ADC}}) + 0.5 ) \cdot \sigma_{\mathrm{ADC}} \; ,
}\label{digitizationConversionEQ} \end{equation}

\noindent and the quotient of the deposited charge with the ADC resolution, $\mc{Q}(q_{dep},\sigma_{\mathrm{ADC}})$, is defined such that

\begin{equation}{
\begin{gathered}
\alpha = \mc{Q}(\alpha,\beta) \cdot \beta + \gamma , \\
0 \le \gamma < \abs{\beta} .
\end{gathered}
}\label{digitizationConversionEQ} \end{equation}

\Autoref{dynamicRange1TABLE} shows the restrictions on the dynamical range of the two modes of operation. Since in the calibration mode the spectrum of MIPs will be measured, the resolution must be better than one MIP. For the choice of a low charge bound, $q_{min} = 0.2 ~ \mathrm{MIPs}$, the high bound will be determined by the digitization constant. The digitization constant will also determine the low bound for the physics mode, once the upper bound is set to $q_{max} = 1540 ~ \mathrm{MIPs}$. In \autoref{dynamicRange2TABLE} are presented the values of $q_{max}$ for the calibration mode and of $q_{min}$ of the physics mode for several choices of the digitization constant.

\begin{table}[htp]
\begin{center} \begin{tabular}{ ccc }
&  Calibration Mode & Physics Mode \\ [2pt]
\hline \hline
$q_{min}$ & 0.8~fC (0.2~MIPs) & $\sigma_{\mathrm{ADC}}$ \\
$q_{max}$ & 0.8~fC $\times 2^{\mc{B}_{digi}}$ & 6~pC (1540~MIPs) \\ \hline
\end{tabular} \end{center}
\caption{\label{dynamicRange1TABLE}Low and high bounds, $q_{min}$ and $q_{max}$, of the dynamical ranges of LumiCal for operation in the calibration (high gain) and in the physics (low gain) modes.}

\vspace{20pt}

\begin{center} \begin{tabular}{ |c|c|c| }
\hline
\multirow {2}{*}{ $\mc{B}_{digi}$~[bits] } & $q_{max}$ of & $q_{min}$ of \\
& Calibration Mode & Physics Mode \\ [2pt]
\hline \hline
6  & 49.9~fC (13~MIPs) & 93.7~fC (24~MIPs) \\ \hline
8  & 199.7~fC (52~MIPs) & 23.4~fC (6~MIPs) \\ \hline
10 & 798.7~fC (205~MIPs) & 5.9~fC (1.5~MIPs) \\ \hline
12 & 3.2~pC (819~MIPs) & 1.5~fC (0.4~MIPs) \\ \hline
14 & 12.8~pC (3277~MIPs) & 0.4~fC (0.1~MIPs) \\ \hline
\end{tabular} \end{center}
\caption{\label{dynamicRange2TABLE}Dependence of the high bound, $q_{max}$, of the calibration mode, and of the low bound, $q_{min}$, of the physics mode on the digitization constant, $\mc{B}_{digi}$.}
\end{table}

\begin{figure}[htp]
\begin{center}
\subfloat[]{\label{digiEngyFIG1}\includegraphics[width=.49\textwidth]{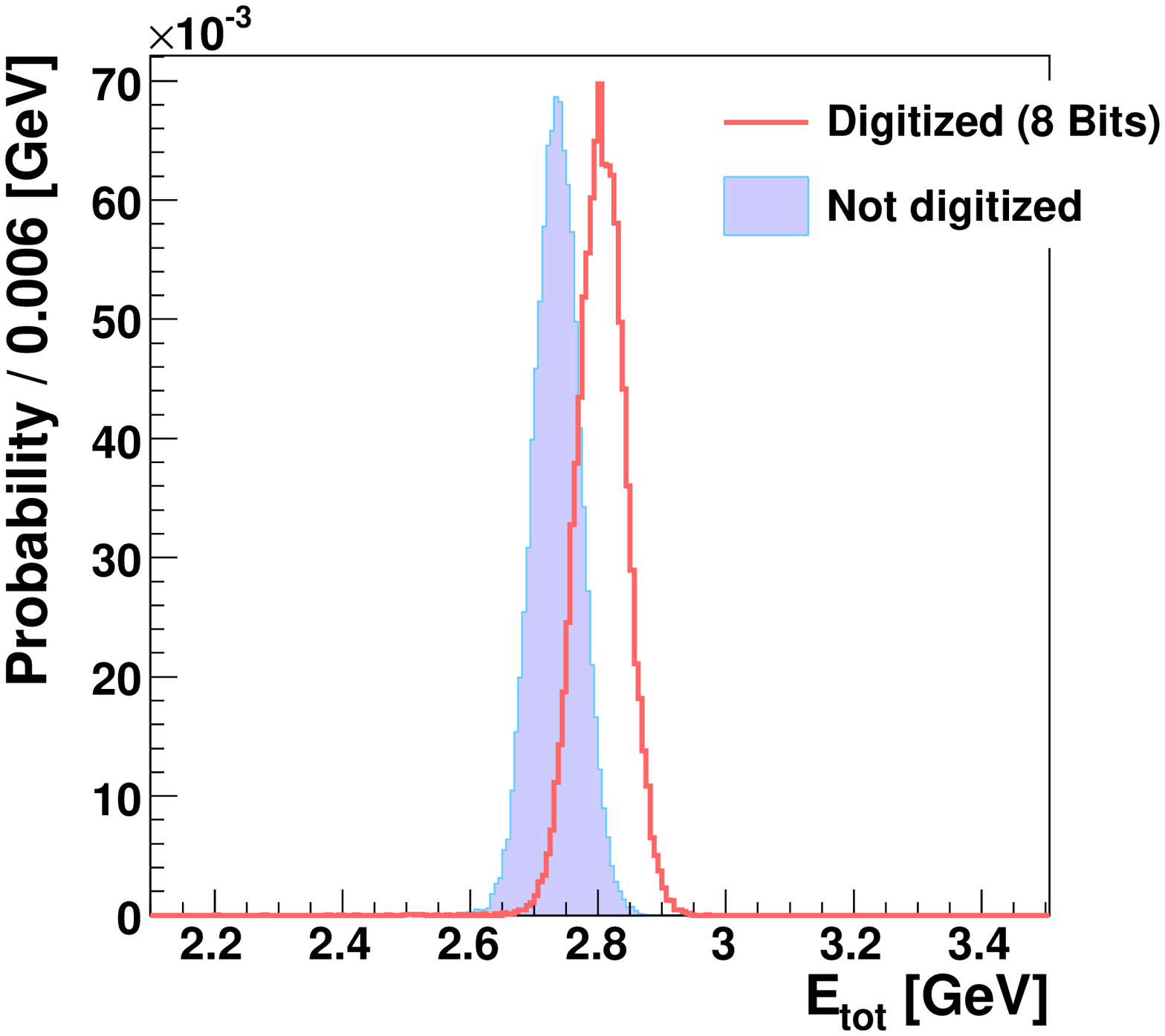} }
\subfloat[]{\label{digiEngyFIG2}\includegraphics[width=.49\textwidth]{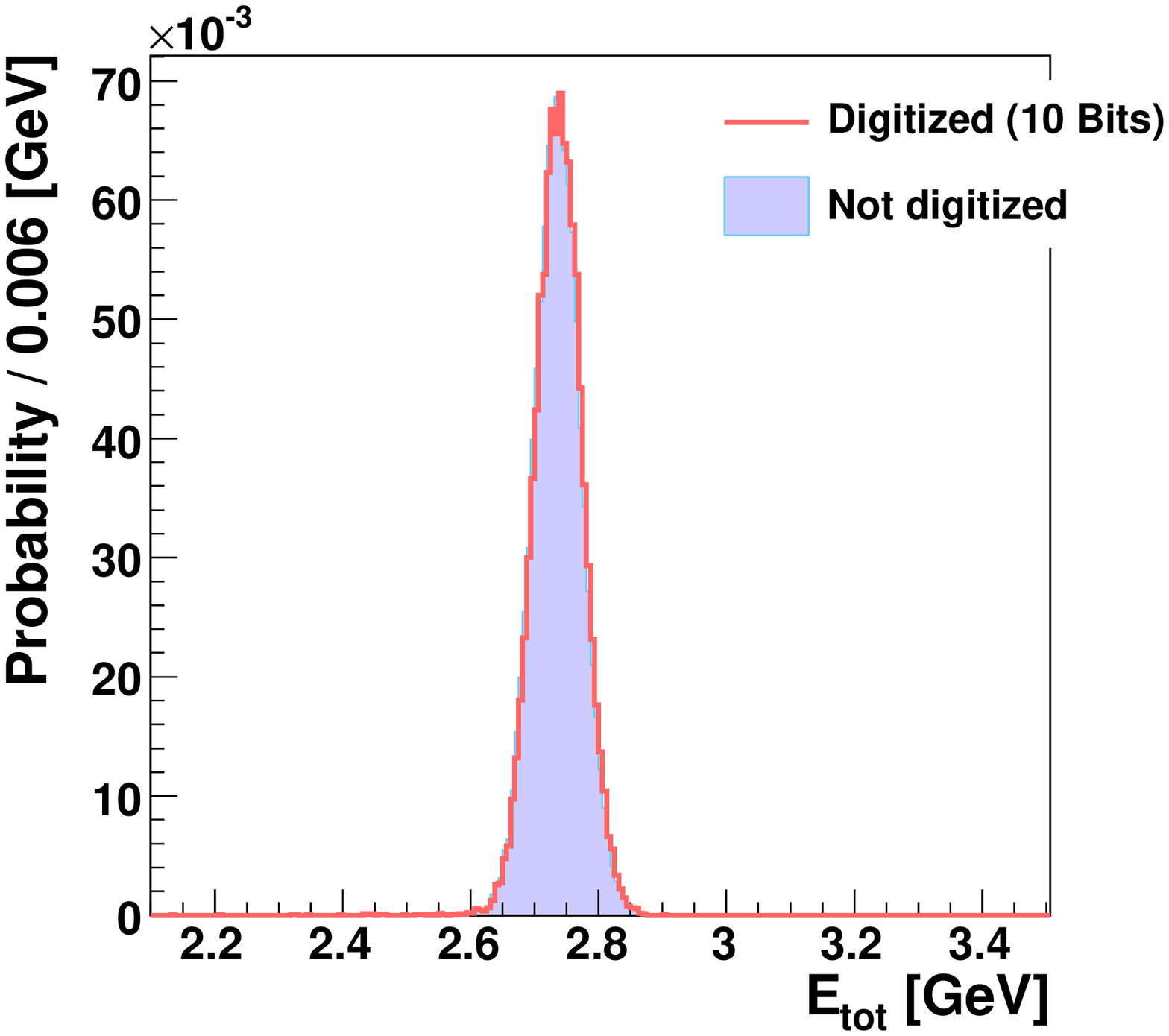}}
\caption{\label{digiEngyFIG}Normalized distribution of the total event energy, $E_{tot}$, of 250~GeV electron showers for a non-digitized, and either an 8 \Subref{digiEngyFIG1} or a 10~bit \Subref{digiEngyFIG2} digitized detector signal, as denoted in the figures.}
\end{center}
\end{figure}

\Autoref{digiEngyFIG} shows the distribution of the total event energy for the case of a non-digitized, and an 8 or 10~bit digitized detector signal. The mean value of the distribution for the 8~bit digitized case is higher by $\mc{O}(1\%)$ compared to the  non-digitized case. This is due to the fact that by far the greatest percentage of contributing cells have small energy, as can be observed in \autoref{maxChargeCell250GeVFIG}. The charge of more hits will, therefore, be overestimated, rather than underestimated. To illustrate this point for the 8~bit case, for the first ADC bin, $44\%$ of the hits in a 250~GeV electron shower belong to the bottom half of the bin, while $21\%$ belong to its top half. Since all of these hits are read-out with a digitized charge of exactly half the bin width, more contributions are overestimated, rather than underestimated, with weights according to the distribution of \autoref{mipSignalFIG}. One, therefore, finds that the total digitized charge is higher by $2.2~\mathrm{\frac{pC}{shower}}$ than the deposited charge, which amounts to a $26\%$ increase in the integrated measured signal in this ADC channel. Since this effect depends on $\sigma_{\mathrm{ADC}}$ the difference in the mean value of the distributions of \autoref{digiEngyFIG} with respect to the non-digitized case decreases for larger values of $\mc{B}_{digi}$.

\begin{figure}[htp]
\begin{center}
\includegraphics[width=.49\textwidth]{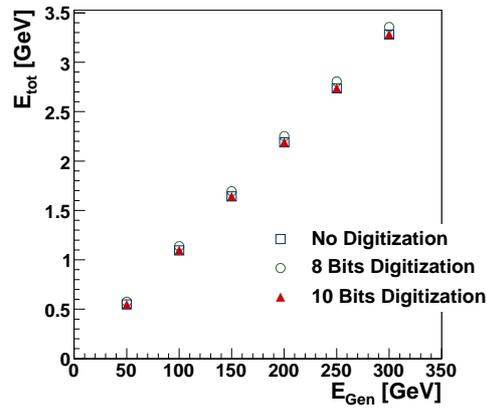}
\caption{\label{engyCalibrationFIG}Dependence of the detector signal, $E_{tot}$, on the energy of the electron which initiated the shower, $E_{Gen}$. The detector signal is either un-digitized, or digitized with an 8 or a 10~bit digitization constant, as denoted in the figure.}
\end{center}
\end{figure}

In practice, the shift in the mean bears no consequence other than the need to adjust the signal to energy calibration. The dependence of the detector signal on the energy of the particle which initiated the shower is shown in \autoref{engyCalibrationFIG} for several digitization schemes. There is no significant change as a result of the digitization, for the values of $\mc{B}_{digi}$ which were used.

\begin{figure}[htp]
\begin{center}
\subfloat[]{\label{digiThetaRecFIG1}\includegraphics[width=.49\textwidth]{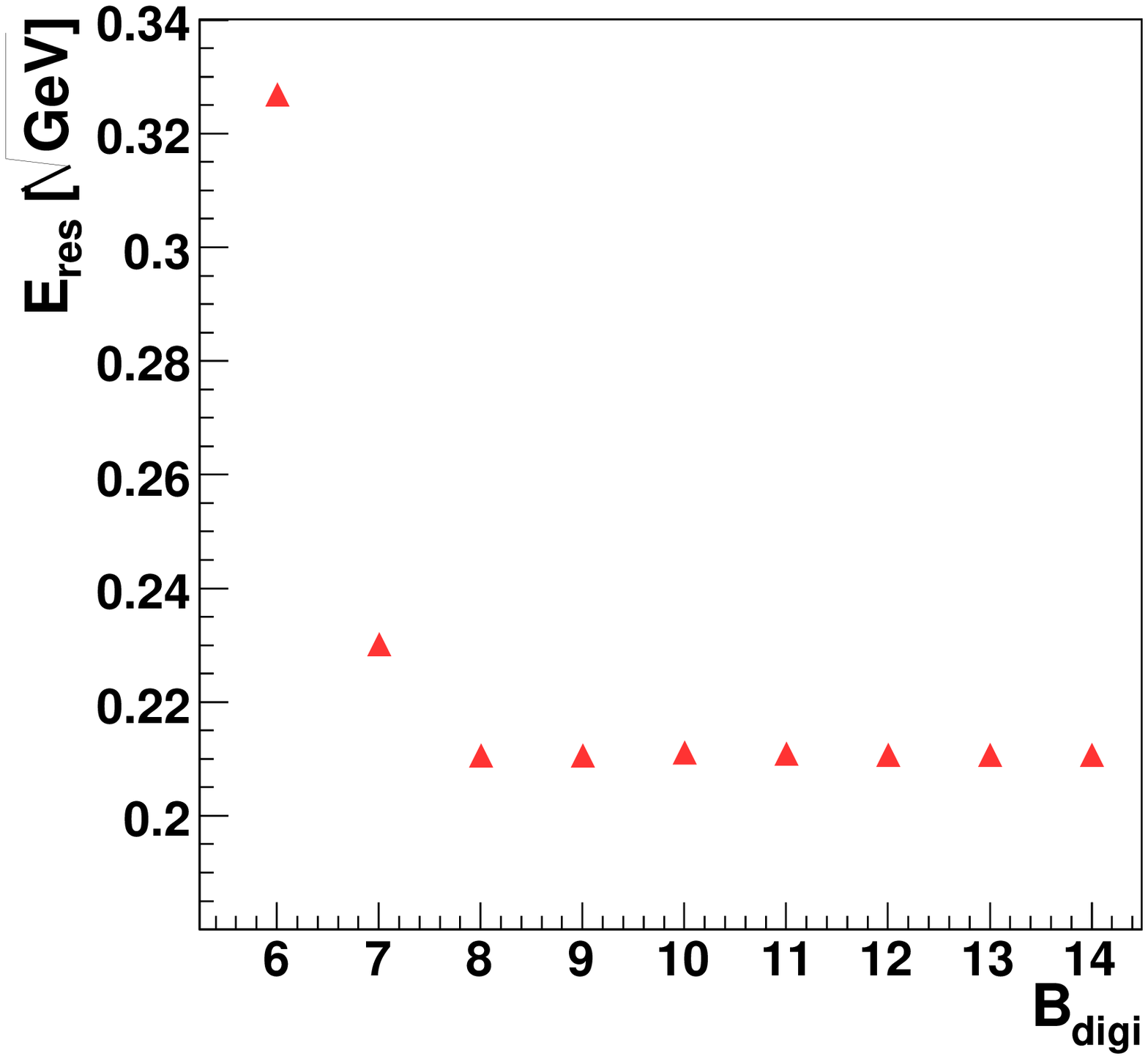}} \\
\subfloat[]{\label{digiThetaRecFIG2}\includegraphics[width=.49\textwidth]{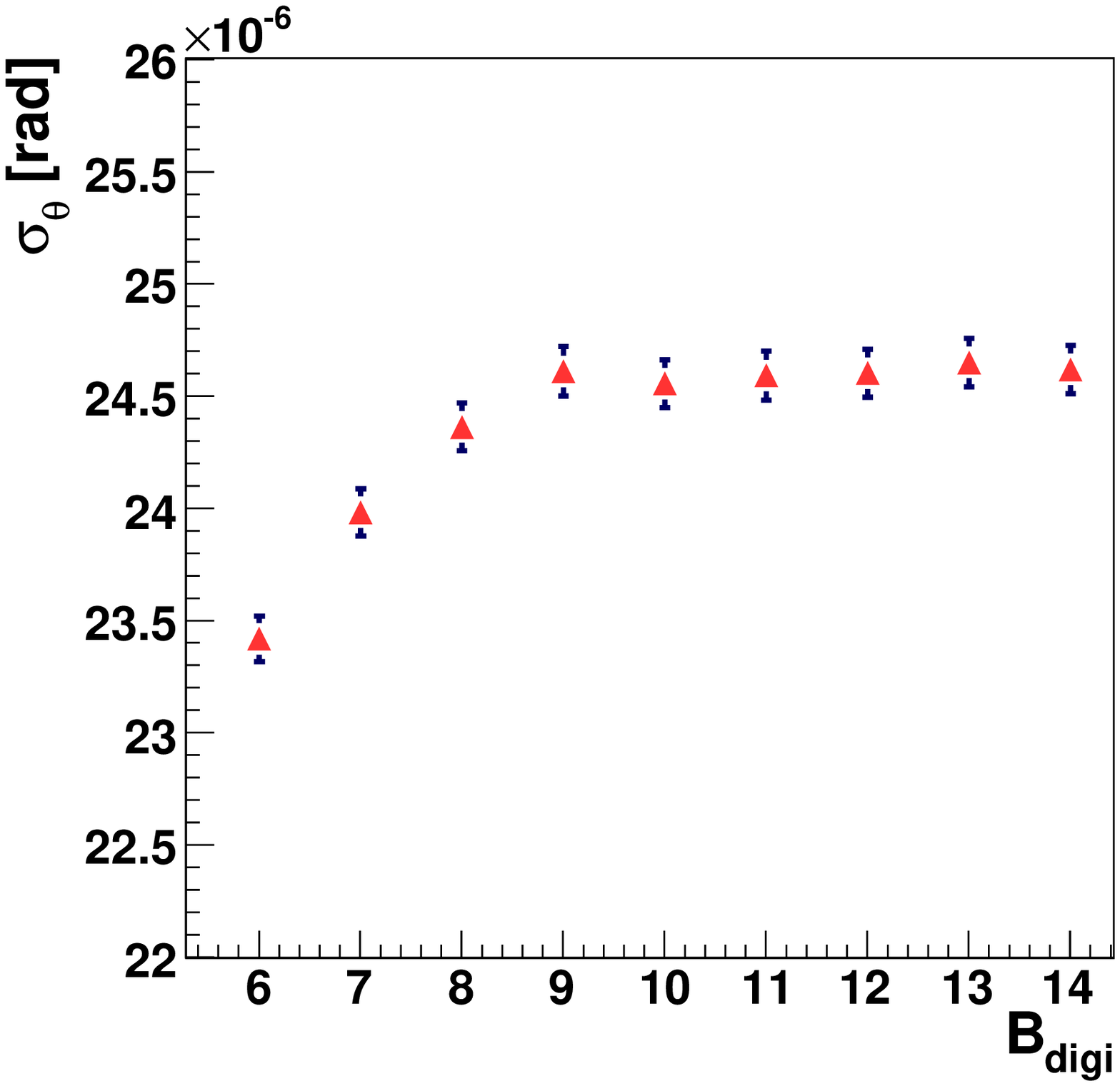}}
\subfloat[]{\label{digiThetaRecFIG3}\includegraphics[width=.49\textwidth]{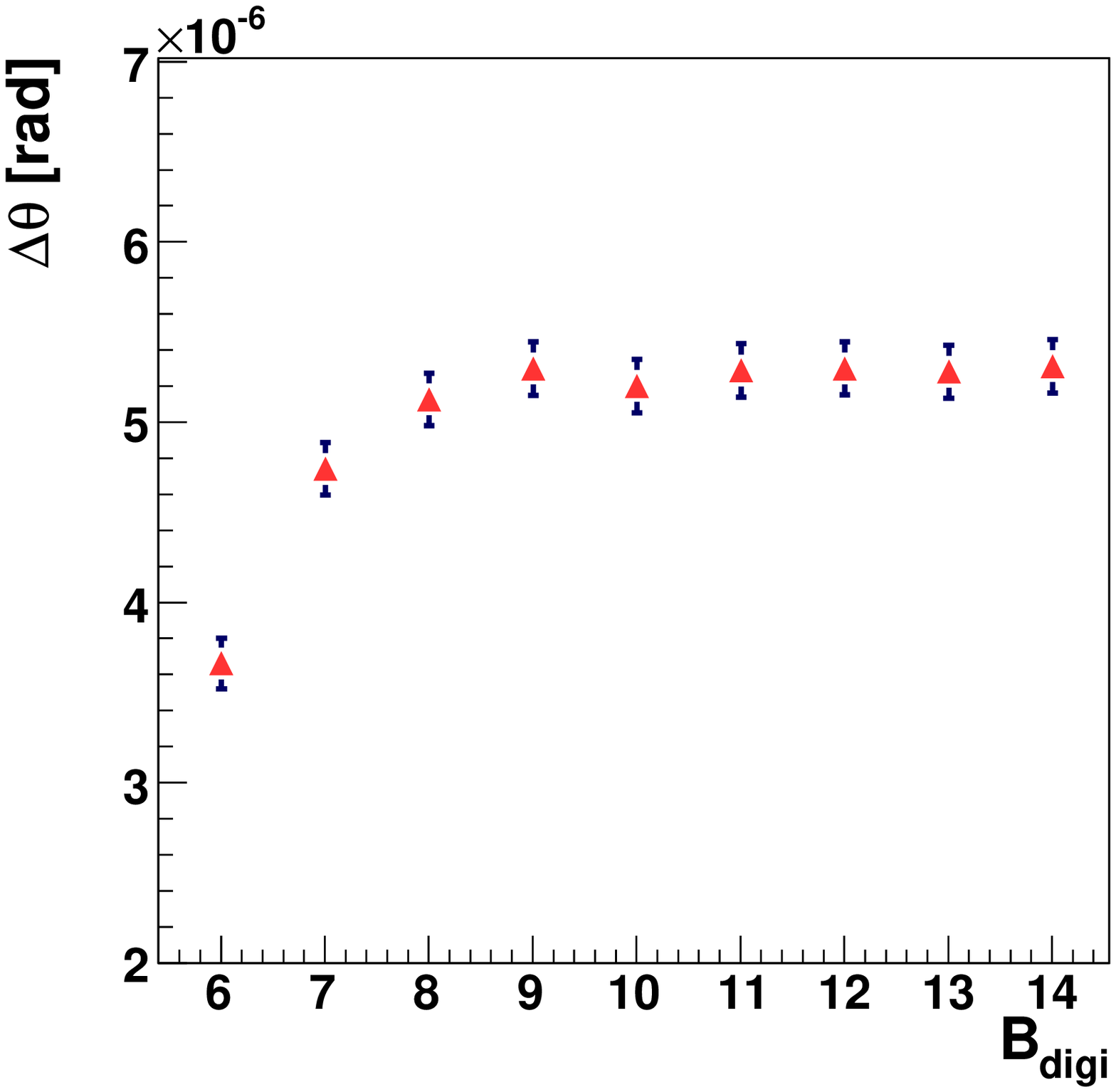}}
\caption{\label{digiThetaRecFIG}Dependence of the energy resolution, $E_{res}$, \Subref{digiThetaRecFIG1} the polar resolution, $\sigma_{\theta}$, \Subref{digiThetaRecFIG2} and the polar bias, $\Delta \theta$, \Subref{digiThetaRecFIG3} on the digitization constant, $\mc{B}_{digi}$.}
\end{center}
\end{figure}

The important quantity that has to be controlled is the energy resolution, which must be the same for the digitized and the non-digitized cases. \autoref{digiThetaRecFIG} shows the dependence of the energy resolution, $E_{res}$, the polar resolution, $\sigma_{\theta}$, and the polar bias, $\Delta \theta$ on the digitization constant. The values shown for $\mc{B}_{digi} = 14$ are equivalent to a non-digitized readout.

For $\mc{B}_{digi} > 8$ it is apparent that the energy resolution, the polar resolution, and the polar bias are all stable. Below this limit there is severe degradation of $E_{res}$ and slight improvement of $\sigma_{\theta}$ and $\Delta \theta$. The reason for this is that $E_{res}$ depends on the accuracy with which each and every detector cell is read-out. This means that fluctuations in $q_{digi}$ of cells with small energy become critical when $\sigma_{\mathrm{ADC}}$ is large in comparison to the cell signals. This effect, naturally, depends on the sizes of the LumiCal cells, since smaller cells have smaller signals, as suggested by \autoref{maxChargeCell250GeVFIG}. On the other hand, the polar angle reconstruction only takes into account contributions from cells with relatively large energy (\autorefs{logWeigh1EQ} and \ref{logWeigh2EQ}), for which $\sigma_{\mathrm{ADC}}$ is small in comparison. Fluctuations, which hinder the polar reconstruction, decrease for low energy hits. With the negative influence on the energy resolution being the driving factor, it is concluded that the minimization of $E_{res}$ requires the digitization constant to be higher than 8~bit.

\section{Geometry Optimization \label{geometryOptimizationSEC}}

The chosen geometrical parameters for the baseline model are given in \autoref{baselineGeometryTABLE}. The following is a systematic study, in which it will be shown that the values given in the table optimize the performance of LumiCal. Different single parameters will be varied, keeping the others constant, and the consequences of each change will be discussed.

\subsection{The Number of Radial Divisions \label{numRadialDivisionsSEC}}

\begin{figure}[htp]
\begin{center}
\subfloat[]{\label{thetaBiasDivisionsFIG1}\includegraphics[width=.49\textwidth]{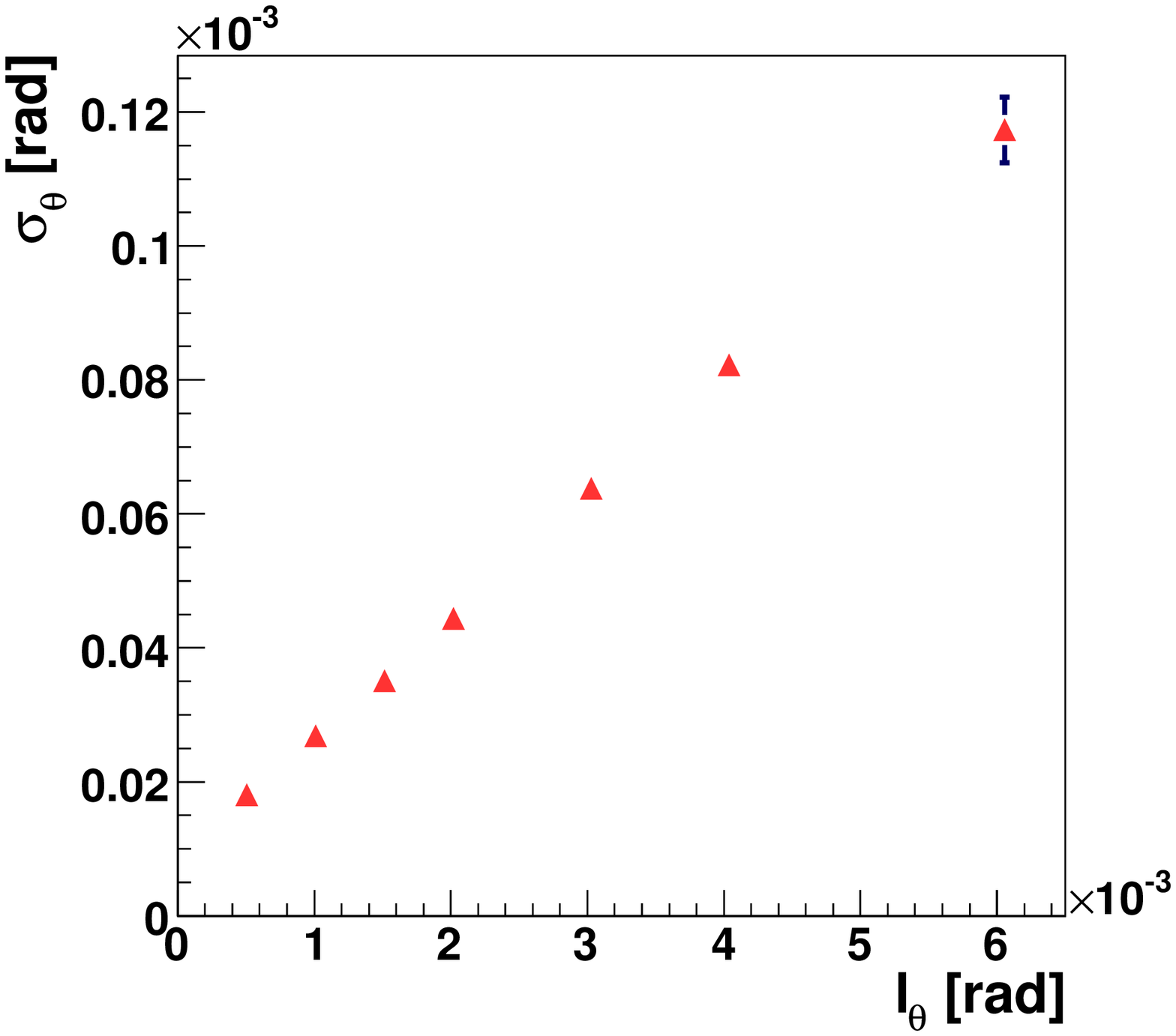}}
\subfloat[]{\label{thetaBiasDivisionsFIG2}\includegraphics[width=.49\textwidth]{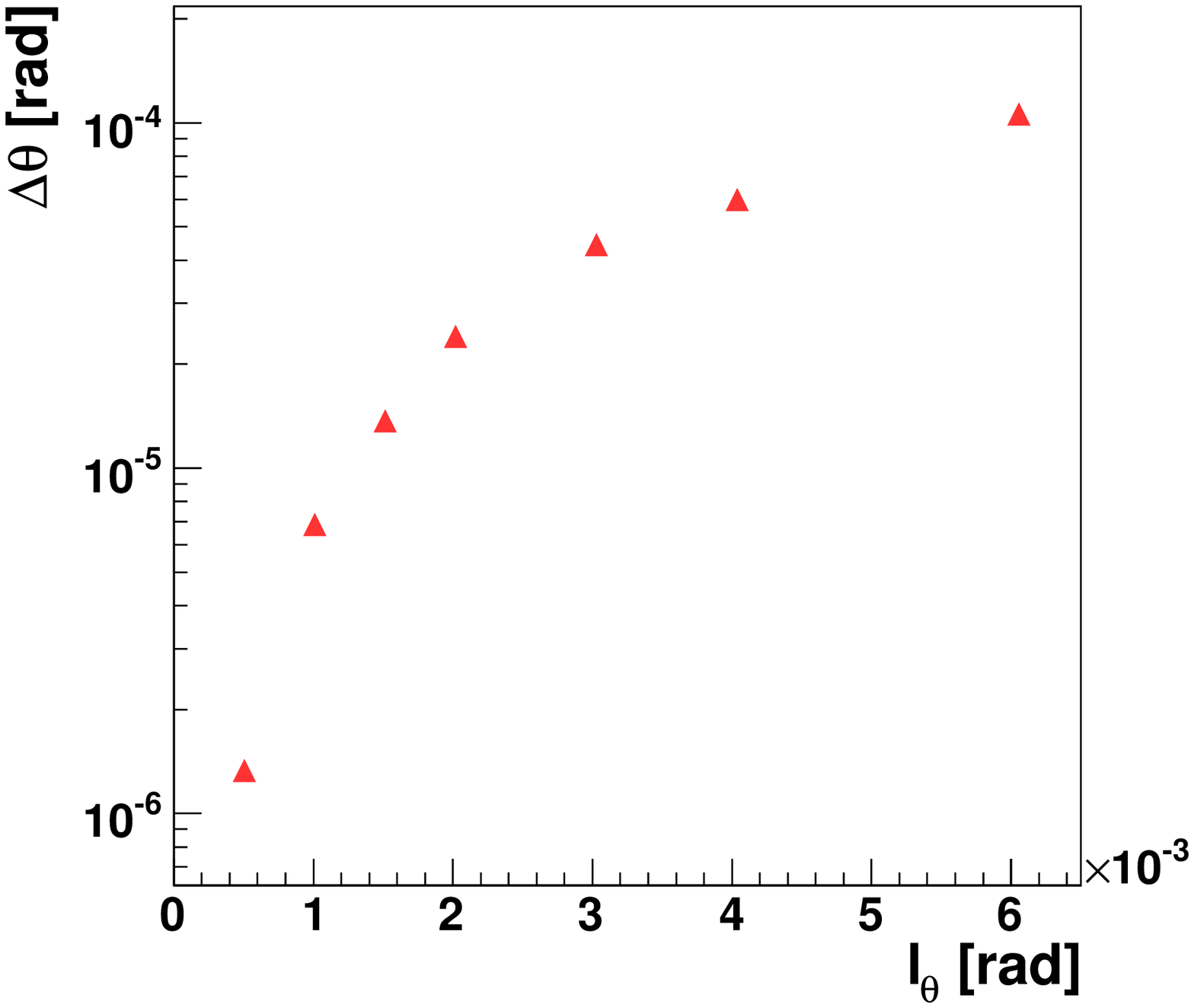}}
\caption{\label{thetaBiasDivisionsFIG}The polar resolution, $\sigma_{\theta}$, \Subref{thetaBiasDivisionsFIG1} and the polar bias, $\Delta \theta$, \Subref{thetaBiasDivisionsFIG2} for the optimal logarithmic weighing constant, as a function of the angular cell size, $\ell_{\theta}$. Electron showers of 250~GeV were used.}
\end{center}
\end{figure}

For different radial granularity one needs to re-optimize the logarithmic weighing constant, $\mc{C}$, of \autoref{logWeigh2EQ}, as the size of cells changes for each case. The polar resolution and bias are plotted in \autoref{thetaBiasDivisionsFIG} as a function of the angular cell size, $\ell_{\theta}$. Their values are presented in \autoref{thetaGranTABLE}, along with the corresponding relative error in the luminosity measurement (according to \autoref{luminosityRelativeErrRec2EQ}).

\begin{table}[htp]
\begin{center} \begin{tabular}{ |c|c|c|c|c| }
\hline
Radial Divisions & ~~$\ell_{\theta}$ [mrad]~~ & ~$\sigma_{\theta}$ [mrad]~  & ~$\Delta \theta$ [mrad]~ & ~~~\begin{large} $\frac{2 \Delta \theta}{\theta_{min}}$ \end{large}~~~ \\ [2pt]
\hline \hline
96 & $0.5 $ &  $1.8  \cdot 10^{-2}$ &  $1.3  \cdot 10^{-3}$ &  $0.6 \cdot 10^{-4}$  \\ \hline
64 & $0.8 $ &  $2.2  \cdot 10^{-2}$ &  $3.2  \cdot 10^{-3}$ &  $1.5 \cdot 10^{-4}$  \\ \hline
48 & $1   $ &  $2.7  \cdot 10^{-2}$ &  $6.9  \cdot 10^{-3}$ &  $3.1 \cdot 10^{-4}$  \\ \hline
32 & $1.5 $ &  $3.5  \cdot 10^{-2}$ &  $13.7 \cdot 10^{-3}$ &  $6.2 \cdot 10^{-4}$  \\ \hline
24 & $2   $ &  $4.4  \cdot 10^{-2}$ &  $24   \cdot 10^{-3}$ &  $10.9 \cdot 10^{-4}$ \\ \hline
16 & $2.5 $ &  $6.4  \cdot 10^{-2}$ &  $44.4 \cdot 10^{-3}$ &  $20.2 \cdot 10^{-4}$ \\ \hline
\end{tabular} \end{center}
\caption{\label{thetaGranTABLE}The polar resolution, $\sigma_{\theta}$, and bias, $\Delta \theta$, for LumiCal with different numbers of radial divisions, corresponding to different angular cell sizes, $\ell_{\theta}$. The corresponding values of the relative error in the measurement of the luminosity are also shown.}
\end{table}

Both $\sigma_{\theta}$ and $\Delta \theta$ become smaller as the angular cell size decreases. The relative error in luminosity follows the same trend. This is due to the fact that the bounds on the fiducial volume do not strongly depend on the number of radial divisions. Consequently the minimal polar angle, $\theta_{min}$, is the same (41~mrad) for all the entries of \autoref{thetaGranTABLE}.

Many problems arise when one increases the number of channels beyond a certain density. One has to resolve such problems as cross-talk between channels, power consumption issues, the need for cooling,  etc. It is, therefore, advisable to keep the number of cells as low as possible. The other important parameter, the energy resolution, does not depend on the number of channels, since the energy is integrated over all cells\footnotemark. The chosen baseline number of 64 radial divisions is, therefore, a compromise between trying to minimize the relative luminosity error, and limiting the number of  channels.

\footnotetext{This statement is true provided that the digitization constant is not too small, as discussed in  \autoref{digitizationSEC}.}

\subsection{The Structure of Layers \label{numberOfLayersSEC}}

\begin{figure}[htp]
\begin{center}
\subfloat[]{\label{engy90LayerFIG1}\includegraphics[width=.49\textwidth]{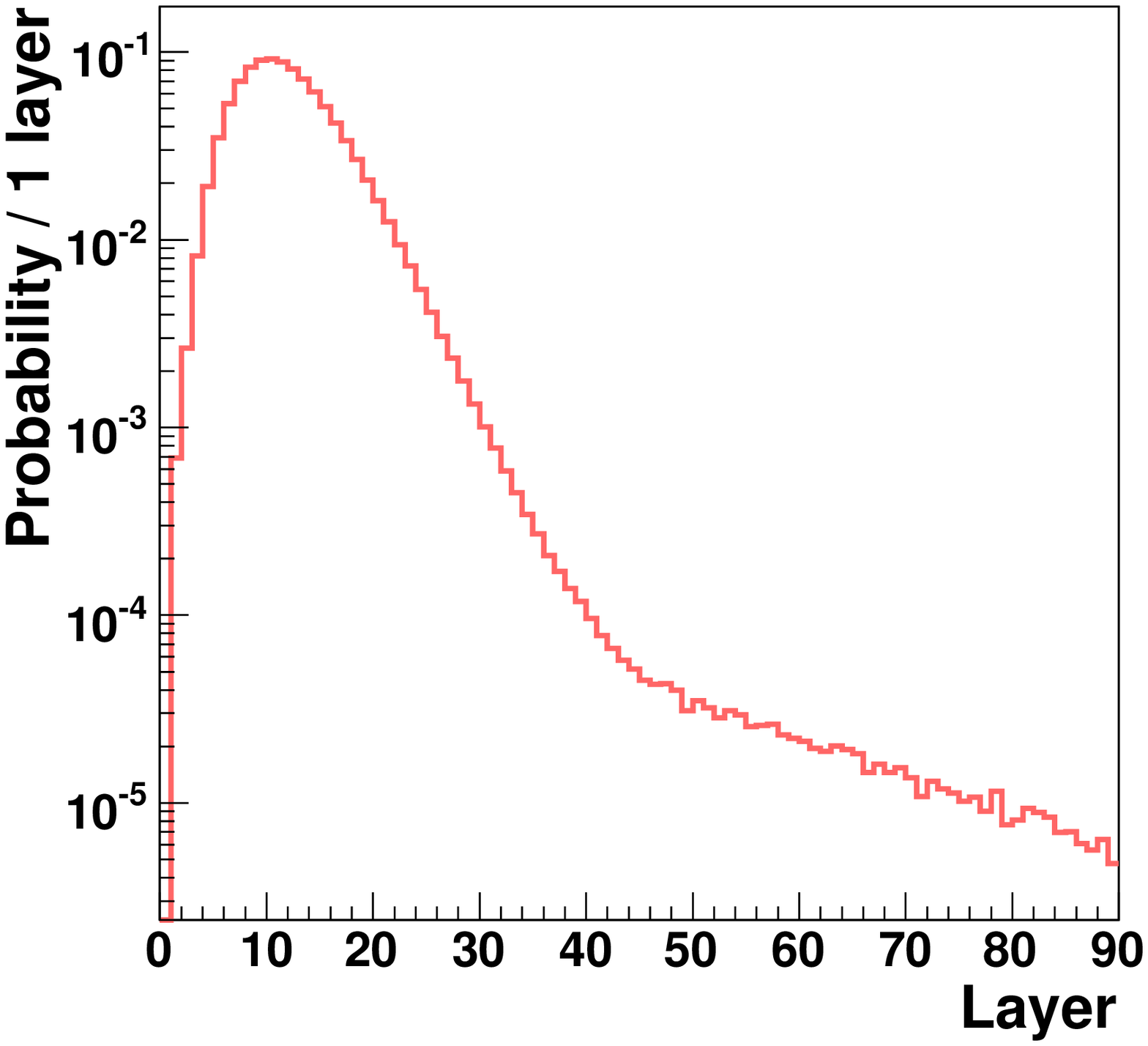}}
\subfloat[]{\label{engy90LayerFIG2}\includegraphics[width=.49\textwidth]{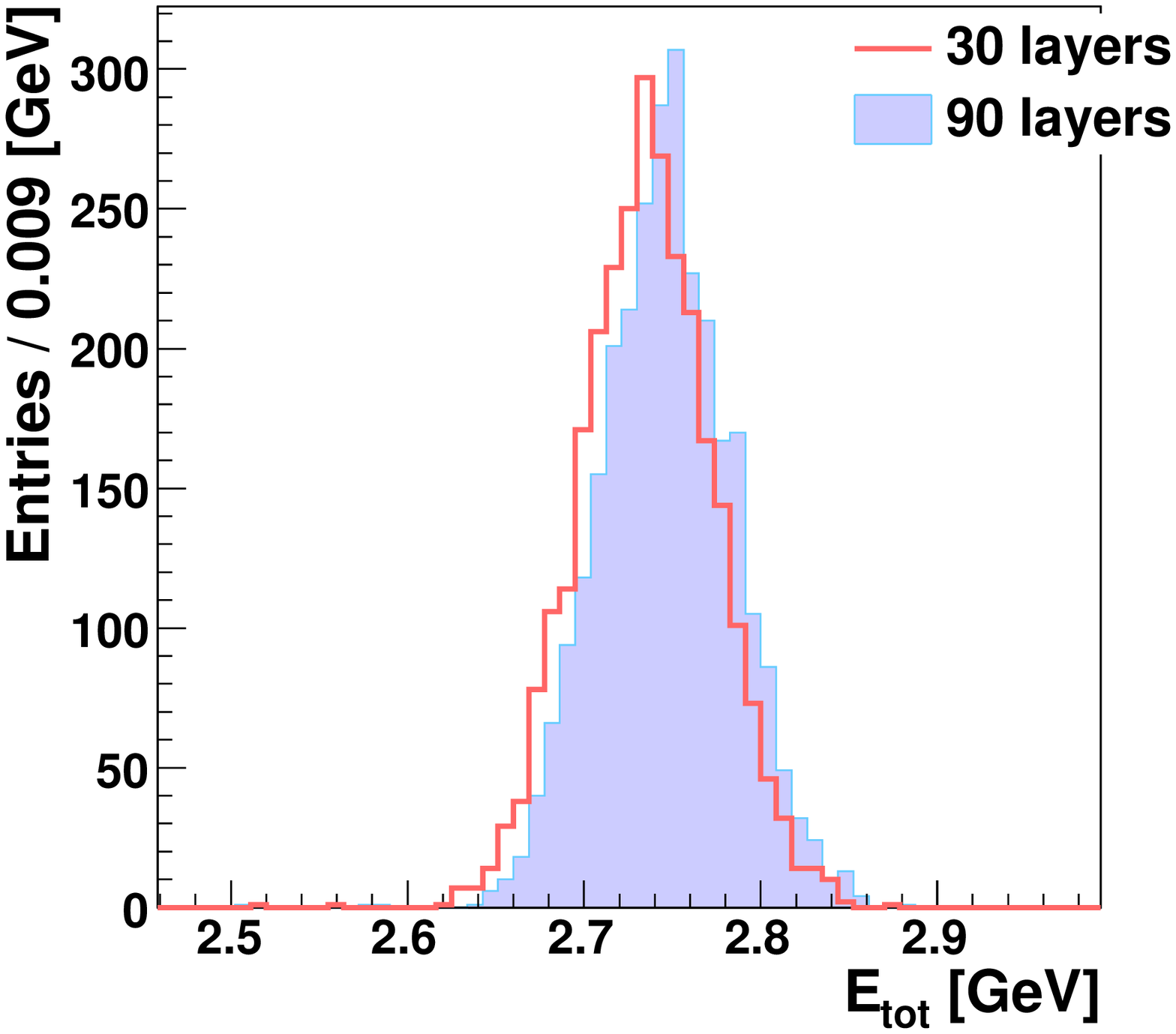}}
\caption{\label{engy90LayerFIG}\Subref{engy90LayerFIG1} Normalized distribution of the energy deposited in LumiCal as a function of the layer number for a detector with 90 layers. \Subref{engy90LayerFIG2} Comparison of the distribution of the total deposited energy, $E_{tot}$, for a 90 layer LumiCal with that of a 30 layer LumiCal, as denoted in the figure.}
\end{center}
\end{figure}

Each layer of LumiCal consists of 3.5~mm of tungsten, which is equivalent to one radiation length. A distribution of the energy deposited in a layer by 250~GeV electrons for a LumiCal of 90 layers is presented in \autoref{engy90LayerFIG1}. Only $0.4\%$ of the event energy is deposited beyond 30 layers. The distribution of the total event energy for 250~GeV electrons is plotted in \autoref{engy90LayerFIG2} for a LumiCal with 90 layers and for a LumiCal with 30 layers. A small difference is apparent in the mean value of the distributions, but this bears no consequence, as was explained in \autoref{digitizationSEC}. The energy resolution is the same for both cases. Since there is no degradation of the energy resolution it is concluded that 30 layers are sufficient for shower containment.

\begin{figure}[htp]
\begin{center}
\subfloat[]{\label{molRadFIG1}\includegraphics[width=.49\textwidth]{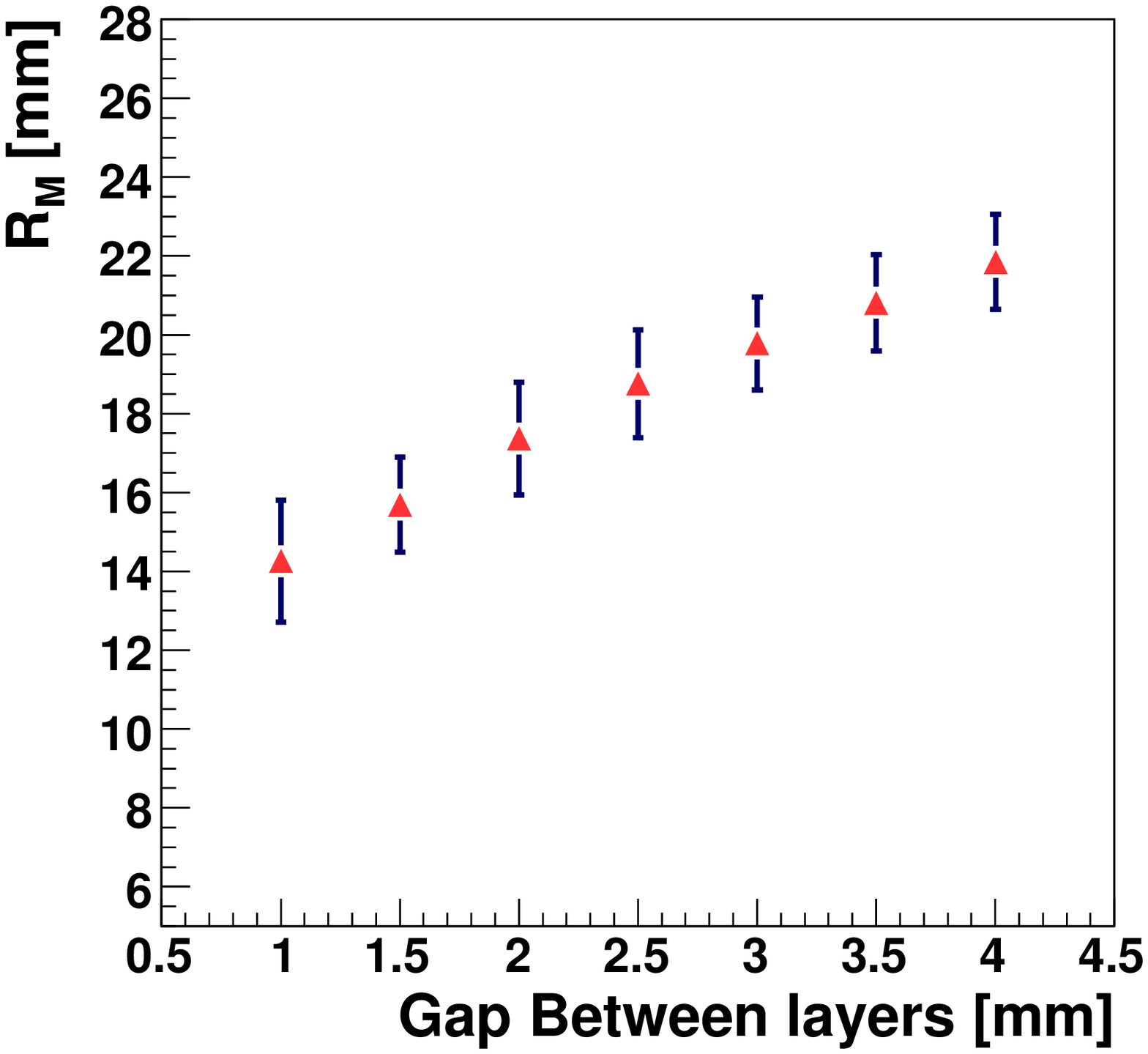}}
\subfloat[]{\label{molRadFIG2}\includegraphics[width=.49\textwidth]{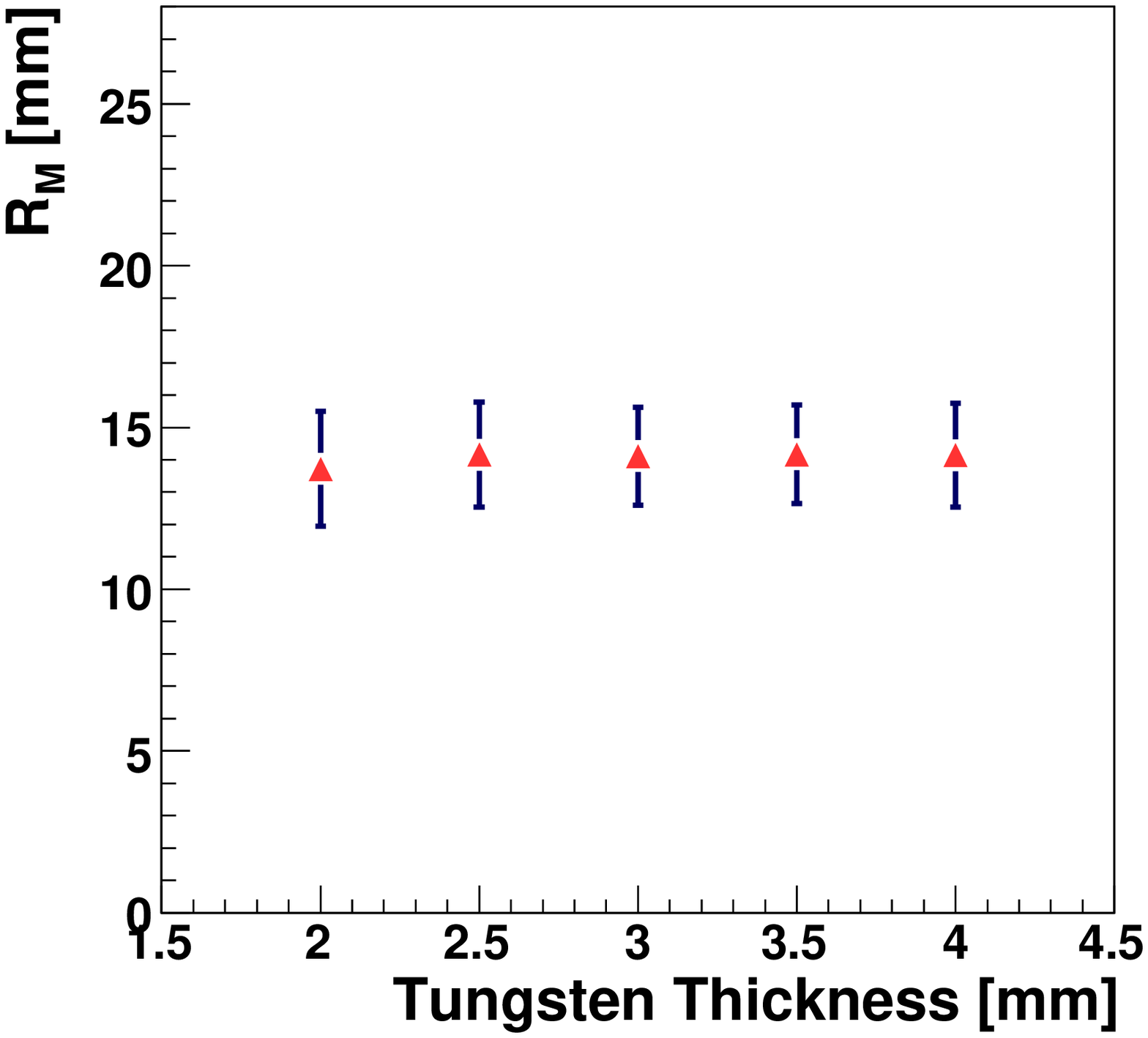}}
\caption{\label{molRadFIG}The Moli\`ere radius of LumiCal, $R_{\mc{M}}$, as a function of the gap between tungsten layers \Subref{molRadFIG1}, and as a function of the thickness of each layer \Subref{molRadFIG2}.}
\end{center}
\end{figure}

The Moli\`ere radius of LumiCal, $R_{\mc{M}}$, is plotted in \autoref{molRadFIG1} as a function of the gap between tungsten layers. Since a smaller $R_{\mc{M}}$ improves both the shower containment and the ability to separate multiple showers, the air gap should be made as small as possible. \autoref{molRadFIG2} shows the dependence of $R_{\mc{M}}$ on the tungsten thickness. It is apparent that there is no significant change in $R_{\mc{M}}$ over the considered range.

Changing the thickness of tungsten layers increases the sampling rate. In order to ensure shower containment, the total number of layers must remain 30 radiation lengths. Consequently, for a thinner tungsten layer length, more layers are needed, as  shown in \autoref{nLayersForTungsThicknessTABLE}.

\begin{table}[htp]
\begin{center} \begin{tabular}{ c|ccccc }
Tungsten Thickness [mm]      & 2.0 & 2.5 & 3.0 & 3.5 & 4 \\ \hline
Number of Required of Layers & 53 & 42 & 35  & 30 & 26 \\ 
\end{tabular} \end{center}
\caption{\label{nLayersForTungsThicknessTABLE}The required number of LumiCal layers as a function of the thickness of each tungsten layer.}
\end{table}

\begin{figure}[htp]
\begin{center}
\subfloat[]{\label{tungstenEngyResFIG1}\includegraphics[width=.49\textwidth]{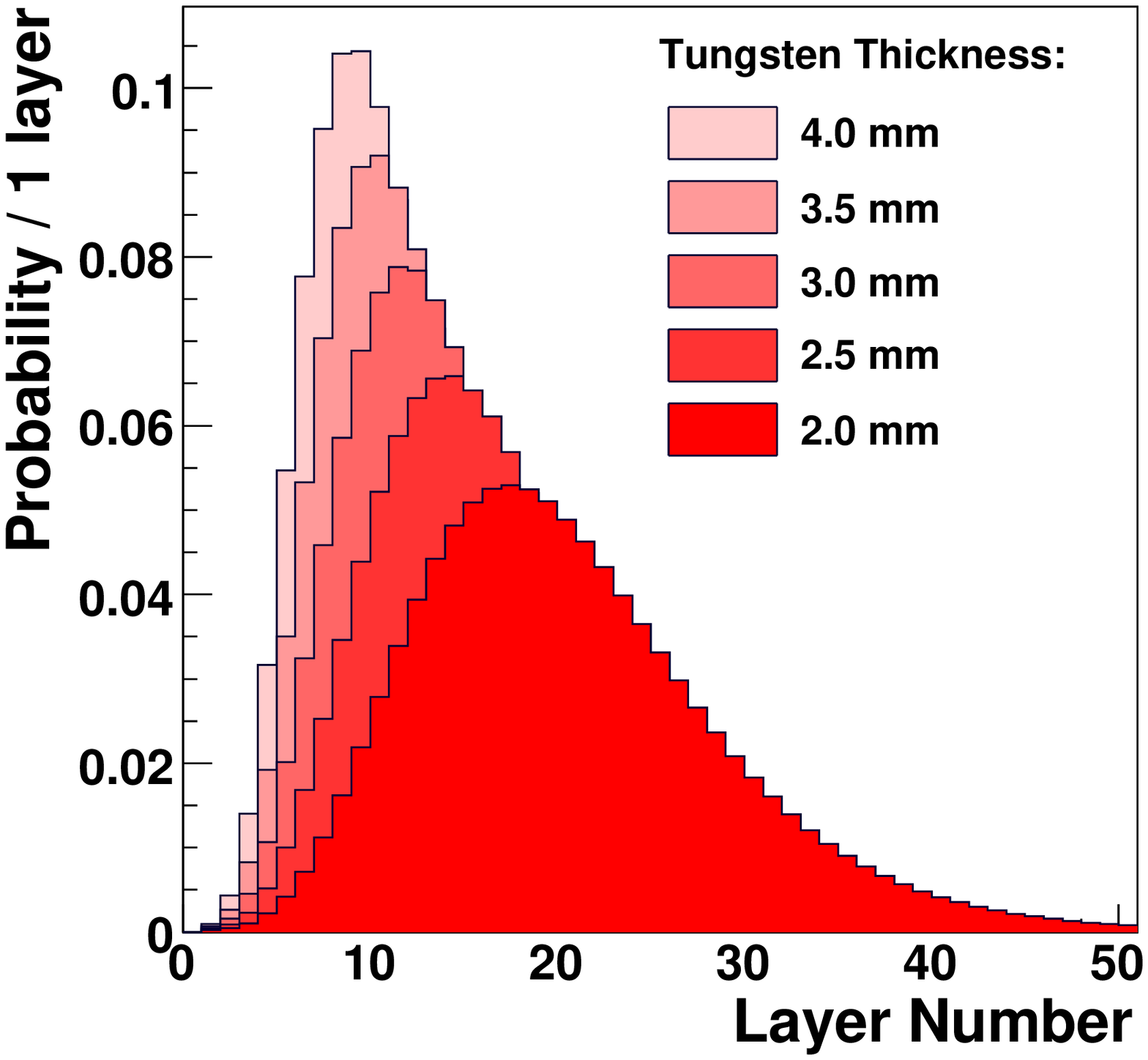}}
\subfloat[]{\label{tungstenEngyResFIG2}\includegraphics[width=.49\textwidth]{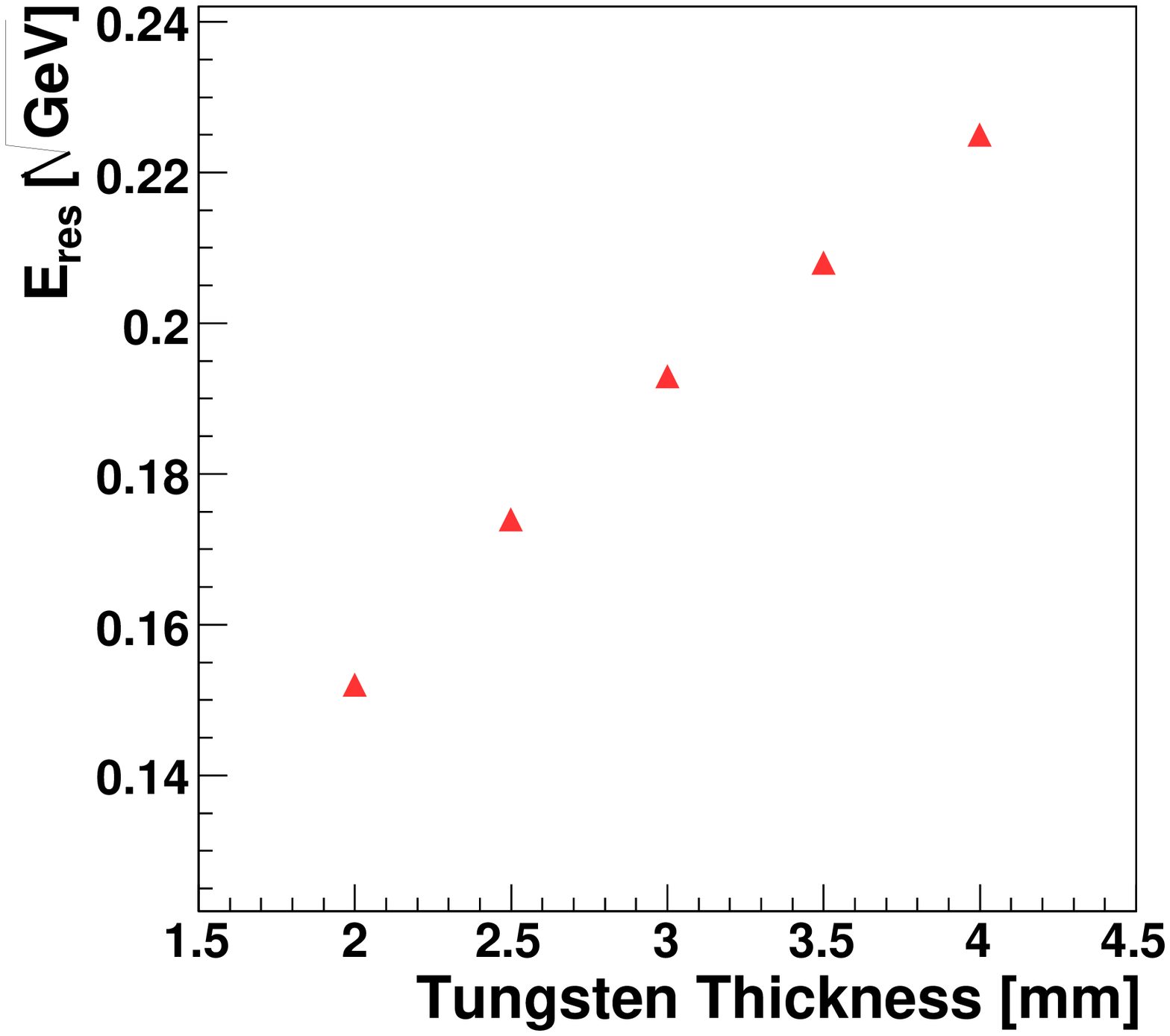}}
\caption{\label{tungstenEngyResFIG}\Subref{tungstenEngyResFIG1} Normalized distribution of the deposited energy for 250~GeV electron showers as a function of the layer number for several layer thicknesses, as denoted in the figure. \Subref{tungstenEngyResFIG2} The energy resolution, $E_{res}$, as a function of the thickness of tungsten layers.}
\end{center}
\end{figure}

\begin{figure}[htp]
\begin{center}
\subfloat[]{\label{tungstenPolarRecFIG1}\includegraphics[width=.49\textwidth]{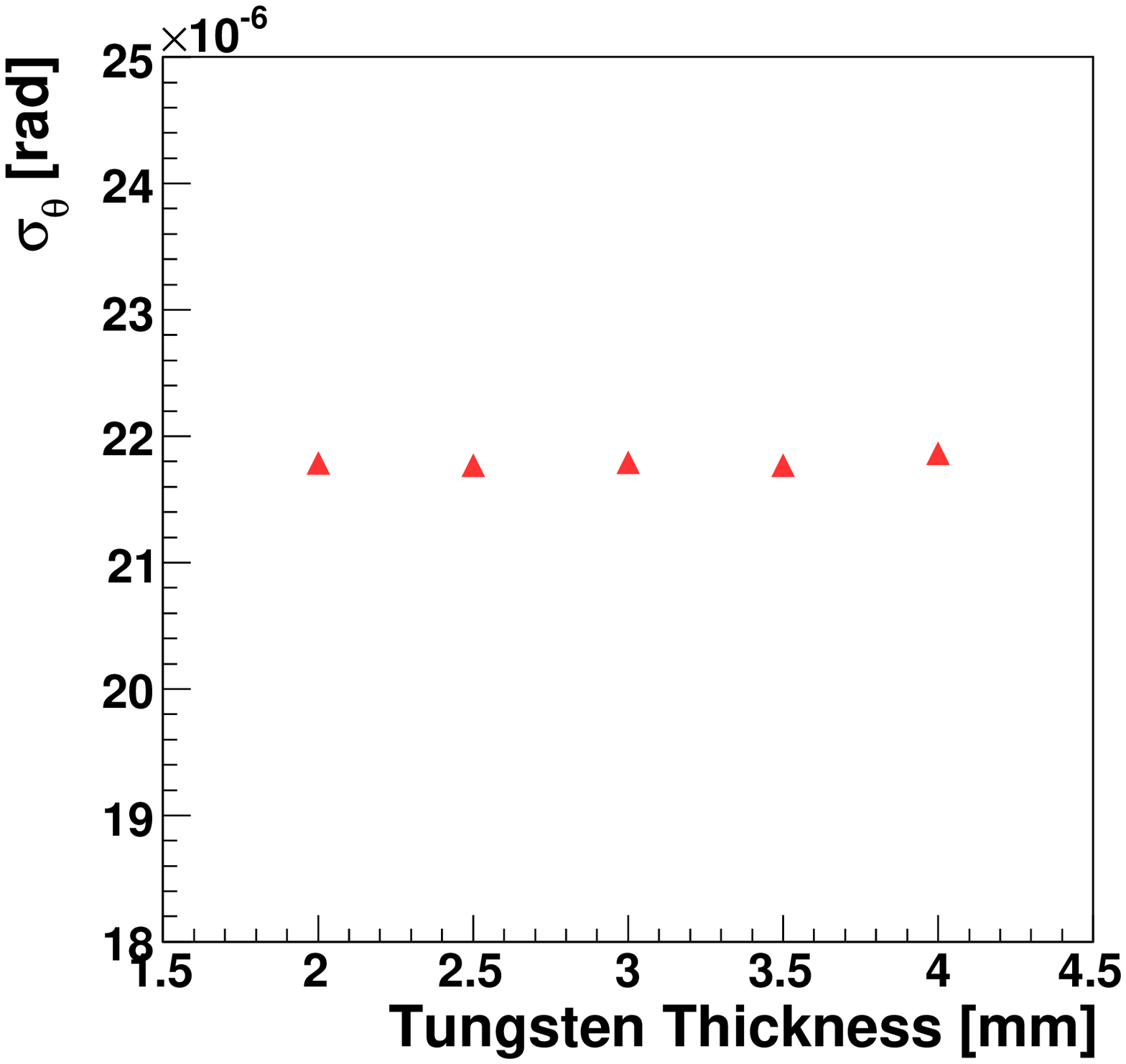}}
\subfloat[]{\label{tungstenPolarRecFIG2}\includegraphics[width=.49\textwidth]{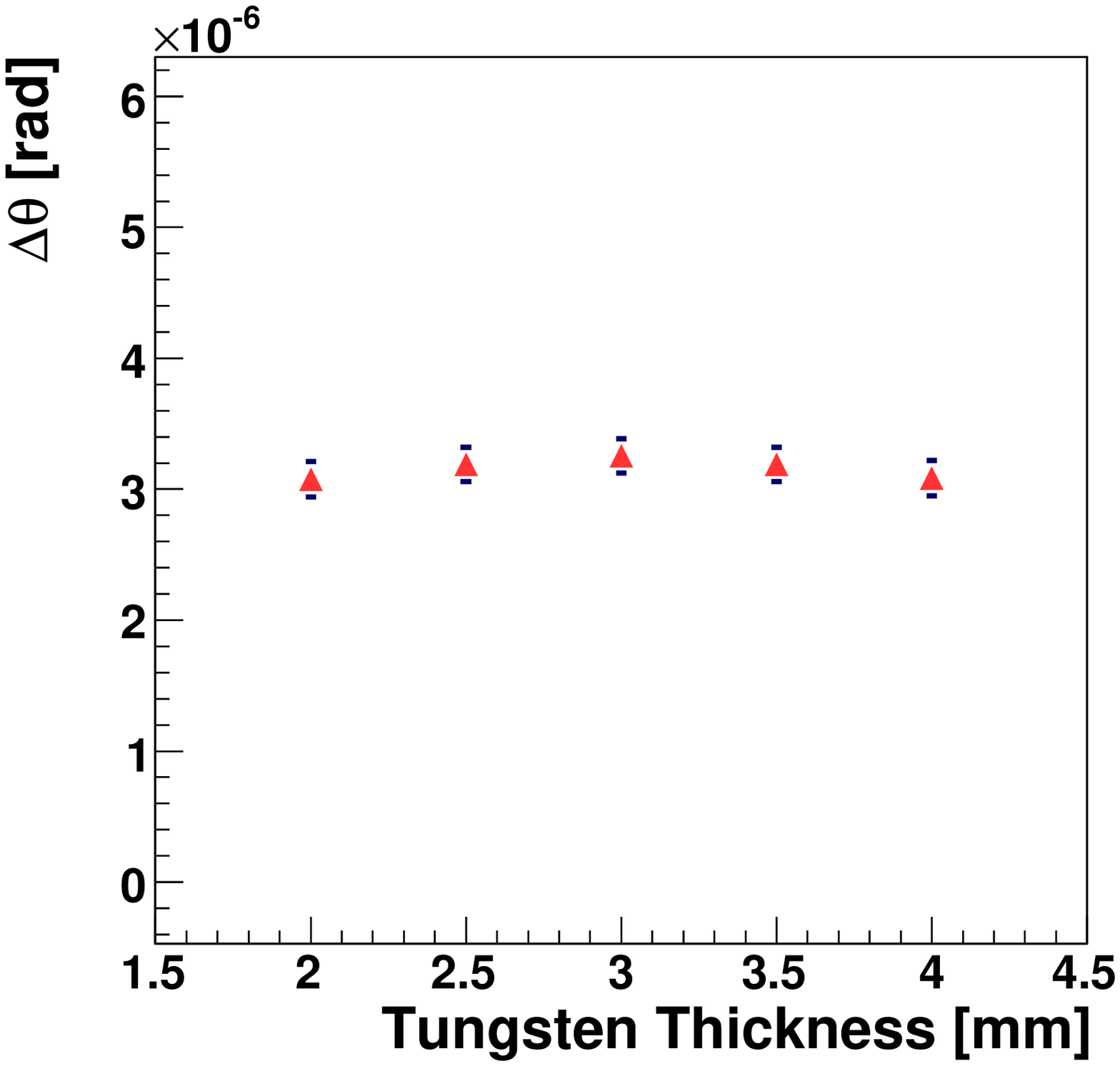}}
\caption{\label{tungstenPolarRecFIG}The polar resolution, $\sigma_{\theta}$, \Subref{tungstenPolarRecFIG1} and bias, $\Delta \theta$, \Subref{tungstenPolarRecFIG2} as a function of the thickness of a LumiCal layer, using 250~GeV electron showers.}
\end{center}
\end{figure}

\Autoref{tungstenEngyResFIG} shows the normalized distribution of the energy deposited per layer as a function of layer thicknesses, and the energy resolution, $E_{res}$, for each configuration. \Autoref{tungstenPolarRecFIG} shows the corresponding polar resolution, $\sigma_{\theta}$, and bias, $\Delta \theta$.

Due to the fact that more layers encompass the shower-peak area for thinner tungsten layers, the energy resolution is improved. The polar reconstruction and the Moli\`ere radius are not affected. The trade-off for choosing a given thickness of tungsten, is then between an improvement in $E_{res}$ and the need to add more layers. Since increasing the number of layers also involves a raise in the cost of LumiCal, a clear lower bound on $E_{res}$ needs to be defined, so as to justify the additional expense. Currently, $E_{res} \approx 0.21 ~ \sqrt{\mathrm{(GeV)}}$ seems sufficient.

\subsection{Inner and Outer Radii \label{innerOuterRadiiSEC}}

The distance from the IP (2.27~m), the radial cell size (0.8~mrad) and the number of radial divisions (64) dictate that the total radial size of LumiCal be 110~mm. Setting the inner and outer radii $R_{min}$ and $R_{max}$, within this limit has several implications. 

\begin{figure}[htp]
\begin{center}
\subfloat[]{\label{pairDistFIG1}\includegraphics[width=.49\textwidth]{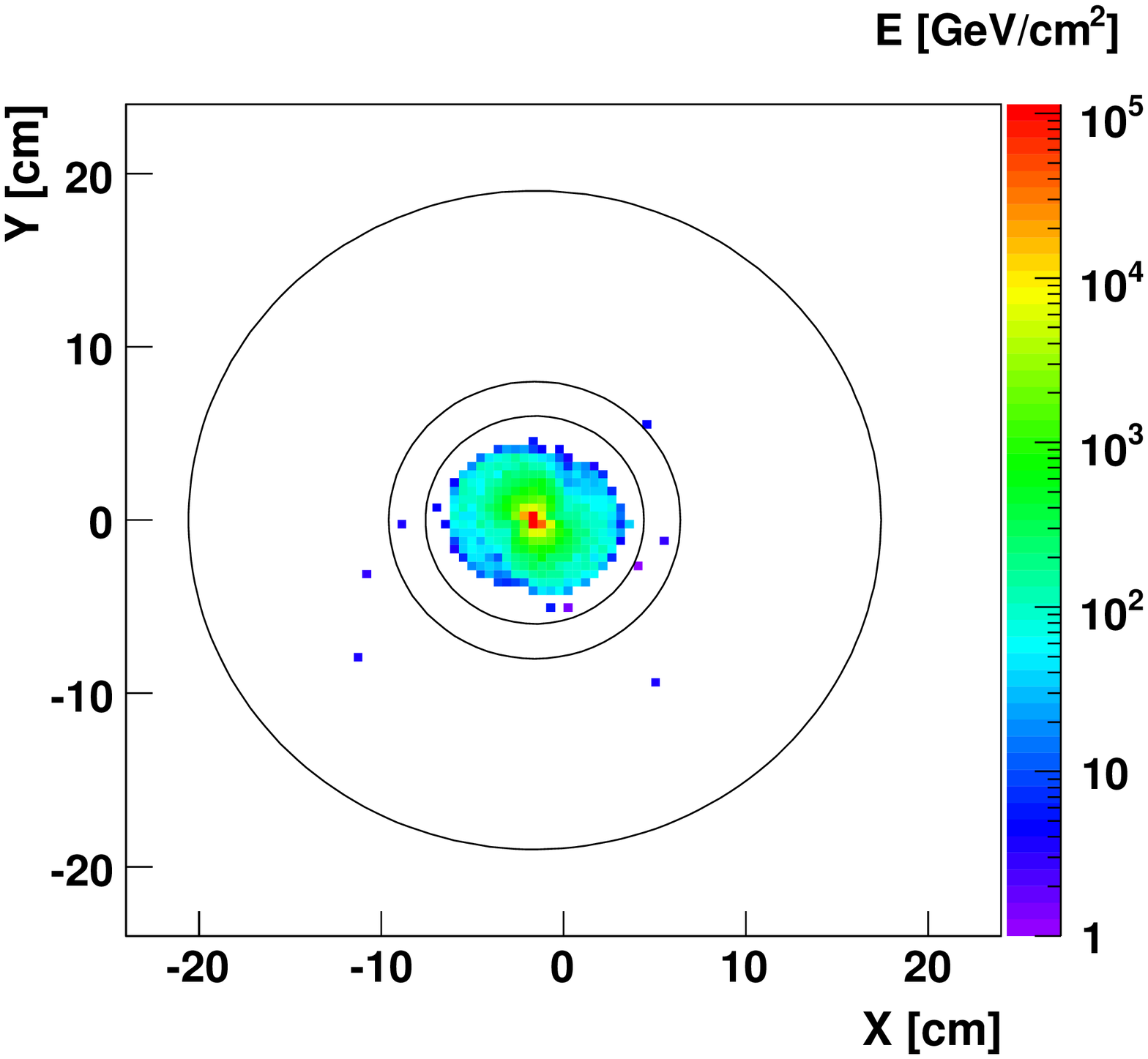}}
\subfloat[]{\label{pairDistFIG2}\includegraphics[width=.49\textwidth]{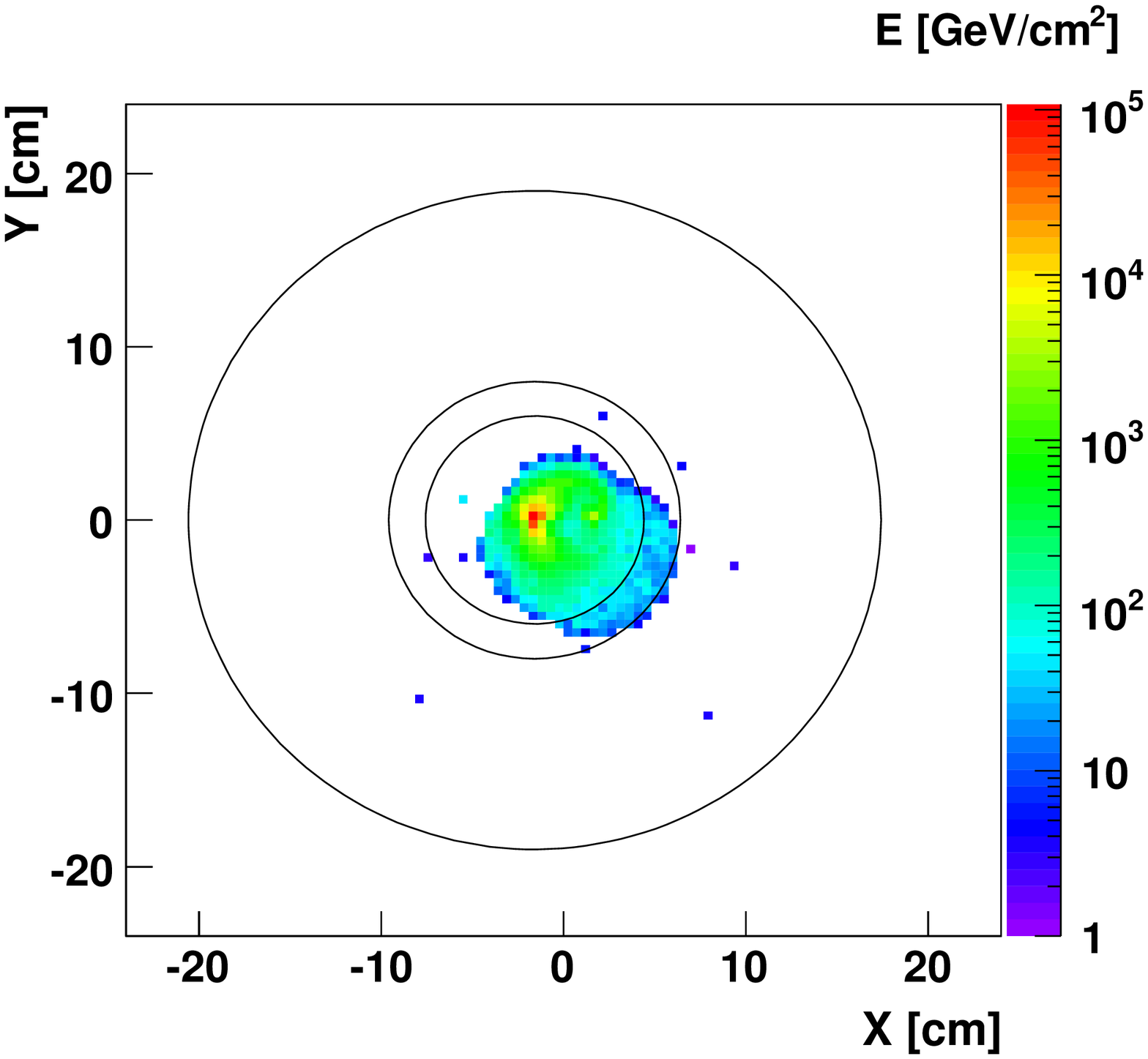}}
\caption{\label{pairDistFIG}Projection of beamstrahlung pair energies on the face of LumiCal for an anti-DID \Subref{pairDistFIG1} and a DID \Subref{pairDistFIG2} magnetic field setup, for the nominal accelerator operational parameters. The concentric circles represent possible inner LumiCal radii of 60 and 80~mm, and an outer radius of 190~mm.}
\end{center}
\end{figure}

In the \textit{detector integrated dipole} (DID)~\cite{revisedDetectorModelBIB7} field configuration, the magnetic field is directed along the incoming beam lines with a kink at the transverse plane containing the IP. Conversely, the magnetic field may also be directed along outgoing beam lines with a kink at the IP plane, a configuration referred to as anti-DID. \Autoref{pairDistFIG} shows a projection of the energy of beamstrahlung pairs on the face of LumiCal for the anti-DID and the DID magnetic field configurations of the accelerator. The two inner concentric black circles represent possible inner radii of 60 and 80~mm, while the outer circle is set at 190~mm. The beamstrahlung spectrum was generated using GUINEA-PIG~\cite{revisedDetectorModelBIB8}.

For a DID field the beamstrahlung pair distribution grazes LumiCal, while for the anti-DID case it does not, though the distribution comes close to the LumiCal inner edge. The affect of exposure to the extremely high energy dose will cause massive damage to the silicon sensors in a matter of months. The anti-DID field is, thus, the better choice. The difference between the two distributions of \autoref{pairDistFIG} suggests that small fluctuations in the magnetic field from the nominal configuration will cause the pair distribution to become wider. It is, therefore, concluded that it would be preferable to add a safety margin to the minimal (60~mm) choice of $R_{min}$.

It has also been shown~\cite{revisedDetectorModelBIB5} that for $R_{min} < 70$~mm there is a significant increase in the amount of backscattered particles from LumiCal to the inner detector (the TPC). This too constitutes a motivation for setting $R_{min}$ at a higher value.

Since the Bhabha cross-section falls off quickly with the polar angle (see \autoref{bhabhaXs2EQ}), it is advantageous to set $R_{min}$ as low as possible in order to increase the number of Bhabha events within the fiducial volume of LumiCal. \Autoref{rMinMaxTABLE} gives the integrated Bhabha cross-section, $\sigma_{\mathrm{B}}$, in the fiducial volume defined by several choices of $R_{min}$ and $R_{max}$. The number of Bhabha events and the relative statistical error is calculated according to \autorefs{luminosityByXsCountingEQ} and \ref{luminosityRelativeErrStatEQ}, respectively. An integrated luminosity of $500~\mathrm{fb}^{-1}$ was assumed. The relative error resulting from the polar reconstruction (\autoref{luminosityRelativeErrRec2EQ}) is also shown in the table, where a polar bias $\Delta \theta = 3.2  \cdot 10^{-3}$~mrad (\autoref{thetaGranTABLE}) and the appropriate minimal polar angles were used in each case.

\begin{table}[htp]
\begin{center} \begin{tabular}{ |c|c|c|c|c|c| }
\hline
$R_{min} \rightarrow R_{max}$ & ~~~$\theta_{min}$~~~ & ~~~$\theta_{max}$~~~ & ~~~~$\sigma_{\mathrm{B}}$~~~~ & \multirow {2}{*}{\begin{large}$\frac{\Delta N_{B}}{N_{B}}$\end{large}} & \multirow {2}{*}{\begin{large}$\frac{2 \Delta \theta}{\theta_{min}}$\end{large} } \\ [-6pt]
 [mm] & [mrad] & [mrad] & [nb] &  & \\ [2pt]
\hline \hline
60 $\rightarrow$ 170 & 33 & 59 & 2.58 & ~$2.8  \cdot 10^{-5}$~ & ~$1.9  \cdot 10^{-4}$~ \\ \hline
70 $\rightarrow$ 180 & 37 & 64 & 1.98 & $3.2  \cdot 10^{-5}$ & $1.7  \cdot 10^{-4}$ \\ \hline
80 $\rightarrow$ 190 & 41 & 69 & 1.23 & $4  \cdot 10^{-5}$   & $1.5  \cdot 10^{-4}$ \\ \hline
90 $\rightarrow$ 200 & 50 & 74 & 0.86 & $4.8  \cdot 10^{-5}$ & $1.3  \cdot 10^{-4}$ \\ \hline
\end{tabular} \end{center}
\caption{\label{rMinMaxTABLE}The fiducial volume, $R_{min} \rightarrow R_{max}$, bound by the minimal and maximal polar angles, $\theta_{min}$ and $\theta_{max}$, and the integrated Bhabha cross-section, $\sigma_{\mathrm{B}}$, for a center of mass energy $\sqrt{s} = 500$~GeV. The two relative errors on the luminosity measurement, the statistical error and the one resulting from reconstruction of the polar angle, are also shown. The number of Bhabha events, $N_{B}$, is computed for an integrated luminosity of $500~\mathrm{fb}^{-1}$, and the polar bias used is $\Delta \theta = 3.2  \cdot 10^{-3}$~mrad.}
\end{table}

As expected from \autoref{bhabhaXs2EQ}, for low values of $R_{min}$ the number of Bhabha events increases, thus decreasing the statistical error. As the polar bias depends on the angular cell size, which was kept constant, and not on the radii, the error resulting from the polar reconstruction decreases slightly with the rise of $\theta_{min}$. Both effects contribute to the overall uncertainty in the luminosity measurement.

It should be noted here that in practice the counting rates of Bhabha events will be lower than presented in \autoref{rMinMaxTABLE}. This is due to the fact that the efficiency for counting Bhabha events is not $100\%$ due to selection cuts~\cite{introductionBIB22}. This in itself does not add to the luminosity error, as long as the efficiency is known to high precision, but it does increases the statistical error\footnote{For instance, for a pessimistic selection efficiency of roughly $50\%$, the error will increase by a factor of $\sqrt{2}$.}.

The contribution of the error due to the polar bias also needs further consideration. In practice it will be possible to determine the polar bias using a test beam, and correct for this effect. The final error will then depend on how well one can correct for the bias, so that the values given in the table are an upper bound on the error.

In conclusion, even given the lower counting rate and no corrections of the polar bias, it is apparent that for the range of $R_{min} \rightarrow R_{max}$ given in \autoref{rMinMaxTABLE}, the relative error in luminosity is well within the design goal. Aiming to increase the available statistics as much as possible, while maintaining a safe distance from the beamstrahlung pairs, $R_{min} = 80~\mathrm{mm}$ was finally chosen.

\subsection{Clustering \label{clusteringGeometryDependanceSEC}}

The effectiveness of the clustering algorithm, which was described in \autoref{clusteringCH}, in measuring the Bhabha cross-section, depends on the granularity of LumiCal. The performance, which was presented in \autoref{clusteringObservablesSEC}, was evaluated for a detector with inner and outer radii,  $R_{min} = 80~\mathrm{mm}$ and $R_{max} = 350~\mathrm{mm}$ respectively, 104 radial divisions, and 96 azimuthal divisions. This geometry corresponds  to a radial cell size of 2.6~mm, and an azimuthal cell size of 8.5~mm at the center of LumiCal. For the optimized  geometry, described in \autoref{baselineGeometryTABLE}, the radial cell size is 3.25 times smaller, and the azimuthal cell size is twice as large.

In order to estimate the dependence  of the performance of the clustering algorithm on these changes, the clustering of a sample  of $10^{4}$ Bhabha events was performed for different LumiCal geometries\footnotemark. In all cases the inner and outer radii were kept at the optimized values of $R_{min} = 80~\mathrm{mm}$ and $R_{max} = 190~\mathrm{mm}$, respectively, and the number of radial and azimuthal divisions was changed. The merging cuts used on the minimal energy of a cluster and on the separation distance between a pair of clusters  (see \autoref{clusteringEventSelectionSEC}) are $E_{low} \ge 20$~GeV and $d_{pair} \ge R_{\mc{M}}$, respectively.

\footnotetext{The clustering  algorithm described in \autoref{clusteringAPP} was utilized. Adjustments of several of the parameters were made, such as the number of near-neighbors and the weighting constant, $\mc{C}$, (\autoref{logWeighClustering2EQ}) in order to accommodate  the changes in cell size.}

The values of the acceptance and purity (\autoref{apeEQ}) are presented in \autoref{apeGeometryTABLE}. Also shown is the total relative error on the measurement of the effective Bhabha cross-section (\autorefs{missCountingRelErr1EQ} and \ref{missCountingRelErr2EQ}) for an integrated luminosity of $500~\mathrm{fb}^{-1}$.

\begin{table}[htp]
\begin{center} \begin{tabular}{ |c|c|c|c|c| }
\hline
\multicolumn {2}{|c|}{Number of Divisions}  &
\multirow {2}{*}{\; $\mc{A} ~[\%]$ \;}  &
\multirow {2}{*}{\; $\mc{P} ~[\%]$ \;}  &
\multirow {2}{*}{\; \begin{large} $\frac{\Delta N_{tot}}{N_{tot}}$ \end{large} \;}  \\  \cline{1-2}
Azimuthal & ~~ Radial ~~ & & & \\ [2pt]
\hline \hline
96  & 128 & 99 &  94 &  $2.9\cdot10^{-5}$  \\ \hline
96  & 32  & 98 &  92 &  $3.5\cdot10^{-5}$  \\ \hline
48  & 128 & 94 &  79 &  $6.6\cdot10^{-5}$  \\ \hline
48  & 64  & 93 &  77 &  $7.5\cdot10^{-5}$  \\ \hline
48  & 32  & 90 &  84 &  $9.1\cdot10^{-5}$  \\ \hline
24  & 64  & 76 &  22 &  $11.1\cdot10^{-5}$ \\ \hline
\end{tabular} \end{center}
\caption{\label{apeGeometryTABLE}The values of the acceptance, $\mc{A}$, and purity, $\mc{P}$, of the algorithm, for several division schemes. Also shown is the total relative error on the measurement of the effective Bhabha cross-section for an integrated luminosity of $500~\mathrm{fb}^{-1}$. The merging-cuts on the minimal energy of a cluster and on the separation distance between a pair of clusters are $E_{low} \ge 20$~GeV and $d_{pair} \ge R_{\mc{M}}$, respectively.}
\end{table}

Changes in the number of azimuthal divisions have a large effect on the final error of the cross-section measurement, compared to changes in the number of radial divisions. This difference is due to the fact that LumiCal is more finely granulated in the radial direction. One may fine-tune the parameters of the algorithm in order to adjust the values of the acceptance and of the purity. In general, an increase in $\mc{A}$ will be followed by a decrease of $\mc{P}$, as one is a measure of over-merging of clusters, and the other of under-merging. The contribution to the statistical error of the number of shower-pairs which are reconstructed as one cluster, far outweighs that of the number of single showers which are reconstructed as two clusters. Since it is advisable to choose parameters, such that the total error is minimal, the acceptance tends to be higher than the purity.

It is apparent that the increase in cell size diminishes from the effectiveness of the clustering. For the optimized geometry, though, the final relative error with which the Bhabha cross-section may be measured, is within the design goal.

\section{Conclusions on the Optimization Procedure\label{optimizationConclusionsSEC}}

In the following, the dependence of the various performance parameters on the geometry of LumiCal are summarized.

\vspace{15pt} \noindent \textbf{Dynamical range of the signal} -
Reducing the maximal signal of a single cell may be accomplished by increasing the number of cells (\autoref{maxChargeCell250GeVFIG}). A low signal size brings about less rounding errors when digitizing the induced charge in a cell, and is also preferable from the readout electronics point of view. This change, however, carries the added complication of increasing the number of channels, which in turn hinders the readout, as discussed above.

\vspace{15pt} \noindent \textbf{Energy resolution} -
The energy resolution depends on the containment of the EM shower, on the precision with which each cell is read-out, and on the sampling rate of the shower.

\begin{list}{-}{}
\item
Containment of showers is achieved by keeping the total number of layers $30~X_{0}$ thick, and imposing fiducial cuts on the polar angle of incident showers.

\item
The accuracy of reading-out cell-energies is guarantied not to degrade the energy resolution, so long as a digitization scheme with high enough resolution (\autoref{digiThetaRecFIG1}) is implemented.

\item
The sampling rate of the shower is determined by the thickness of each tungsten layer (\autoref{tungstenEngyResFIG}). The best way to improve the energy resolution is to decrease the thickness of layers. One must increase the number of layers accordingly.
\end{list}

\vspace{15pt} \noindent \textbf{Moli\`ere radius} -
Keeping the Moli\`ere radius small improves both the shower containment, by relaxing the fiducial cuts, and the ability to resolve multiple showers. This may be done by decreasing the gap between tungsten layers to the minimal possible value.

\vspace{15pt} \noindent \textbf{Inner and outer radii of LumiCal} -
The inner radius of LumiCal determines the minimal polar angle, which is accessible by the calorimeter. Since the Bhabha event-rate falls off rapidly with the polar angle (\autoref{bhabhaXs2EQ}), it is preferable to decrease $R_{min}$ as much as possible. The lower bound on $R_{min}$ must be set such that the beamstrahlung pairs do not enter LumiCal, and the backscattering from LumiCal into the inner detector is acceptable. The outer radius of LumiCal is less important in terms of a gain in statistics.

\vspace{15pt} \noindent \textbf{Relative error on the luminosity measurement} -
Possible future changes in the position and size of LumiCal may affect its angular coverage and its angular cell size. This will influence the accuracy of the polar angle reconstruction, and accordingly the relative error on the luminosity measurement (\autoref{thetaBiasDivisionsFIG} and \autoref{thetaGranTABLE}). The number of radial divisions must therefore be adjusted, so that the angular cell size remains constant. Decreasing the number of azimuthal cells does not affect the reconstruction of the polar angle, yet a measurement of the Bhabah cross-section through clustering, requires that cell sizes not be large in comparison  to the Moli\`ere radius. The total fiducial volume of LumiCal must also be large compared to $R_{\mc{M}}$, so as to achieve a high rate of Bhabha events, which pass the merging-cut on the separation distance between cluster pairs.

\chapter{Summary \label{summaryCH}}

The International Linear Collider will provide physicists a new doorway to explore energy regimes beyond the reach of today's accelerators. A proposed electron-positron collider, the ILC will complement the Large Hadron Collider, a proton-proton collider at the European Center for Nuclear Research in Geneva, Switzerland, together unlocking some of the deepest mysteries in the universe. With LHC discoveries pointing the way, the ILC, a true precision machine, will provide the missing pieces of the puzzle. In order to achieve its goal, the luminosity of the ILC will have to be know with a precision of $10^{-4}$, which poses a significant challenge. Luminosity in the ILC is measured by counting Bhabha scattering events, a task which will require pattern recognition in the main detector, and in the luminosity calorimeter.

It has been shown here that it is possible to resolve the distribution of radiative Bhabha photons on top of the electron distribution, and thus measure an effective Bhabha cross-section directly. Using this measurement, it will be possible to verify the influence of the beam-beam effects, and of the energy spread of the collider, on the Bhabha cross-section.

A study has also been presented, in which the design of LumiCal was optimized, with the goal of reducing the error in the luminosity measurement to the required threshold. While this study holds merit in and of itself, it also serves as a template for future optimizations. Such procedures are foreseen to be needed, in light of the expected changes to the detector-concept, as a result of ongoing R$\&$D efforts.

\addcontentsline{toc}{chapter}{Acknowledgments}
\chapter*{Acknowledgments}

I would like to thank my supervisors, Prof.~Halina Abramowicz and Prof.~Aharon Levy, for introducing me to the field of high energy physics, and for all of their help and advice along the way. A special thanks also goes to Dr.~Wolfgang Lohmann, who adopted me into the DESY-Zeuthen ILC group during the last several months of my work. I am grateful to my colleagues, Ronen Ingbir, Dr.~Sergey Kananov and Dr.~Christian Grah, for the discussions on physics, luminosity and hardware, and also to Dr.~Zhenya Gurvich, whose help on Linux related issues has been invaluable. Finally, I would like to acknowledge my office-mates, Amir Stern, Ori Smith, Martin Ohlerich and Ringo Schmidt for all the mental support they provided, and for keeping the coffee flowing.

This work is partly supported by the Commission of the European Communities under the $6^{\mathrm{th}}$ Framework Programme ``Structuring the European Research Area'', contract number RII3-026126, and by the Israeli Science Foundation.

\appendix

\chapter{The Clustering Algorithm in Full Detail \label{clusteringAPP}}

The clustering algorithm, with was developed for LumiCal, operates in three main phases;

\begin{list}{-}{}

\item
selection of shower-peak layers, and two-dimensional clustering therein;

\item
fixing of the number of global (three-dimensional) clusters, and collection of all hits onto these;

\item
testing of the global-clusters, by means of the evaluation of their energy density.

\end{list}

The algorithm is described in detail in the following subsections.

\section{Clustering in the Shower-Peak Layers \label{nn2DclustersSPlayersAPPSUBSEC}}

\subsection{Near-Neighbor Clustering}

As the initiating particle traverses LumiCal, more and more secondary particles are created and the shower spreads out (see \autoref{clorimetryCH}). The greatest density of energy per individual layer is concentrated around a local center-of-gravity. The method of near-neighbor clustering (NNC) exploits the gradient of energy around the \textit{local shower-center}, by assuming that in first order, the further a hit is relative to the shower center, the lower its energy. By comparing the energy distribution around the center at growing distances, one may check whether the energy is increasing or decreasing. An increase in energy for growing distance from the shower-center would then imply that the hit should be associated with a different shower.

It is helpful to consider at this point only cells which have a relatively high energy content, and so a cut on cell energy (given below in \autoref{NNCenergyCutEQ}) is made before the NNC begins. Hits of low energy are, therefore, associated with clusters at a later stage.

\begin{figure}[htp]
\begin{center}
\includegraphics[width=.49\textwidth]{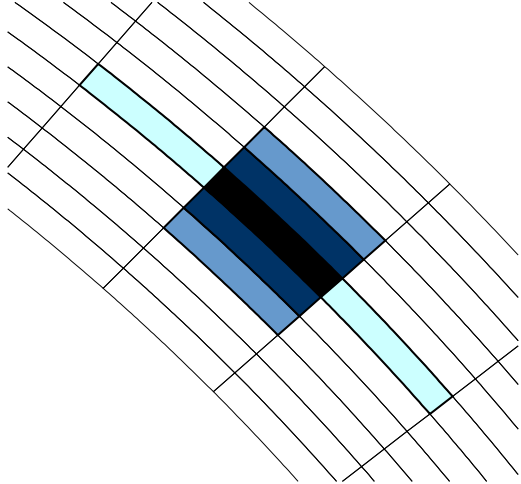}
\caption{\label{NNidsFIG}Schematic representation of the relative locations of the near-neighbors which are considered in the NNC algorithm. Darker shadings represent shorter distances between the cells in question and the principle cell, which is at the center of the drawing, painted black.}
\end{center}
\end{figure}

The radial cell length is constant and has a value of 2.6~mm (for the geometry described in \autoref{lumiCalDescriptionSEC}), while the azimuthal cell length, which is radially dependent, is 5~mm at its lowest and 23~mm at its highest. The Moli\`ere radius of LumiCal is $R_{\mc{M}} = 14$~mm, which is comparable with the cell lengths in the middle of the detector. This means that the cell size of LumiCal hinders the effectiveness of the procedure described above, and therefore the NNC method is used in a mode, in which only six near-neighbors are considered for each hit. One and two displacements are considered in the radial direction, and one in the azimuthal direction. A schematic representation of the relative locations of the near-neighbors is shown in \autoref{NNidsFIG}. The lighter the shade of the cell, the further away is its center, relative to the center of the principle cell. The ratio of the radial to azimuthal cell size is representative of the middle region of LumiCal.

For each shower-peak layer separately, the algorithm associates each cell which has an energy deposit with its highest-energy near-neighbor. The result of the NNC phase is a collection of clusters in each layer, centered around local maxima, as illustrated for a single layer in \autoref{NNclustersFIG1} (\autoref{nn2DclustersSPlayersSUBSEC}). In this example the algorithm produces six clusters, which are enumerated in the figure. The different clusters are also distinguished by different color groups, where darker shadings indicate a higher energy content of the cell in question.

\subsection{Cluster-Merging}

The next phase of the algorithm is cluster-merging. The NNC method only connects cells which are relatively close, while showers tend to spread out over a large range of cells, as indicated in \autoref{longitudinalProfileFIG2} (\autoref{characteristicsOfShowersSEC}). The result of the algorithm after the cluster-merging phase (explained below) is illustrated for a single layer in \autoref{NNclustersFIG2} (\autoref{nn2DclustersSPlayersSUBSEC}).

The first step in the cluster-merging phase is the assignment of a center-position for each cluster. This is done by averaging over all the hits of each cluster, using the hit cell centers, $\left( x_{i} , y_{i} \right)$, and a weight function, $\mc{W}_{i}$. For the $x$-coordinate of the shower-center one has

\begin{equation}{
<x> = \frac{\sum_{i}x_{i} \cdot \mc{W}_{i}}{\sum_{i} \mc{W}_{i}} \; ,
}\label{logWeighClustering1EQ} \end{equation}

\noindent and equivalently for the $y$-coordinate. Weights are determined by the so-called logarithmic weighting~\cite{revisedDetectorModelBIB4}, for which

\begin{equation}{
\mc{W}_{i} = \mathrm{max} \{~ 0 ~,~ \mc{C} + \mathrm{ln} \frac{E_{i}}{E_{cl} ~} \},
}\label{logWeighClustering2EQ} \end{equation}

\noindent where $E_{i}$ is the individual cell energy, $E_{cl}$ is the sum of the energy of all the cluster cells, and $\mc{C}$ is a constant.  The constant, $\mc{C}$, acts as an effective cut on energy, so that only cells which contain a high percentage of the cluster energy contribute to the reconstruction of the position of the cluster. Setting different values for $\mc{C}$ influences the performance of the algorithm. It was found that the value of $\mc{C}$, determined for a single shower of 250~GeV (\autoref{polarAngleRecSEC}) also gives the best results for the case of multiple showers with the same total energy. It was also found that one should use the same weighting constant for all of the showers.

Inverting \autoref{logWeighClustering2EQ} allows to finally define the cut on cell energy, $C_{min}$, which was used before the NNC began,

\begin{equation}{
C_{min}( \mc{P}_{cut}) = \mathrm{e}^{- \mc{C}} E_{all} \cdot \mc{P}_{cut} \, ,
}\label{NNCenergyCutEQ} \end{equation}

\noindent where $E_{all}$ is the total energy deposited in all the layers of LumiCal, and $\mc{P}_{cut}$ is some fraction. The best performance of the algorithm was achieved for $\mc{P}_{cut} = 1\%$.

Once a center-position is determined for each cluster, the merging process begins. A vector of weights, $\vec{\mc{W}}_{merge}$, is computed for each cluster, the elements of which are the weights, $\mc{W}_{merge}^{j}$, for the cluster, $i$, to merge with a cluster $j$ (for $i,j \in [1,...,n]$ , $i \neq j$). Weights are defined such that

\begin{equation}{
\mc{W}_{merge}^{j}(\alpha,\beta) = \begin{cases}
(E_{j})^{\alpha} (d_{i,j})^{\beta} & \text{if $\Omega_{reject} = 0$},\\
0 & \text{otherwise}.
\end{cases} \quad ,
}\label{clusterMergingWeights1EQ} \end{equation}

\noindent where $E_{j}$ is the total energy of cluster $j$, and $d_{i,j}$ is the distance between the two clusters. The symbols $\alpha$ and $\beta$ are parameters of the weighting process and $\Omega_{reject}$ is a rejection condition. Clusters are arranged according to their energy in an ascending order, and $\vec{\mc{W}}_{merge}$ is then computed for each cluster. A cluster is merged with the partner ${j}$ for which $\mc{W}_{merge}^{j} \in \vec{\mc{W}}_{merge}$ is maximal. In the case that $\left\{ \forall j \in n : \mc{W}_{merge}^{j} = 0 \right\}$, no merging is performed.

The rejection condition is determined according to two criteria. The first is that the two candidates must not be too far apart, and the second is that the energy density of the merged cluster must be high, close to its cluster-center. \autoref{NNclustersFIG2} (\autoref{nn2DclustersSPlayersSUBSEC}) illustrates this point well. Had clusters one and two been mistakenly merged, the cluster-center would have ended up being in the space between the two, where the energy density is zero. The rejection condition may be written as

\begin{equation}{
\begin{gathered}
\Omega_{reject}(d_{min},d_{sum}) = 0\\
\mathrm{if} \quad d_{i,j} < d_{min} \quad \mathrm{or} \quad \mc{E}_{cl}(d_{sum}) \ge E_{merged} \, ,
\end{gathered}
}\label{clusterMergingRejectionConditionEQ}\end{equation}

\noindent where $d_{min}$ is the minimal separation distance between two clusters, $E_{merged}$ is the total energy of the merged cluster, and $\mc{E}_{cl}$ is the amount of energy contained within a distance $d_{sum}$ of the merged cluster-center. 

The best performance of the algorithm was achieved using:

\begin{equation}
\begin{array}{ccl}
d_{min}           & = & R_{\mc{M}} \quad \mathrm{and}\\
d_{sum}           & = & \frac{1}{2} R_{\mc{M}} \, .
\end{array} 
\label{clusterMergingRejectionParametersEQ} \end{equation}

Each cluster is defined as either \emph{small} or \emph{large}, according to the number of cells which the cluster holds, compared to the total number of hits in the layer. Small clusters are defined as having no more than $10\%$ of the hits in the layer, and large clusters as having no less than $15\%$. Merging is attempted twice. In the first pass, small clusters attempt to merge with large clusters, and in the second, large clusters attempt to merge with the small. In this way the merging process becomes smoother, as the addition of large clusters to small tends to change the cluster-center of the small cluster much more than that of the large. It was also found that for the merging of small clusters with large, the distance between the pair is more important than the energy of the large cluster. For the case of the merging of large clusters to small the conclusion was that the two factors are equally important. The optima of values for the weighting constants, $\alpha$ and $\beta$, for merging small clusters with large ($\mathrm{1^{st}}$ pass) and for merging large clusters with small ($\mathrm{2^{nd}}$ pass) are, therefore,

\begin{align}
\mathrm{\underline{1^{st}~pass:}} \quad & \left\{
\begin{array}{ccc}
\alpha & = & 1 \\
\beta  & = & -3
\end{array} \right. \quad \mathrm{and} \label{clusterMergingWeightConstants1EQ} \\
\mathrm{\underline{2^{nd}~pass:}} \quad &
\left\{
\begin{array}{ccc}
\alpha & = & 1 \\
\beta  & = & -1
\end{array} \right. \quad . \label{clusterMergingWeightConstants2EQ}
\end{align}

\subsection{Clustering of the Low-Energy Cells}

Once the cluster-merging is over, the low-energy cells, which were discarded previously, are incorporated into existing clusters. This is done using weights computed according to the inverse of the distance between the hit in question and each of the clusters in the layer. Weights are, therefore, determined such that

\begin{equation}{
\mc{W}_{merge}^{j} = (d_{i,j})^{-1},
}\label{clusterMergingWeights2EQ} \end{equation}

\noindent which is equivalent to \autoref{clusterMergingWeights1EQ} with parameters $\alpha =  0$ and $\beta = -1$, and no rejection condition.

\section{Global Clustering}

The basic actions taken in the global-clustering phase are as follows:

\begin{enumerate}

\item
fixing the number of global-clusters according to a frequency count in the shower-peak layers;

\item
re-clustering of hits from excess clusters where there are too many layer-clusters, and discarding of all clusters from layers where there are too few layer-clusters;

\item
creation of virtual-clusters (with a given center-position and size) in non-shower-peak layers, and in those where the layer-clusters were discarded;

\item
clustering of all remaining hits into the virtual-clusters, which are then identified with global-clusters.

\end{enumerate}

The most important stage of the clustering algorithm is the determination of the number of reconstructed showers. The aftermath of  clustering in the shower-peak layers is several collections of two-dimensional hit aggregates, the number of which, $n_{cl}(\ell)$, varies from layer to layer. The final number of showers, $N_{cl}$, is then determined as the most frequent value of $n_{cl}(\ell)$ from the collections in the shower-peak layers.

The first step in the global-clustering procedure is calculating the center-position of the global-clusters. Since layers for which $n_{cl}(\ell) \neq N_{cl}$ introduce ambiguities, they are temporarily not considered. In all \emph{accepted-layers}, where $n_{cl}(\ell) = N_{cl}$,  matching ensues between the layer-clusters and the global-clusters. This is done by comparing the two-dimensional distance between the centers of layer-cluster pairs in different layers, and choosing the minimally separated pairs. With all clusters matched, the position of the global-clusters is computed using \autorefs{logWeighClustering1EQ} and \ref{logWeighClustering2EQ}, where $E_{cl}$ now refers to the total energy of the global-cluster.

Turning back to the shower-peak layers for which $n_{cl}(\ell) \neq N_{cl}$, two situations are possible. For layers where $n_{cl}(\ell) > N_{cl}$, the $N_{cl}$ best matches are made between the layer clusters and the global clusters. The remaining $\left (n_{cl}(\ell) - N_{cl} \right )$ clusters are disbanded, and their hits are associated with one of the $N_{cl}$ remaining clusters. Once again, this is done using \autoref{clusterMergingWeights2EQ}.

Regarding shower-peak layers where $n_{cl}(\ell) < N_{cl}$, all information gained by the algorithm is ignored. For these and for the non shower-peak layers clustering is performed in the following manner: firstly, shower-centers are fixed in all un-clustered layers. This is done by fitting a straight line through the centers of layer-clusters in the accepted layers, and extrapolating the position onto the rest. In this way, each of the non-accepted-layers now has $N_{cl}$ \emph{virtual-clusters}, which correspond to the different global-clusters. Secondly, in each of the non-accepted layers separately, individual hits are associated with one of the virtual-clusters.

Adding hits to the virtual-clusters is done in several steps. In the first stage, the \emph{virtual-cluster range} is estimated to be the area around the virtual-center, which is spanned by the effective layer-radius of the layer, $r_{eff}(\ell)$ (\autoref{effectiveLayerRadiusEQ} in \autoref{characteristicsOfShowersSEC}). All of the hits inside virtual-cluster ranges are added to their respective virtual-clusters. Cases where a hit is within the range of more than one virtual-cluster are resolved by a proximity test to the virtual-center positions. Once a core of hits has been added to each virtual-cluster, the center-position of each one is computed in the usual manner. All of the remaining un-clustered hits are finally associated with one of the now \emph{real-clusters} by means of \autoref{clusterMergingWeights2EQ}. It is important to perform this process in these two phases, i.e., first to gather hits from the close proximity of the virtual-center, and only then to merge all of the remaining hits. The reason for this is that the real-cluster center, the position of which is mostly determined by its core hits, may be slightly different than the extrapolated virtual-center. This difference tends to bias the entire procedure in a given layer, if it is not accounted for.

A schematic representation of the global-clustering phase is depicted in \autoref{globalClusteringFIG} (\autoref{globalClusteringSUBSEC}).

\section{Corrections Based on the Energy Distribution}

By this point all of the hits in the calorimeter have been integrated into one of the global-clusters. Before moving on, it is beneficial to make sure that the clusters have the expected characteristics.

\subsection{Energy Density Test}

EM shower development in LumiCal has been described in \autoref{characteristicsOfShowersSEC}. Accordingly, one would expect that $90\%$ of the energy of a cluster would be found within one Moli\`ere radius, $R_{\mc{M}}$, of its center. While statistically this is true, on a case-by-case basis fluctuations may happen, and thus this should not be taken as a hard rule. One can define, instead, the following general conditions:

\begin{gather}{
\sum_{cl}^{N_{cl}}{\mc{E}_{cl}(R_{\mc{M}})} \ge \sum_{cl}^{N_{cl}}{E_{cl}} \cdot \mc{P}_{tot} \, , \quad \mathrm{and} \label{molRadCondition2EQ} \\
\mc{E}_{cl}(R_{\mc{M}} \cdot \mc{P}_{cl}) \ge E_{cl} \cdot \mc{P}_{tot}  \, , \label{molRadCondition1EQ}
}\end{gather}

\noindent where $N_{cl}$ is the number of clusters in the layer, $\mc{E}_{cl}(d)$ is the energy in a cluster, $cl$, within a distance  $d$ of its center, $E_{cl}$ is the total energy of cluster $cl$, and $\mc{P}_{tot}$ and $\mc{P}_{cl}$ are fractions.

As an initial check, we evaluate \autoref{molRadCondition2EQ} with the constant $\mc{P}_{tot} = 0.9$, meaning that the energy around each of the reconstructed clusters is integrated around $R_{\mc{M}}$ of its center, and the sum of these is compared to $90\%$ of the total energy found in the detector. If the condition is not upheld, an attempt is made to re-cluster and produce \textit{profile-clusters}, based on the profile image of the energy distribution (see \autoref{showerEnergyProfileFIG} in \autoref{characteristicsOfShowersSEC}). The idea is that the core of the hits of a shower tends to stand-out against the background of low-energy deposits. By stripping the noise away, one is able to trace the propagation of the particle, which initiated the shower, through the different layers, and conduct a rough position reconstruction. The following steps are taken:

\begin{enumerate}
\item \label{profileSteps1}
Hit energies from all layers are integrated along the longitudinal direction, resulting in a two-dimensional energy distribution.

\item \label{profileSteps2}
The NNC method (\autoref{nn2DclustersSPlayersSUBSEC}) is utilized in order to build the profile-clusters using all of the available hits.

\item \label{profileSteps3}
The condition of \autoref{molRadCondition1EQ} is evaluated with the constants $\mc{P}_{cl} = 0.4$ and $\mc{P}_{tot} = 0.8$.

\item \label{profileSteps4}
If the condition of step \ref{profileSteps3} is not upheld, then a cut on hit-energy is made, removing some percentage of the lowest-energy hits.

\item \label{profileSteps5}
Steps \ref{profileSteps2}-\ref{profileSteps4} are repeated until either the condition of step \ref{profileSteps3} is met, or the number of remaining hits after the progressive energy cuts drops below a certain threshold, in which case the construction of profile-clusters has failed.

\item \label{profileSteps6}
If the condition of step \ref{profileSteps3} is met, then all of the discounted low-energy hits are taken back into consideration, and \autoref{molRadCondition1EQ} is evaluated with $\mc{P}_{tot} = 0.9$. If this is upheld, the profile-clusters are accepted.
\end{enumerate}

If the building of profile-clusters is successful, the resulting shower-centers of the profile-clusters are used in order to construct global-clusters. This is done on a layer-to-layer basis, by means of the weights given in \autoref{clusterMergingWeights2EQ}. The new global-clusters are then compared to the original global-clusters by computing for each cluster set the energy density, $\rho_{\mc{E}}$, which is defined as

\begin{equation}{
\rho_{\mc{E}} \equiv \frac{\sum_{cl}{\mc{E}_{cl}(R_{\mc{M}})}}{\sum_{cl}{E_{cl}}} \, .
}\label{molRadCondition3EQ} \end{equation}

\noindent If the profile-clusters have a higher energy density, then they are an improvement on the original clusters, and are kept instead of the originals.

In addition to the procedure described above, each cluster is also examined separately, with a relaxed condition \ref{molRadCondition1EQ}, using $\mc{P}_{cl} = 1$ and $\mc{P}_{tot} = 0.5$. Clusters that fail the test are disbanded, and their hits are assigned to other existing clusters using  \autoref{clusterMergingWeights2EQ}.

\subsection{Unfolding of Mixed Clusters \label{clusterMixingSEC}}

\begin{figure}[htp]
\begin{center}
\includegraphics[width=.49\textwidth]{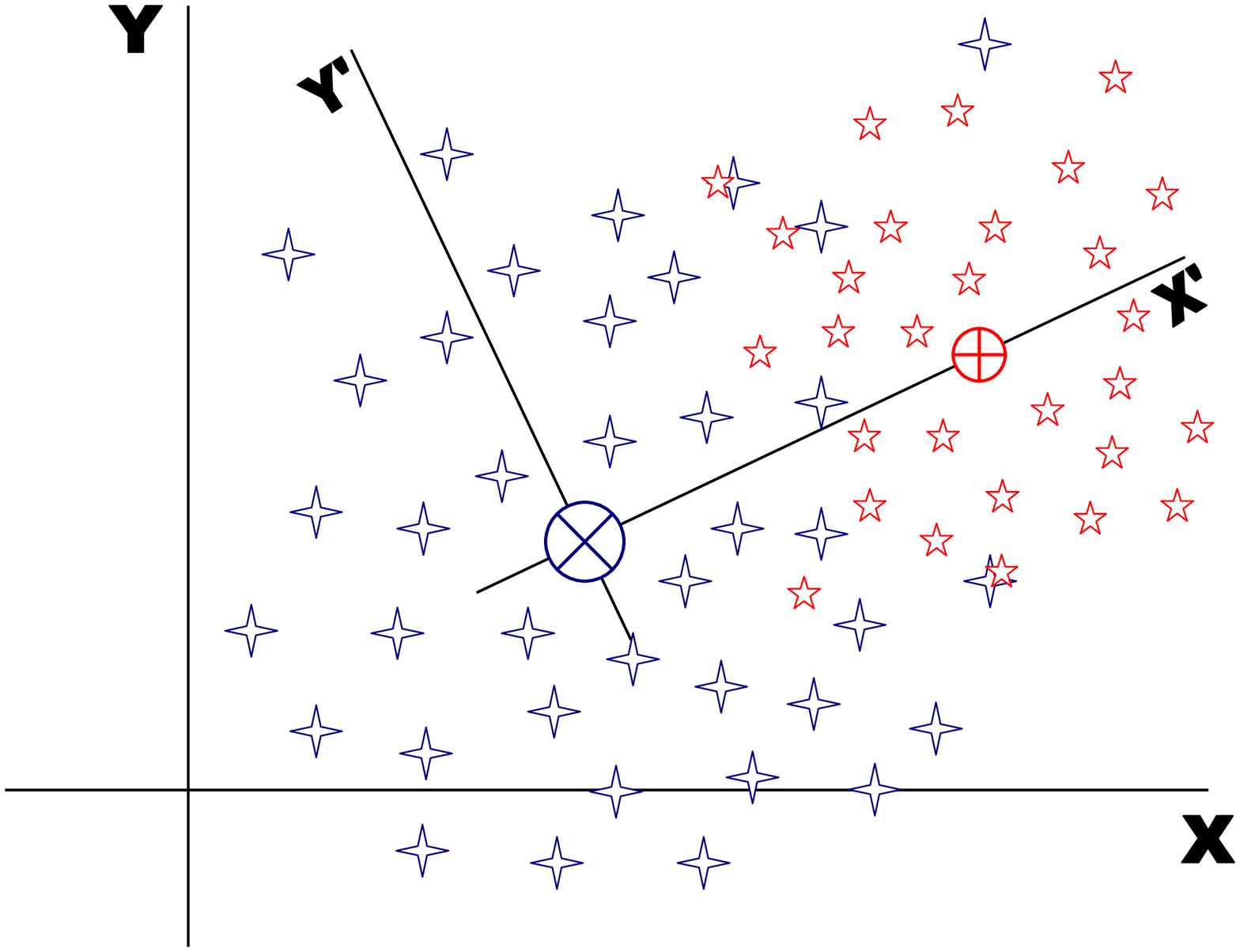}
\caption{\label{clusterMixingAxisFIG} Schematic representation of the definition of the projection coordinate system, $(x',y')$, between a pair of intermixed clusters.}
\vspace{20pt}
\includegraphics[width=.49\textwidth]{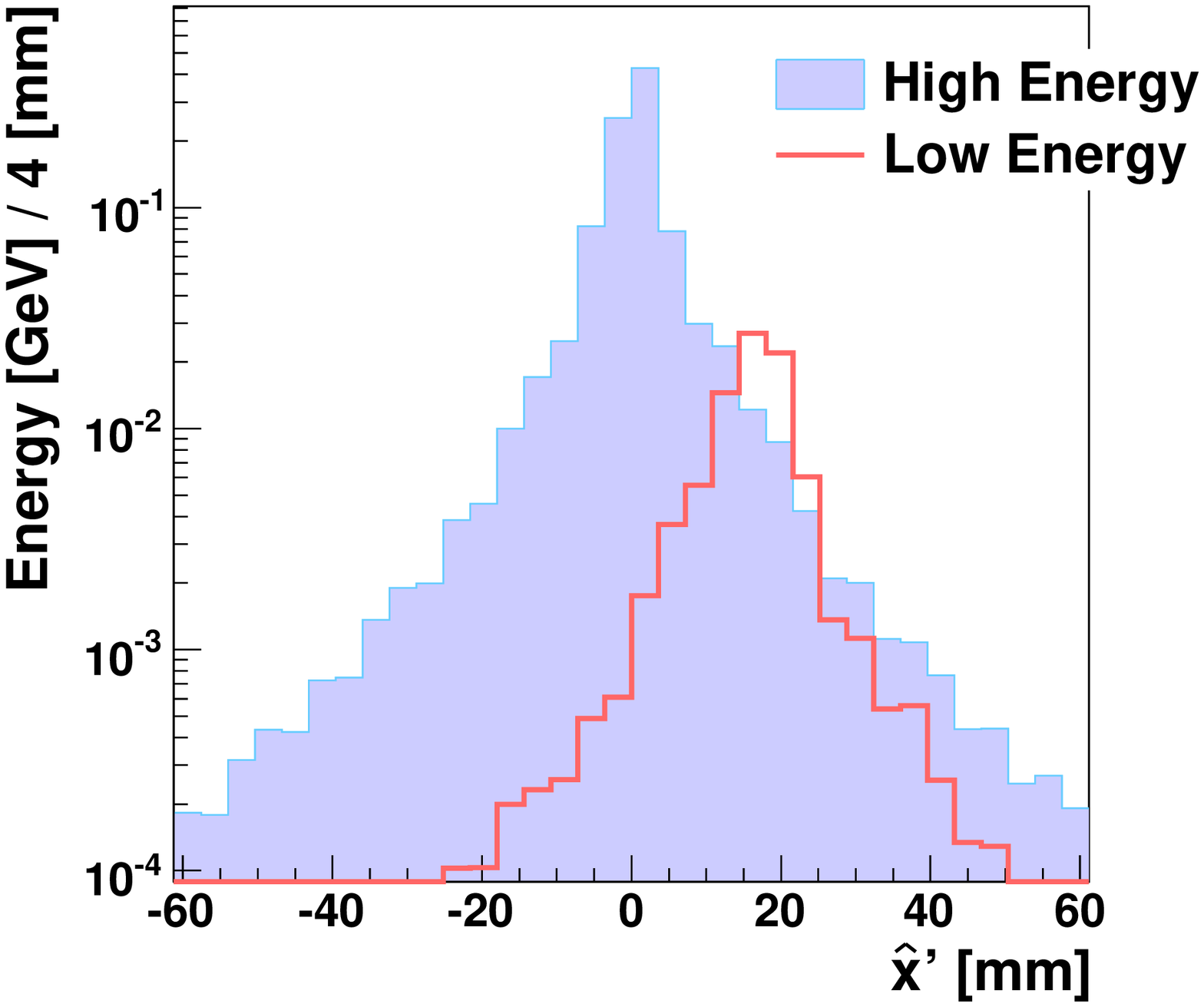}
\caption{\label{clusterMixing1FIG} Projected energy distribution for a sample of high and low-energy showers, initiated by electrons of 230 and 20~GeV, respectively. The distributions are normalized according to the high-energy shower distribution. The two particles are evenly distributed on the face of LumiCal, and are generated with a constant separation distance of 20~mm on the first layer. The coordinates of the high-energy shower-center are translated to $(x',y') = (0,0)$, and those of the low-energy shower to a positive value along $\hat{x}'$, according to the separation between the two showers.}
\end{center}
\end{figure}

Another modification that can be made in the aftermath of the clustering procedure, is allocation of hits for mixed cluster pairs. A convenient way to define the intermixing of a pair of clusters, is to project the hits belonging to each cluster on the axis defined by the cluster-centers. The projection axis is schematically represented in \autoref{clusterMixingAxisFIG}. All of the hits which were represented in the original coordinate system, $(x,y)$, are integrated along the $\hat{y}'$ direction of the \textit{projection coordinates}, $(x',y')$.

The distribution of the projected energy for a sample of cluster pairs is shown in \autoref{clusterMixing1FIG}. Several hundred events were simulated, and the projected energy distributions are all normalized according to the high-energy shower distribution, so as to keep the ratio between distributions constant. The energy of the initiating particles for the large and small-energy showers is 230 and 20~GeV, respectively. Both showers were simulated according to a flat distribution along the detector face, while keeping the separation distance between the two at a constant $d_{pair} = 20$~mm ($\sim 1.5 \cdot R_{\mc{M}}$) on the first layer. This way, the effects of different inclinations with regard to cell pitch, as well as the shower-shape distortions that are expected due to changing cell size, are smoothed out. In order to produce a clean distribution, the coordinates of the high-energy shower-center were always translated to $(x',y') = (0,0)$, and those of the low-energy shower to a positive value along the projection axis, $\hat{x}'$.

The distributions in \autoref{clusterMixing1FIG} share a \textit{mixing-range}, that extends from the peak of one shower to the next $(0 < x' < d_{pair})$. This, of course, is due to the fact that the un-projected distance between the pair is one and a half Moli\`ere radius, beyond which the energy content of each shower is negligible. This fact poses an opportunity for resolving the mixing of the hits of the two showers. The area of negative $x'$ holds a clean sample of hits, which have energy contributions from the high-energy shower alone. For $x' > d_{pair}$, only the low-energy shower contributes significantly. It is, therefore, possible to estimate the amount of energy  deposits owing to each shower inside the mixing-range, by distracting from the mixed distribution either one of the \textit{clean distributions}, $(x' < 0)$ or $(x' > d_{pair})$.

For this to work, one has to assume that the shape of the original (un-projected) energy distributions is smooth. This is not always the case, though, as fluctuations in the development of showers tend to distort their shape. Other sources of distortions are the inclination of the projected axis, $\hat{x}'$, with relation to the cells of LumiCal, and the radial dependence of cell sizes, as mentioned above.

The fact that the reference for corrections is the projected energy distribution also poses problems, as one would like to know the original two-dimensional distribution of energy for each cluster. The proper assignment of individual cell energies, as opposed to a global correction on the integrated cluster energy, is important. The reason for this is that the amount of energy in each individual cell determines its weight in the position reconstruction procedure (\autoref{logWeighClustering2EQ}). For the low-energy cluster especially, wrongly assigned cells of high energy, which comprise a significant amount of the shower energy, may bias the position of the reconstructed cluster-center.

While the ambiguities mentioned here prevent from drawing global correction factors directly from the distributions of \autoref{clusterMixing1FIG}, on a case-by-case basis, some degree of unfolding is possible. The way this is done is by projecting the hits of every cluster pair in the $(x',y')$ system, as described above, and then determining weights according to the ratio of the energy distributions at $x' > 0$ and $x' < 0$. These weights are defined as

\begin{equation}{
\mc{W}_{mix}(b_{i}) \equiv \frac{E_{proj}(b_{i}^{p}) - E_{proj}(b_{i}^{n})}{E_{proj}(b_{i}^{p})} ,
}\label{clusterMixingWeights1EQ} \end{equation}

\noindent where

\begin{equation}{
E_{proj}(b_{i}^{p \, , \, n}) = \sum_{j \, \in \, b_{i}^{p \, , \, n}}{\mc{E}_{proj}^{j}} \, .
}\label{clusterMixingWeights2EQ} \end{equation}

\noindent The weights, $\mc{W}_{mix}(b_{i})$, are calculated for bins, $b_{i}$, along the projection axis for $(0 < x' < d_{pair})$, where $d_{pair}$ is the distance between the centers of the two clusters. Energy is integrated from all hit contributions, $\mc{E}_{proj}^{j}$, in bin pairs of equal distance from the origin $(x' = 0)$ along the positive ($b_{i}^{p}$) and negative ($b_{i}^{n}$) directions. Bin sizes were chosen to be $5\%$ of $R_{\mc{M}}$.

Returning to the un-projected energy distributions, each hit that would have satisfied $(0 < x' < d_{pair})$ had a projection been made, is split. A fraction of the hit energy, determined by $\mc{W}_{mix}(b_{i})$, is associated with each of the two clusters. The low-energy cluster looses $\mc{W}_{mix}(b_{i})\%$ of the energy of each hit, and the high-energy cluster gains the same amount. Splitting of hits is only performed for cases where $\mc{W}_{mix}(b_{i}) > 0$, i.e., where more energy is projected in the mixing-range of the two clusters, than in the respective clean range of the high-energy cluster.

The option of comparing the distribution of energy in the mixing-range to that of the clean range of the low-energy cluster ($x' > d_{pair}$) was also considered. This did not produce stable results, though, due to the fact that low-energy clusters tend to fluctuate more in shape, and are also more susceptible to cell geometry changes.

\begin{figure}[htp]
\begin{center}
\subfloat[]{\label{clusterMixing2FIG1}\includegraphics[height=0.36\textheight]{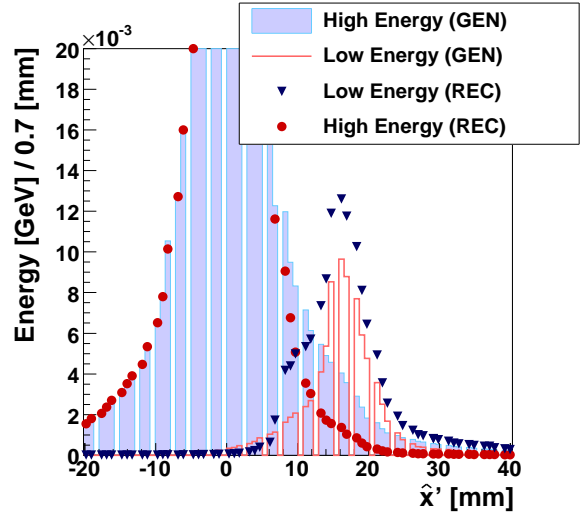}} \\
\subfloat[]{\label{clusterMixing2FIG2}\includegraphics[height=0.36\textheight]{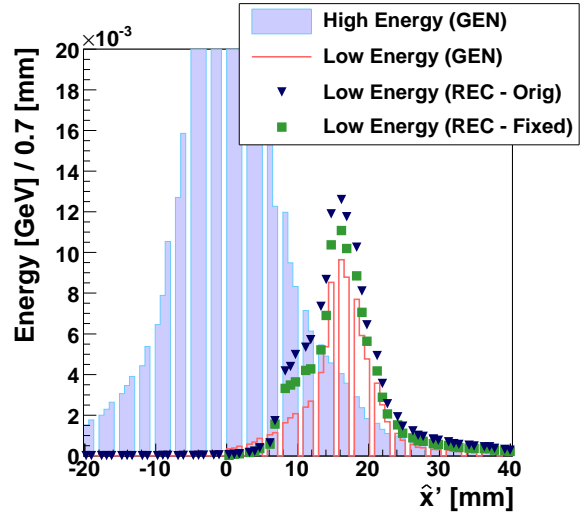}}
\caption{\label{clusterMixing2FIG}\Subref{clusterMixing2FIG1} Distributions of the energy in the projection coordinate system for the high-energy and low-energy showers/clusters. Both the simulation-truth shower (GEN) and the reconstructed cluster (REC) distributions are shown. No unfolding of the distributions has been performed. \Subref{clusterMixing2FIG2} Distributions of the projected energy of the original (REC - Orig), and of the unfolded (REC - Fixed) low-energy reconstructed clusters. The distributions of the simulation-truth (GEN) are also shown for the low and high-energy showers. In both \Subref{clusterMixing2FIG1} and \Subref{clusterMixing2FIG2} the distributions are normalized according to the high-energy shower distribution.}
\end{center}
\end{figure}

\Autoref{clusterMixing2FIG} shows the projected energy distributions of the generated (simulation-truth) and reconstructed (results of the clustering) energy distributions before and after the unfolding of hits. The sample is the same as the one which was shown in \autoref{clusterMixing1FIG}, and is normalized in a similar manner. It is apparent that the splitting of hit energies improves the distribution of energy between the two clusters. It may also be noticed from the figure that the energy of low-energy clusters tends to be overestimated when mixing is in play, which explains why no splitting of hits is performed for $\mc{W}_{mix}(b_{i}) < 0$. 

\begin{figure}[htp]
\begin{center}
\subfloat[]{\label{corectedAndUncorectedEngyFIG1}\includegraphics[width=.49\textwidth]{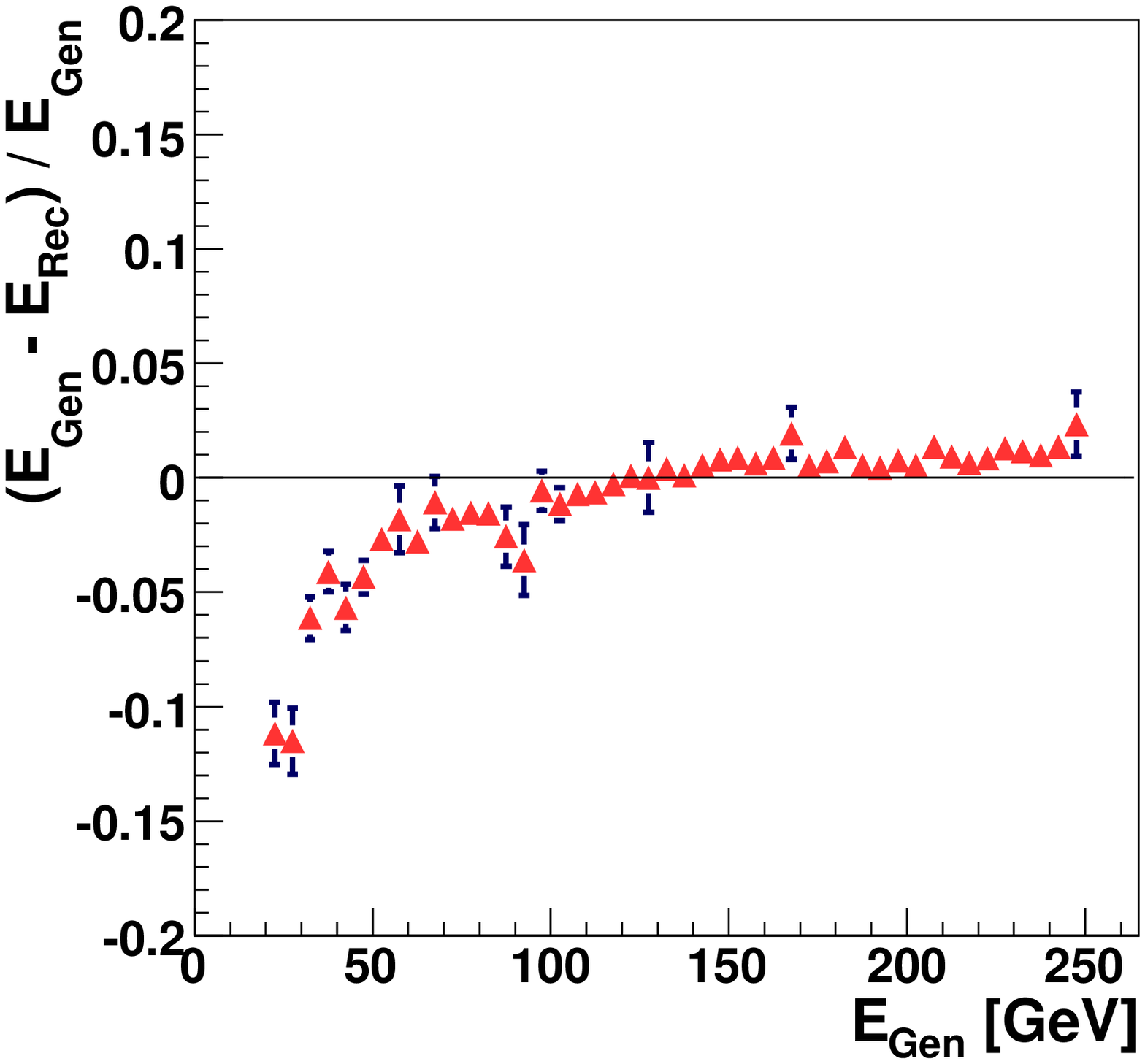}} 
\subfloat[]{\label{corectedAndUncorectedEngyFIG2}\includegraphics[width=.49\textwidth]{pics/clustering/engyRecMinGenEvtByEvt}}
\caption{\label{corectedAndUncorectedEngyFIG}The normalized difference between the energy of generated showers (GEN) and their respective reconstructed clusters (REC), as a function of the energy of the generated shower, before \Subref{corectedAndUncorectedEngyFIG1} and after \Subref{corectedAndUncorectedEngyFIG2} the unfolding procedure.}
\end{center}
\end{figure}

The improvement of the energy reconstruction, due to the unfolding procedure, can be quantified by studying its affect on a physics sample of Bhabha events (see \autoref{clusteringPhysicsSample}). \Autoref{corectedAndUncorectedEngyFIG} shows the normalized difference between the energy of reconstructed clusters and their respective generated showers, as a function of the energy of the generated shower, before and after the unfolding procedure. The reconstructed clusters, which are taken into account here, belong to the effective Bhabha cross-section for which $E_{low} \ge 20$~GeV and $d_{pair} \ge R_{\mc{M}}$.

\bibliographystyle{ieeetr}  \bibliography{a}   

\begin{thebibliography}{10}

\bibitem{introductionBIB6}
``{T}he {I}nternational {L}inear {C}ollider.'' {URL:}
  \url{http://www.linearcollider.org/cms/}.

\bibitem{introductionBIB1}
J.~{B}agger \textit{et al.}, ``Discovering the quantum universe: The role of
  particle colliders,'' {\em Report of the {DOE/NSF} High Energy Physics
  Advisory Panel}, 2006.

\bibitem{introductionBIB2}
{ATLAS Collaboration}, ``{ATLAS} physics {TDR},'' {\em
  CERN-LHCC-99-14~,~CERN-LHCC-99-15}, 2006.
\newblock {URL:}
  \url{http://atlas.web.cern.ch/Atlas/GROUPS/PHYSICS/TDR/TDR.html}.

\bibitem{introductionBIB3}
{CMS Collaboration}, ``{CMS} physics {TDR},'' {\em CERN/LHCC/2006-021}, 2006.
\newblock {URL:}
  \url{http://atlas.web.cern.ch/Atlas/GROUPS/PHYSICS/TDR/TDR.html}.

\bibitem{introductionBIB7}
R.~{H}euer \textit{et al.}, ``{P}arameters for the {L}inear {C}ollider.''
  {URL:} \url{http://www.fnal.gov/directorate/icfa/LC_parameters.pdf}, 2003.

\bibitem{introductionBIB8}
{N}an~{P}hinney \textit{et al.}, ``{ILC} {R}eference {D}esign {R}eport {V}olume
  3 - {A}ccelerator,'' {\em arXiv:0712.2361v1}.

\bibitem{revisedDetectorModelBIB4}
T.~C. {A}wes \textit{et al.}, ``A simple method of shower localization and
  identification in laterally segmented calorimeters,'' {\em Nucl. Inst.
  Meth.}, vol.~A311, p.~130, 1992.

\bibitem{introductionBIB25}
``{The Large Detector Concept}.'' {URL:} \url{http://www.ilcldc.org/}.

\bibitem{introductionBIB26}
``{The Iternational Large Detector}.'' {URL:} \url{http://www.ilcild.org/}.

\bibitem{introductionBIB5}
H.~{A}bramowicz \textit{et al.}, ``Instrumentation of the very forward region
  of a linear collider detector,'' {\em IEEE transactions of Nuclear Science},
  vol.~51, p.~2983, 2004.

\bibitem{introductionBIB27}
S.~{J}adach, ``{Theoretical error of the luminosity cross section at LEP},''
  {\em {hep-ph/0306083}}, 2003.
\newblock {URL:} \url{http://arxiv.org/abs/hep-ph/0306083}.

\bibitem{introductionBIB9}
G.~{A}ltarelli \textit{et al.} {\em {P}hys. {L}ett. {B}}, vol.~349, p.~145,
  1995.

\bibitem{introductionBIB10}
R.~{H}awkings and K.~{M\"o}nig {\em {EPJ}direct}, vol.~C 8, p.~1, 1999.

\bibitem{introductionBIB28}
K.~{M\"o}nig, ``{Physics Needs for the Forward Region}.'' Talk given at the
  Zeuthen FCAL meeting, Aug. 2004.

\bibitem{introductionBIB15}
M.~{C}affo \textit{et al.}, ``{B}habha {S}cattering,'' {\em in Z Physics at
  LEP1}, vol.~CERN Report 89-08, p.~1, 1989.
\newblock {URL:}
  \url{http://documents.cern.ch/cgi-bin/setlink?base=cernrep&categ=Yellow_Repo%
rt&id=89-08_v1}.

\bibitem{introductionBIB11}
T.~{B}echer and K.~{M}elnikov, ``Two-loop {QED} corrections to {B}habha
  scattering,'' {\em arXiv:0704.3582}.

\bibitem{introductionBIB12}
S.~{A}ctis, M.~{C}zakon, J.~{G}luza, and T.~{R}iemann, ``Two-loop fermionic
  corrections to massive {B}habha scattering,'' {\em arXiv:0704.2400}.

\bibitem{introductionBIB13}
A.~A. {P}enin, ``Two-loop photonic corrections to massive {B}habha
  scattering,'' {\em Nucl. Phys. B}, vol.~734, p.~185, 2006.
\newblock (arXiv:hep-ph/0508127).

\bibitem{introductionBIB14}
M.~{C}zakon, J.~{G}luza, and T.~{R}iemann, ``The planar four-point master
  integrals for massive two-loop {B}habha scattering,'' {\em Nucl. Phys. B},
  vol.~751, p.~1, 2006.
\newblock (arXiv:hep-ph/0604101).

\bibitem{introductionBIB16}
C.~{R}imbault \textit{et al.}, ``Impact of beam-beam effects on precision
  luminosity measurements at the {ILC},'' {\em JINST 2 P09001}, 2007.
\newblock {URL:} \url{http://www.iop.org/EJ/abstract/1748-0221/2/09/P09001}.

\bibitem{introductionBIB17}
M.~N. {F}rary and D.~J. {M}iller, ``{M}onitoring the {L}uminosity {S}pectrum,''
  {\em DESY-92-123A}, p.~379, 1991.
\newblock {URL:} \url{http://www.hep.ucl.ac.uk/lc/documents/frarymiller.pdf}.

\bibitem{introductionBIB18}
K.~{M\"o}nig, ``Measurement of the {D}ifferential {L}uminosity using {B}habha
  events in the {F}orward-{T}racking region at {TESLA},'' {\em
  LC-PHSM-2000-60-TESLA}, 2000.
\newblock {URL:} \url{http://www-flc.desy.de/lcnotes/}.

\bibitem{introductionBIB19}
``{TELSA} {T}echnical {D}esign {R}eport,'' {\em ECFA-2001-209}, 2001.
\newblock {URL:} \url{http://arxiv.org/PS_cache/hep-ph/pdf/0106/0106315v1.pdf}.

\bibitem{introductionBIB21}
A.~{S}tahl, ``{Luminosity Measurement via Bhabha Scattering: Precision
  Requirements for the Luminosity Calorimeter},'' {\em {LC-DET-2005-004}},
  2005.
\newblock {URL:}
  \url{http://www-flc.desy.de/lcnotes/notes/LC-DET-2005-004.ps.gz}.

\bibitem{introductionBIB22}
{H.Abramowicz, R. Ingbir, S. Kananov, A. Levy}, ``{A Luminosity Detector for
  the International Linear Collider},'' {\em {LC-DET-2007-006}}, 2007.
\newblock {URL:}
  \url{http://www-flc.desy.de/lcnotes/notes/LC-DET-2007-006.pdf}.

\bibitem{introductionBIB23}
J.~{B}locki \textit{et al.}, ``{The proposed design of the silicon sensors for
  the LumiCal, $\&$ Silicon sensors for the LumiCal - prototype design},'' {\em
  {EUDET-Memo-2007-09 , $\&$ EUDET-Memo-2007-47}}, 2007.
\newblock {URL:} \url{http://www.eudet.org/e26/e28}.

\bibitem{introductionBIB24}
J.~{B}locki \textit{et al.}, ``{A proposal for the mechanical design of the
  LumiCal detector},'' {\em {Report No. 1990/PH , IFJ PAN, Cracow, Poland}},
  2006.

\bibitem{calorimetryBIB1}
E.~{S}egre {\em {Nuclei and Particles, New York, Benjamin}}, p.~65 ff, 1964.

\bibitem{introductionBIB20}
W.-M.~t. {Yao}, ``{Review of Particle Physics},'' {\em {Journal of Physics G}},
  vol.~33, 2006.
\newblock {URL:} \url{http://pdg.lbl.gov}.

\bibitem{calorimetryBIB2}
B.~{R}ossi {\em {High Energy Particles, Prentice-Hall, Inc., Englewood Cliffs,
  NJ}}, 1952.

\bibitem{revisedDetectorModelBIB1}
``{MOKKA} - a detailed {GEANT4} detector simulation for the {F}uture {L}inear
  {C}ollider.'' {URL:}
  \url{http://polywww.in2p3.fr/geant4/tesla/www/mokka/mokka.html}.

\bibitem{revisedDetectorModelBIB2}
J.~{A}llison \textit{et al.}, ``{GEANT4} developments and applications,'' {\em
  IEEE Transactions on Nuclear Science}, vol.~53, No. 1, pp.~270--278, Feb.
  2006.

\bibitem{revisedDetectorModelBIB3}
``{MARLIN} - a {C}++ software framework for {ILC} software.'' {URL:}
  \url{http://ilcsoft.desy.de/portal/software_packages/marlin/index_eng.html}.

\bibitem{revisedDetectorModelBIB6}
S.~{J}adach, W.~Placzek, and B.~F.~L. {W}ard, ``{BHWIDE} 1.00: O(alpha) {YFS}
  exponentiated {M}onte {C}arlo for {B}habha scattering at wide angles for
  {LEP1/SLC} and {LEP2},'' {\em Phys. Lett.}, vol.~B390, pp.~298--308, 1997.
\newblock {URL:} \url{http://arxiv.org/pdf/hep-ph/9608412}.

\bibitem{constraintsElectronicsBIB1}
M.~{B}attaglia \textit{et al.}, ``Contrasting {S}upersymmetry and {U}niversal
  {E}xtra {D}imensions at colliders.'' {URL:}
  \url{http://arxiv.org/pdf/hep-ph/0507284}.

\bibitem{revisedDetectorModelBIB7}
A.~{S}eryi and B.~{P}arker, ``{Compensation of the effects of detector solenoid
  on the vertical beam orbit in NLC},'' {\em {Phys.Rev.ST Accel.Beams}},
  vol.~8, p.~041001, 2005.
\newblock {URL:}
  \url{http://www-project.slac.stanford.edu/lc/ilc/TechNotes/LCCNotes/PDF/LCC-%
143.pdf}.

\bibitem{revisedDetectorModelBIB8}
D.~{S}chulte {\em {TESLA-97-08}}, 1996.

\bibitem{revisedDetectorModelBIB5}
C.~{G}rah, ``Pair background and the forward region,'' {\em Talk given at the
  October 2007 {FCAL Collaboration Meeting, LAL-Orsay, France}}.
\newblock {URL:} \url{http://events.lal.in2p3.fr/conferences/FCAL07/}.

\end{thebibliography}

\end{document}